\documentclass[fleqn,usenatbib]{mnras}

\usepackage{newtxtext,newtxmath}

\usepackage[T1]{fontenc}
\usepackage{ae,aecompl}

\usepackage{graphicx}	
\usepackage{amsmath}	
\usepackage{amssymb}	

\usepackage{textcomp}
\usepackage{hyperref}
\usepackage{breakurl}
\usepackage[center]{caption}
\usepackage{xcolor}
\usepackage{tablefootnote}

\newcommand{\urlwofont}[1]{\urlstyle{same}\url{#1}}
\usepackage{color,xcolor,ulem}

\newcommand{\lara}[1]{{\langle#1\rangle}}
\newcommand{\kms}{km~s$^{-1}$}

\title[TDE candidate AT~2017gbl in a LIRG]{AT~2017gbl: a dust obscured TDE candidate in a luminous infrared galaxy}

\author[E. C. Kool et al.]{
E. C. Kool$^{1,2}$\thanks{E-mail: erik.kool@astro.su.se},
T. M. Reynolds$^{3}$,
S. Mattila$^{3}$,
E. Kankare$^{3}$,
M. A. P\'{e}rez-Torres$^{4}$,
\newauthor
A. Efstathiou$^{5}$,
S. Ryder$^{2}$,
C. Romero-Ca\~nizales$^{6, 7}$,
W. Lu$^{8}$,
T. Heikkil\"a$^{3}$,
\newauthor
G. E. Anderson$^{9}$,
M. Berton$^{10,11}$,
J. Bright$^{12}$,
G. Cannizzaro$^{13,14}$,
D. Eappachen$^{13,14}$,
\newauthor
M. Fraser$^{15}$,
M. Gromadzki$^{16}$,
P. G. Jonker$^{13,14}$,
H. Kuncarayakti$^{3,10}$,
P. Lundqvist$^{1}$,
\newauthor
K. Maeda$^{17}$,
R. M. McDermid$^{2}$,
A. M. Medling$^{18,19}$,
S. Moran$^{3}$,
A. Reguitti$^{20,21,22}$,
\newauthor
M. Shahbandeh$^{23}$,
S. Tsygankov$^{3,24}$,
V. U$^{25}$,
T. Wevers$^{26}$
\\
$^{1}$The Oskar Klein Centre, Department of Astronomy, Stockholm University, AlbaNova, SE-10691, Stockholm, Sweden\\
$^{2}$Department of Physics and Astronomy, Macquarie University, NSW 2109, Sydney, Australia\\
$^{3}$Tuorla observatory, Department of Physics and Astronomy, University of Turku, FI-20014 Turku, Finland\\
$^{4}$Instituto de Astrof\'{i}sica de Andaluc\'{i}a (CSIC), Glorieta de la Astronom\'{i}a s/n, E-18080 Granada, Spain\\
$^{5}$School of Sciences, European University Cyprus, Diogenes Street, Engomi, 1516 Nicosia, Cyprus\\
$^{6}$N\'ucleo de Astronom\'{\i}a de la Facultad de Ingenier\'{\i}a y Ciencias, Universidad Diego Portales, Av. Ej\'ercito 441, Santiago, Chile \\
$^{7}$IAA, Academia Sinica, 11F of Astronomy-Mathematics Building, AS/NTU No. 1, Sec. 4, Roosevelt Rd, Taipei 10617, Taiwan, R.O.C\\
$^{8}$TAPIR, Walter Burke Institute for Theoretical Physics, Mail Code 350-17, Caltech, Pasadena, CA 91125, USA\\
(continued at end of manuscript)
}

\date{Accepted XXX. Received YYY; in original form ZZZ}

\pubyear{2020}

\begin{document}
\label{firstpage}
\pagerange{\pageref{firstpage}--\pageref{lastpage}}
\maketitle

\begin{abstract}

We present the discovery with Keck of the extremely infrared (IR) luminous transient AT~2017gbl, coincident with the Northern nucleus of the luminous infrared galaxy (LIRG) IRAS~23436+5257. Our extensive multi-wavelength follow-up spans $\sim$900 days, including photometry and spectroscopy in the optical and IR, and (very long baseline interferometry) radio and X-ray observations. Radiative transfer modelling of the host galaxy spectral energy distribution and long-term pre-outburst variability in the mid-IR indicate the presence of a hitherto undetected dust obscured active galactic nucleus (AGN). The optical and near-IR spectra show broad $\sim$2000 km s$^{-1}$ hydrogen, He\,{\sc i} and O\,{\sc i} emission features that decrease in flux over time. Radio imaging shows a fast evolving compact source of synchrotron emission spatially coincident with AT~2017gbl. We infer a lower limit for the radiated energy of $7.3\times10^{50}$ erg from the IR photometry. An extremely energetic supernova would satisfy this budget, but is ruled out by the radio counterpart evolution. Instead, we propose AT~2017gbl is related to an accretion event by the central supermassive black hole, where the spectral signatures originate in the AGN broad line region and the IR photometry is consistent with re-radiation by polar dust. Given the fast evolution of AT~2017gbl, we deem a tidal disruption event (TDE) of a star a more plausible scenario than a dramatic change in the AGN accretion rate. This makes AT~2017gbl the third TDE candidate to be hosted by a LIRG, in contrast to the so far considered TDE population discovered at optical wavelengths and hosted preferably by post-starburst galaxies. 
\end{abstract}



\section{Introduction} \label{sec:intro}
Nuclear variability in galaxies is often attributed to the presence of an active galactic nucleus (AGN), where matter is accreted by a central supermassive black hole (SMBH). AGN are known to be intrinsically variable, where the amplitude and time-scale depend on the wavelength of observation. AGN typically show small-amplitude stochastic variability in brightness of <40\% in the optical \citep{kelly2009}, whereas in the mid-infrared (IR) AGN show larger and smoother variability on longer timescales of years to decades \citep{kozlowski2016}. However, an increasing number of nuclear outbursts are being observed that do not fit in this picture, showing large amplitude variability on a short timescale, both by galaxies with an AGN and by inactive galaxies. The interpretations of these events have included a tidal disruption event (TDE) of a star by a SMBH, or major changes in the accretion rate of a SMBH that result in changes in AGN spectra (changing look AGN, or CLAGN). Supernovae (SNe) have also been suggested, since (core-collapse) SNe are expected in the nuclear regions of starburst and luminous infrared galaxies (LIRGs, $L_{\text{IR}}$ > $10^{11}$ L$_{\odot}$) at rates a couple of orders of magnitude higher than in normal field galaxies \citep{perez2009,kankare2012,mattila2012,kool2018}.

TDEs were theoretically predicted over forty years ago \citep{hills1975,rees1988}, and are expected to give rise to a luminous flare typically peaking in the X-ray/UV/optical. Over the past ten years a number of optical TDE candidates have been discovered, often with concurrent detections at X-ray, UV, or radio wavelengths \citep{bade1996,gezari2012,zauderer2011,vanvelzen2020}. \citet{vanvelzen2018} found a roughly constant volumetric rate for these events for BHs with masses below 10$^{7.5}$ M$_{\odot}$, followed by a sharp drop for more massive black holes, which directly capture stars without a luminous flare. One notable feature of the population of optically discovered TDEs is that they seem to show a preference for E+A galaxies \citep{arcavi2014,french2016}, a class of post-starburst galaxies that are thought to be the result of a merger that occurred approximately 10$^{9}$ years previously \citep{dressler1983}. It has been suggested, based on simulations, that this overabundance is a result of the presence of a secondary (in-spiralling) SMBH enhancing the TDE rate by several orders of magnitude for a period of 10$^{4}$--10$^{5}$ years \citep{chen2009,cen2019} and as such is intrinsic to post-merger galaxies. However, based on a study of four TDE host galaxies, \citet{french2020} suggested high central stellar densities are a more important driver for increased TDE rates.

Most discovered TDEs seem to suffer from negligible host galaxy extinction, which could suggest a selection bias in the predominantly optical/UV discovered TDE sample. The discovery of (candidate) TDEs in LIRGs, galaxies that exhibit high star formation (SF) rates and host copious amounts of dust, supports this suggested bias. The transient Arp 299B-AT1 in the LIRG Arp 299 was shown to have arisen from a TDE based on the detection of a resolved and expanding off-axis radio jet \citep{mattila2018}. The transient was only marginally detected in the optical, and as such was missed by optical surveys despite the distance of only $\sim$45 Mpc. Also, the serendipitous discovery of a TDE candidate in the LIRG IRAS F01004-2237 led \citet{tadhunter2017} to suggest that LIRGs may have an elevated TDE rate, although the TDE nature of this transient is debated \citep{trakhtenbrot2019}. Both of these events showed prominent and long-lasting IR emission, attributed to the absorption and re-radiation of the UV/optical light from the transient by the surrounding dust, with total radiated energies in the IR exceeding 10$^{52}$ erg \citep{mattila2018,dou2017}. Although less energetic, such IR echoes have also been observed for a number of optically discovered TDEs \citep{dou2016,vanvelzen2016a,jiang2016}.

A second phenomenon involving accretion by a SMBH that can result in large amplitude outbursts is observed in CLAGN. In the unified model, AGN are classified based on the presence of emission lines in their optical spectra, originating from the broad line region (BLR) close to the SMBH and narrow line region (NLR) further away. Type 1 AGN show both broad (typically full width at half maximum (FWHM) of a few thousand km/s) and narrow (typically FWHM of a few hundred km/s) lines, whereas Type 2 show only narrow lines. This dichotomy is interpreted as a result of viewing angle, where in Type 2 the line of sight to the BLR is obscured by a dusty torus surrounding the AGN. In the case of CLAGN, the AGN type is observed to change between Type 1 and 2 or vice-versa in optical spectra. This phenomenon is poorly understood, but is commonly attributed to either a sudden change in accretion by the SMBH ionising the BLR \citep[e.g.,][]{macleod2016,sheng2017}, or variable obscuration, where dusty clouds passing across our line of sight cause the disappearance or appearance of broad emission lines \citep{goodrich1989}. The expected dynamical timescales associated with variable obscuration are of the order of 10--70 years \citep{mcelroy2016,sheng2017}, which excludes such CLAGN as an explanation for events on timescales of a few years.

As TDE and CLAGN are both related to accretion by a central SMBH, there is a lack of clear observables to distinguish the two scenarios. It is even argued that CLAGN may be the direct result of TDEs \citep{eracleous1995,merloni2015}. Furthermore, observational biases likely affect our current understanding of these extreme and rare outbursts. Thus, careful study of individual nuclear events across multiple wavelengths is required \citep[e.g.,][]{mattila2018} to constrain their nature and establish observational tracers to aid future classification.

In this paper we report the discovery and the multi-wavelength follow-up campaign of AT~2017gbl, an extremely IR-bright transient coincident with the nucleus of the LIRG IRAS~23436+5257. The paper is organised as follows: Section \ref{sec:obs} describes the discovery and follow-up campaign of AT~2017gbl, including the data reduction and photometry. Section \ref{sec:analysis} reports the analysis of the observed properties of AT~2017gbl and its host galaxy. This section includes SED fitting of the photometric data on the host and the transient, fitting of the spectral features, and analysis of the radio and X-ray properties. In Section \ref{sec:discussion} we discuss the key observations of the transient in the context of three different scenarios; a SN, a CLAGN and a TDE. Finally, in Section \ref{sec:conclusion} we present a summary of our findings. Throughout this paper we assume $H_0$ = 70 km s$^{-1}$ Mpc$^{-1}$, $\Omega_{\Lambda}$ = 0.7, and $\Omega_M$ = 0.3.

\begin{figure}
\centering
	\includegraphics[trim=0 0 0 0, clip, width=\columnwidth]{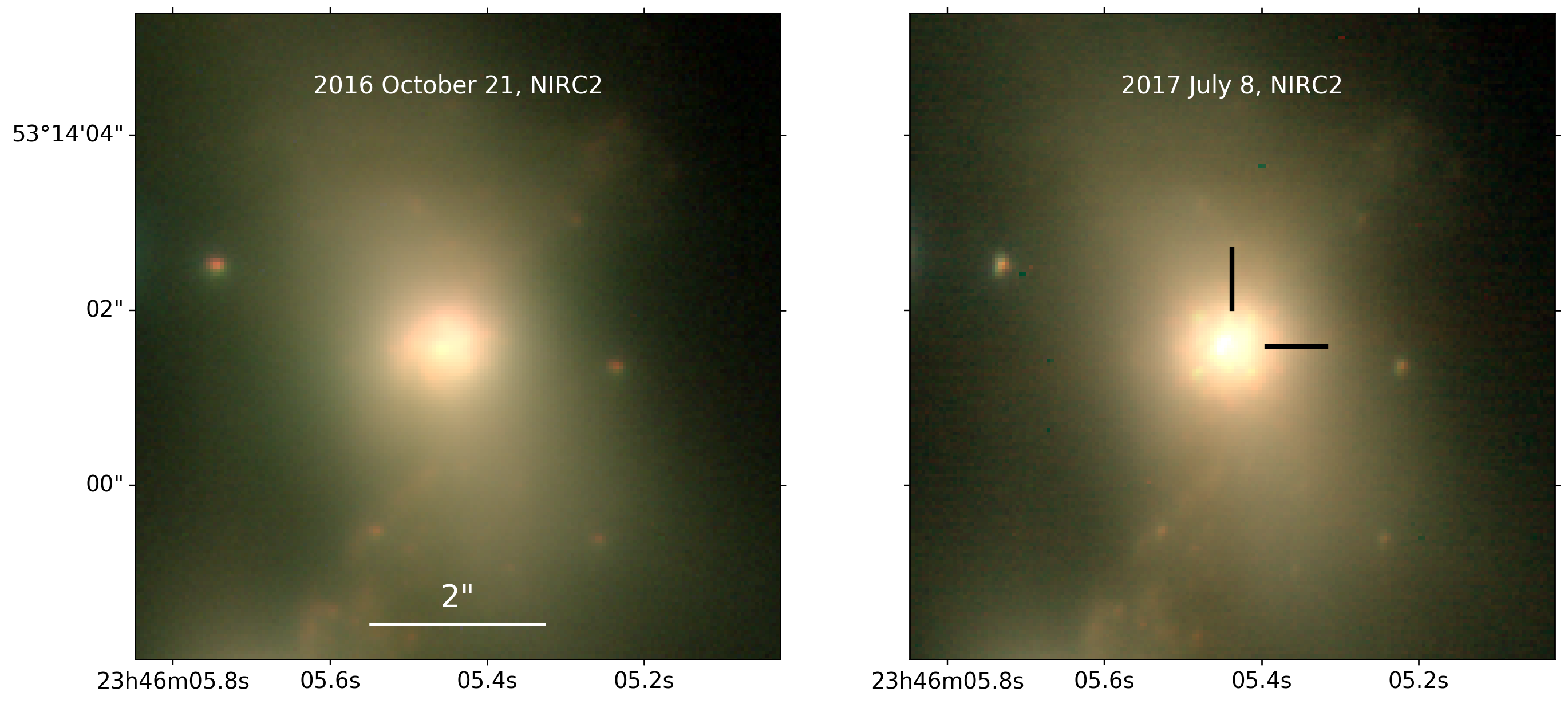}
    \caption{AT~2017gbl coincident with the Northern nucleus of IRAS~23436+5257, discovered with NIRC2. Left panel shows a $JHK_s$ colour composite from the NIRC2 template epoch in October 2016. Right panel shows a $JHK_s$ colour-composite of the discovery epoch of AT~2017gbl in July 2017. Image cutouts are $\sim$7.5\arcsec\ across.}
    \label{fig:at2017gbl}
\end{figure}

\section{Observations and results} \label{sec:obs}

\subsection{Discovery}
\begin{figure}
\centering
	\includegraphics[trim=0 0 0 0, clip, width=\columnwidth]{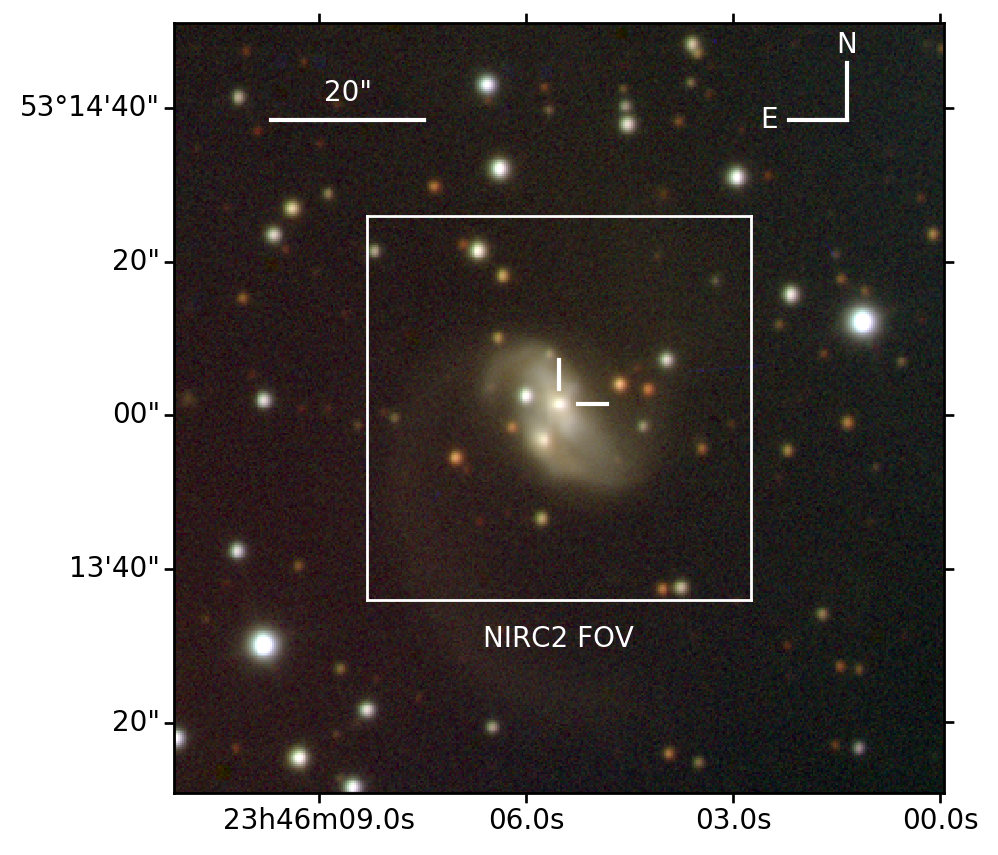}
    \caption{A $gri$-colour composite image of IRAS~23436+5257 as observed with ACAM on the WHT on 2017 July 10, three days after the discovery of AT~2017gbl. The transient was coincident with the Northern nucleus, indicated by tick marks. The FOV of the Keck NIRC2 discovery image is also shown. The ACAM image cutout shown here is 100\arcsec\ across.}
    \label{fig:at2017gbl_acam}
\end{figure}

AT~2017gbl was discovered as part of a systematic search for dust obscured SNe in the nuclear regions of LIRGs called project SUNBIRD \citep[Supernovae UNmasked By Infra-Red Detection;][]{kool2018}. The transient was discovered in IRAS~23436+5257 \citep[$z= 0.034134$,][luminosity distance of 146 Mpc]{strauss1992} in the near-IR $K_s$-band using Laser Guide Star Adaptive Optics \citep[LGS-AO, ][]{wizinowich2006} and the NIRC2 camera on the Keck II telescope on 2017 July 8.5 UT \citep[MJD 57942.5;][]{2017gbl}; see Fig.~\ref{fig:at2017gbl} and Fig.~\ref{fig:at2017gbl_acam}. Follow-up in $J$- and $H$-band was obtained on the same night. Subtractions with $J$, $H$ and $K_s$ observations from NIRC2 on 2016 October 21.4 UT showed an extremely bright residual coincident with the Northern nucleus of IRAS~23436+5257. Registering the image with 20 sources from the Pan-STARRS1 Data Release 1 archive \citep{chambers2016,flewelling2016} yielded R.A. = $23^{\textrm{h}}46^{\textrm{m}}05.52^{\textrm{s}}$ and Decl. = $+53^\circ 14\arcmin 01.29\arcsec$, with 0.03\arcsec\ and 0.05\arcsec\ uncertainty in R.A. and Decl., respectively.

\subsection{Near-IR and optical data reduction and photometry}
Follow-up near-IR imaging of AT~2017gbl was obtained at an approximately monthly cadence with NOTCam on the Nordic Optical Telescope \citep[NOT, ][]{djupvik2010} by the NOT Unbiased Transient Survey (NUTS) collaboration\footnote{\urlwofont{http://csp2.lco.cl/not/}}, from the discovery of the transient until the last detection in February 2019, at +590 days after discovery. The transient was revisited with NIRC2 on Keck in $K_s$-band on 2017 December 5.2 UT, at +150 days. In the optical, AT~2017gbl was observed with ACAM on the William Herschel Telescope (WHT) in $g$, $r$, $i$ and $z$ and with ALFOSC\footnote{The data presented here were obtained in part with ALFOSC, which is provided by the Instituto de Astrofisica de Andalucia (IAA-CSIC) under a joint agreement with the University of Copenhagen and NOTSA.} on the NOT in $i$ and $z$.

The NIRC2 and ACAM data were reduced using \textsc{theli} \citep{erben2005,schirmer2013}, following the steps outlined in \citet{schirmer2013} and \citet{schirmer2015}. \textsc{theli} uses \textit{Scamp} \citep{bertin2006} to calibrate the astrometry of the individual exposures to a reference catalogue to correct for image distortion before the final coaddition. The limited field of view (FOV) of NIRC2 did not contain enough 2MASS \citep{skrutskie2006} astrometric reference sources, so for the first NIRC2 epoch, image quality was optimized by calibrating the astrometry of the individual exposures to a catalogue extracted from a simple image stack of the same data set before coadding the resulting aligned exposures. Final astrometry was obtained by registering the coadded image to Pan-STARRS1 sources using \textsc{iraf}\footnote{\textsc{iraf} is distributed by the National Optical Astronomy Observatory, which is operated by the Association of Universities for Research in Astronomy (AURA) under cooperative agreement with the National Science Foundation \citep{iraf}} tasks. Subsequent NIRC2 images were calibrated using a catalogue extracted from the first $K_s$-band image. The NOTCam data were reduced using a version of the NOTCam Quicklook v2.5 reduction package\footnote{\urlwofont{http://www.not.iac.es/instruments/notcam/guide/observe.html}} with a few functional modifications (e.g. to increase the FOV of the reduced image).

Photometry of AT~2017gbl in the near-IR and optical images was carried out after image subtraction \citep[e.g.,][]{kool2018}, using a slightly modified (to accept manual stamp selection) version of the image subtraction package ISIS 2.2 \citep{alard1998, alard2000}. A NIRC2 image from 2016 October 21.5 UT, 260 days before discovery, was available as a transient-free reference image for the near-IR NIRC2 discovery image and the NIRC2 epoch at +150 days. However, due to the large difference in pixel scale and image quality between the NIRC2 reference image and the seeing-limited follow-up imaging with NOTCam, the NIRC2 template image was not suitable as a reference for NOTCam. Instead, reference NOTCam templates were obtained after the transient had faded below the detection limit, at epoch +744 days in $J$ and $H$ and at +798 days in $K_s$. In the optical, the WHT/ACAM observations from +570 days were used as transient-free reference images for the WHT/ACAM data. Similarly, ALFOSC imaging was obtained at +427 days to act as a reference for the optical NOT data.

Point-spread function (PSF) photometry of the transient was carried out using \textsc{snoopy}\footnote{SNOoPy is a package for SN photometry using PSF fitting and/or template subtraction developed by E. Cappellaro. A package description can be found at \urlwofont{http://sngroup.oapd.inaf.it/ecsnoopy.html}} from template subtracted images. The photometry of the transient in the seeing-limited near-IR (NOTCam) and optical (ALFOSC and ACAM) images were calibrated against five field stars from 2MASS and Pan-STARRS1, respectively. The photometry in the NIRC2 data was calibrated against five field stars in its small FOV, which in turn were calibrated with NOTCam imaging.

In case of a non-detection, a local detection threshold was determined by injecting sources of increasing brightness at the position of the transient, using the task \textit{mkobjects} in \textsc{iraf}, before performing image subtraction. The transient was considered recovered if the signal-to-noise ratio of the aperture flux at the position of the transient in the subtracted image was $>$ 5, compared to 24 empty positions in the immediate vicinity of the transient in the subtracted image.

\begin{figure*}
\centering
	\includegraphics[trim=0 0 0 0, clip, width=\textwidth]{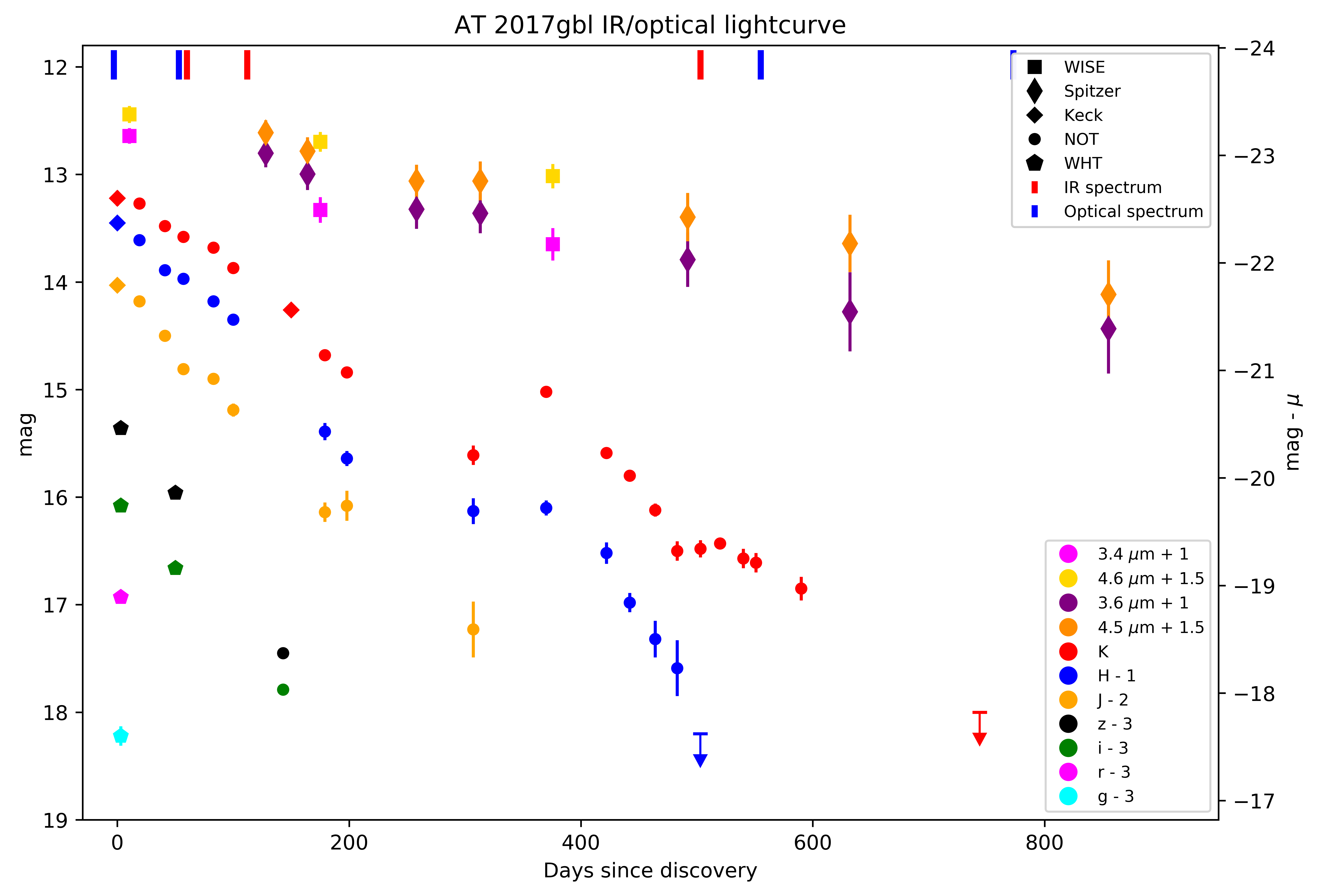}
    \caption{IR and optical evolution of AT~2017gbl. For clarity, the lightcurves have been offset as indicated in the figure legend. IR magnitudes are in the Vega system, optical magnitudes in the AB system. Pre-discovery non-detections are not shown, and consist of a $JHK_s$ near-IR epoch at -260 days, and a 3.4/4.6 \micron\ \textit{WISE} epoch at -188 days.}
    \label{fig:at2017gbl_lc}
\end{figure*}

The resulting host-subtracted lightcurve of AT~2017gbl in the optical and near-IR is shown in Fig.~\ref{fig:at2017gbl_lc} and the photometry is listed in Tables \ref{tab:nir_mags} and \ref{tab:opt_mags}, where the near-IR photometry is in the Vega system and the optical photometry in the AB system. The consistent evolution between the NIRC2 and NOTCam magnitudes supports the assumption that the NOTCam template epochs can be considered transient-free.

\subsection{Mid-IR photometry}
\label{sec:mid-ir}

\subsubsection{Spitzer}
\begin{figure}
\centering
	\includegraphics[trim=0 0 0 0, clip, width=\columnwidth]{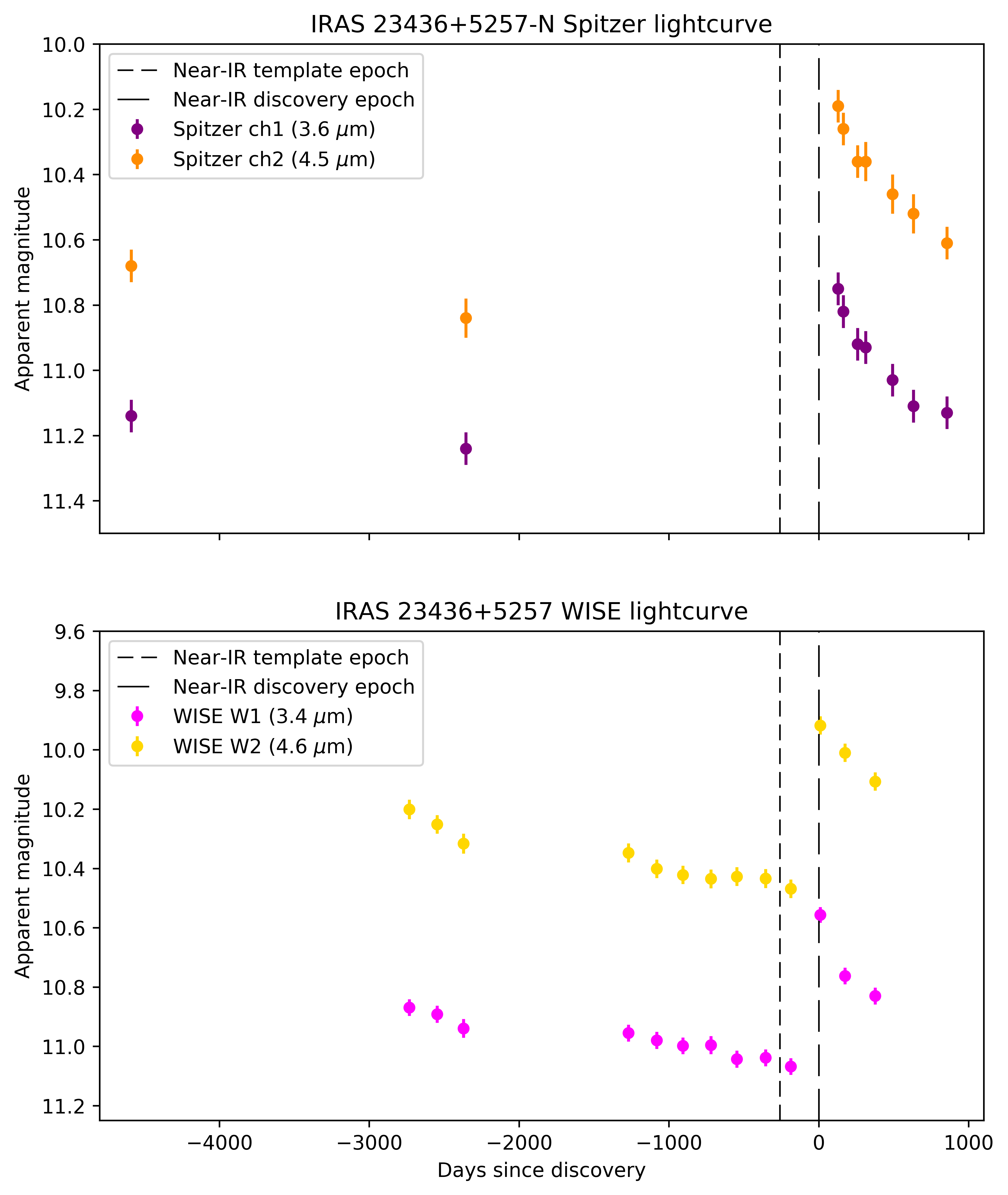}
    \caption{Top panel shows the \textit{Spitzer} lightcurve at 3.6 \micron\ and 4.5 \micron\ of the Northern nucleus of IRAS~23436+5257. Bottom panel shows the archival \textit{WISE} lightcurve at 3.4 \micron\ and 4.6 \micron\ of the full host galaxy. In the \textit{Spitzer} data the host nucleus of AT~2017gbl is resolved, whereas in the \textit{WISE} data the host galaxy is not resolved. The vertical lines indicate the epochs of the near-IR reference data set (relevant for the construction of the SED of the transient, see Section \ref{sec:AT2017gbl_single_comp_fit}) and the near-IR discovery epoch, respectively.}
    \label{fig:iras23436_wise_spitzer}
\end{figure}

Follow-up imaging in the mid-IR was obtained with the \textit{Spitzer Space Telescope} at 3.6 \micron\ and 4.5 \micron\ at seven different epochs, between 2017 November 13.1 UT at epoch +128 days and 2019 November 10.3 UT at epoch +855 days. Two archival epochs of the host galaxy were available from 2004 and 2011. The magnitude of the resolved Northern nucleus, host of AT~2017gbl, was determined at all epochs through relative photometry using a 3.8\arcsec\ aperture with five isolated field stars, for which magnitudes were based on catalog fluxes from the \textit{Spitzer} Heritage Archive. The \textit{Spitzer} lightcurve of the Northern nucleus not only showed the brightening in the mid-IR due to AT~2017gbl, but also suggested evidence for a decline in magnitude between the two archival epochs from 2004 and 2011 of 0.1 $\pm$ 0.07 and 0.16 $\pm$ 0.08 magnitudes at 3.6 \micron\ and 4.5 \micron, respectively. The \textit{Spitzer} magnitudes in the Vega system of the Northern nucleus of IRAS~23436+5257 are listed in Table \ref{tab:iras_23436_spitzer_mags} and the lightcurve is shown in Fig.~\ref{fig:iras23436_wise_spitzer}.

\subsubsection{WISE}
In addition to \textit{Spitzer}, archival observations from the Wide-field Infrared Survey Explorer (\textit{WISE}) were available covering the pre-outburst host galaxy from 2010 until just after the 2016 near-IR reference epoch, as well as three post-outburst epochs.

\textit{WISE} surveyed the full sky at 3.4, 4.6, 12, and 22 \micron\ (channels $W1$ - $W4$) in 2010 during its initial cryogenic mission, followed by the post-cryogenic and NEOWISE surveys in channels $W1$ and $W2$ upon depletion of its cryogen \citep{mainzer2011}. The data from the initial and post-cryo missions have been made available as the AllWISE catalog. Between 2011 February and 2013 October  \textit{WISE} was put in hibernation, after which it was reactivated for the NEOWISE Reactivation \citep[NEOWISE-R;][]{mainzer2014} survey. In total IRAS~23436+5257 was observed twice during the cryogenic mission, once during the post-cryo mission and ten more epochs were obtained at regular intervals from 2013 until 2018 as part of the NEOWISE-R survey. Due to the observing strategy of \textit{WISE}, each epoch consists of $\sim$12--18 exposures across $\sim$2 days, each with profile-fitted magnitudes reported in the AllWISE and NEOWISE-R catalogs. The host galaxy IRAS~23436+5257 consists of two nuclei, see Fig.~\ref{fig:at2017gbl_acam}, which were unresolved in the \textit{WISE} data and well fit by a single profile with no deblending performed. In order to derive a single magnitude for each epoch, we averaged the magnitudes of all exposures of each epoch, after verifying no significant intra-day variability occurred during an epoch, and excluding poor quality exposures (\textit{qual\_frame} > 0). The photometric \textit{WISE} errors were taken as the standard error of the mean in each epoch and added in quadrature a flux error term of 2.4 and 2.8\% in $W1$ and $W2$, respectively, to reflect uncertainty between epochs \citep[e.g.,][]{jarrett2011}. The \textit{WISE} magnitudes in the Vega system of IRAS~23436+5257 at 3.4 \micron\ and 4.6 \micron\ are listed in Table \ref{tab:iras_23436_wise_mags} and the lightcurve shown in Fig.~\ref{fig:iras23436_wise_spitzer}. As can be seen in the lightcurve, the pre-outburst \textit{WISE} observations confirm the long-term decline in mid-IR of the host galaxy suggested by the archival \textit{Spitzer} observations. 

\subsubsection{Mid-IR photometry of AT~2017gbl}
\label{sec:mir_photom}
The mid-IR lightcurves from both \textit{Spitzer} and \textit{WISE} clearly show that the system is not constant in flux, which means it is not appropriate to use a single pre-outburst epoch as a template for the image subtraction. Therefore the magnitude of AT~2017gbl in the mid-IR was determined by arithmetic magnitude subtraction, by subtracting the flux of a reference epoch from the post-outburst epochs. The mid-IR reference epoch was chosen to coincide with the near-IR reference epoch from 21 October 2016 in order to be able to construct a consistent host-subtracted spectral energy distribution (SED) of AT~2017gbl. Assuming any further decline of the host galaxy between the reference epoch and outburst epoch is similar in the near-IR and the mid-IR, this would affect the transient flux by a small constant offset across the SED.

The magnitude of the host galaxy in the \textit{WISE} data at 2016 October 21.4 UT, -260 days before discovery, was determined by interpolating between the magnitudes of the 2016 July 17 and 2017 January 1 epochs. Using the \textit{Spitzer} data, we established that all pre- and post-outburst variability of IRAS~23436+5257 originated in the Northern nucleus, by subtracting at each epoch the flux of the Northern nucleus from the flux in a larger aperture encompassing the full galaxy. The remainder, consisting of the flux of the Southern nucleus and faint structure between the nuclei, was constant within errors across all \textit{Spitzer} epochs from 2004 until 2019. Therefore we concluded that the residual flux after magnitude subtraction of the interpolated reference \textit{WISE} epoch from the post-outburst \textit{WISE} epochs can be fully attributed to AT~2017gbl.

The magnitude of the Northern nucleus of IRAS~23436+5257 in \textit{Spitzer} bands at the time of the near-IR reference epoch was determined by subtracting the flux of AT~2017gbl from the 2017 December 19 epoch, at +164 days. The flux of AT~2017gbl at 3.6 \micron\ and 4.5 \micron\ at 2017 December 19 was inferred from a blackbody, fitted to the (interpolated) \textit{WISE} and near-IR $K_s$-band fluxes, see Section \ref{sec:AT2017gbl_single_comp_fit}. The quiescent magnitude of the Northern nucleus of IRAS~23436+5257 derived in this way was magnitude subtracted from all post-outburst \textit{Spitzer} observations to determine the magnitude of AT~2017gbl in the \textit{Spitzer} data. Table \ref{tab:mir_mags} shows the mid-IR photometry of AT~2017gbl from the \textit{Spitzer} and \textit{WISE} observations, and the mid-IR lightcurve of AT~2017gbl is shown in Fig.~\ref{fig:at2017gbl_lc}. It must be noted that the arithmetic magnitude subtraction of the host introduces a systematic uncertainty related to the magnitude error of the template epoch, which is not included in the lightcurve or table. However, they are included in the SED fits of AT~2017gbl across all filters, discussed in Section \ref{sec:AT2017gbl_single_comp_fit}.

\subsection{Spectroscopy}
\label{sec:spectroscopy}
\subsubsection{Data reduction}
Spectroscopic follow-up was performed in the optical with ALFOSC on the NOT and ISIS on the WHT, and in the near-IR with GNIRS on Gemini North and SpeX on IRTF. A log of the spectroscopic observations is reported in Table \ref{tab:Spectral log}. All spectra will be made available via WISeREP \citep{yaron2012}. Here we briefly summarise the observations and data reduction steps for each data set.

The GNIRS cross-dispersed spectra were reduced using version 2.0 of the {\sc \mbox{XDGNIRS}\footnote{\urlwofont{http://drforum.gemini.edu/topic/gnirs-xd-reduction-script}}} pipeline \citep{mason15}, which provides a convenient wrapper to a series of {\sc PyRAF} tasks provided as part of the Gemini GNIRS data reduction package. Both epochs were observed in a similar manner, nodding the object on and off the 7\arcsec slit in an ABA pattern, with telluric reference stars (A0V spectral type) observed immediately before or after the object. The \mbox{XDGNIRS} pipeline was used to detect and trace the slit orders; extract the science and calibration data; apply the flat field, spatial rectification, and wavelength calibrations; and correct the object spectra for telluric absorption. The latter was done accounting for intrinsic absorption features in the telluric reference star spectrum, and applying modest adaptive rescaling of the telluric spectrum to minimise the residual absorption residuals in the galaxy spectrum. Finally, aperture spectra were extracted using the {\sc apall} task to trace and sum a fixed aperture about the galaxy centre.

The ALFOSC spectra were reduced using the {\sc alfoscgui} package, which uses standard {\sc iraf} tasks to perform overscan, bias, and flat$-$field corrections as well as removal of cosmic ray artifacts using  {\sc lacosmic}  \citep{VanD01}. Extraction of the one$-$dimensional spectra was performed with the {\sc apall} task and wavelength calibration was done by comparison with arc lamps and corrected if necessary by measurement of skylines. The spectra were flux calibrated against photometric standard stars observed on the same night.

The ISIS data were reduced with the same standard {\sc iraf} tasks and steps described above for ALFOSC.

The SpeX data were reduced using the publicly available {\sc spextool} software package \citep{Cushing2004}. This reduction proceeded in a standard way, with image detrending, order identification, and sky subtraction. Corrections for telluric absorption utilized the {\sc XTELLCOR } software and A0V star observations \citep{vacca2003}. After extraction and telluric correction, the 1D spectra from the six orders were rescaled and combined into a single spectrum.

\subsubsection{Line identification}
\begin{figure*}
\centering
	\includegraphics[trim=0 0 0 0, clip, width=1\textwidth]{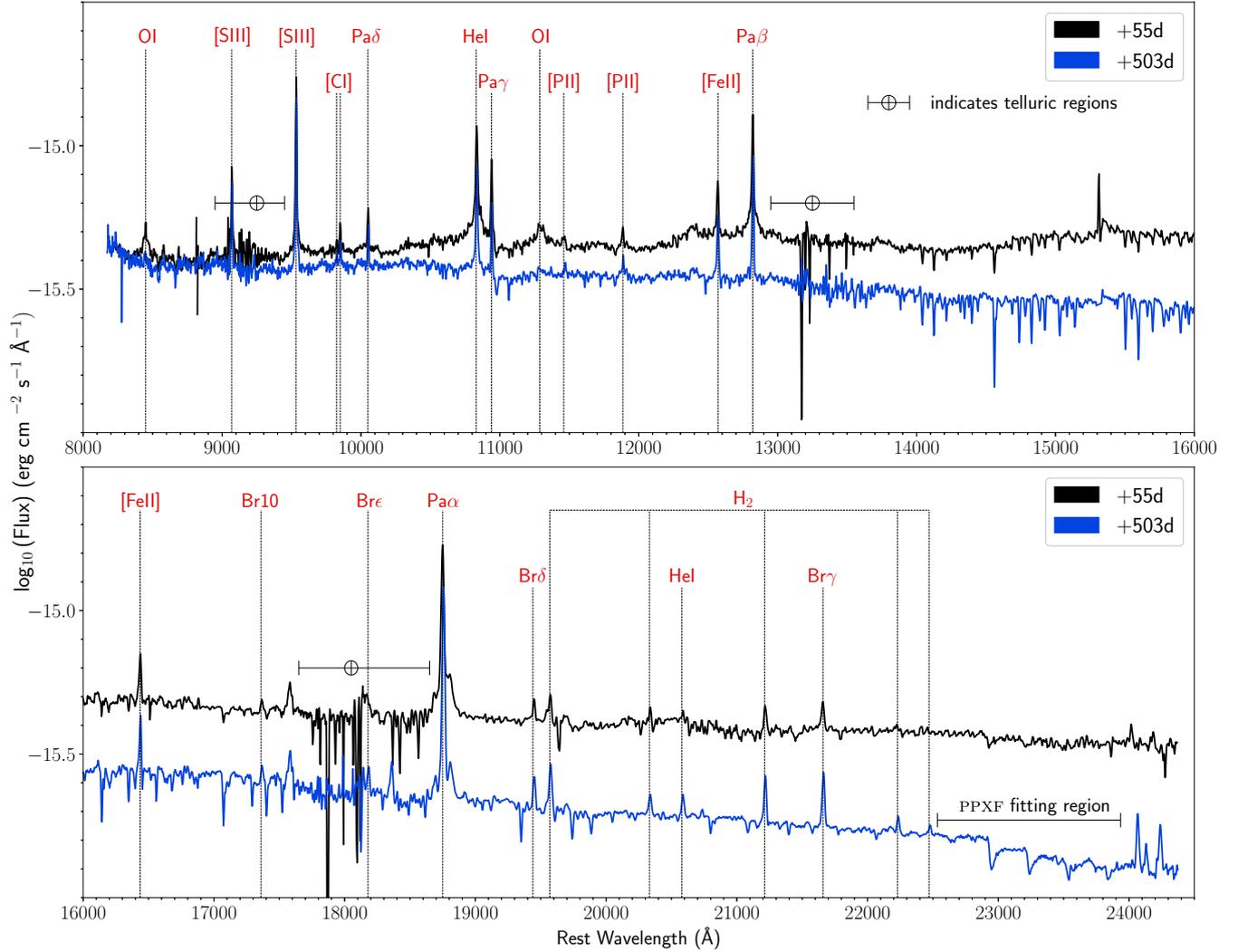}
    \caption{Near-IR follow-up spectra of AT~2017gbl from 2017 September 1 and 2018 November 23, taken with GNIRS. Spectra are corrected for Milky Way reddening. Prominent emission features are indicated. The region used to measure the velocity dispersion with {\sc ppxf} fitting in Section \ref{sec:vel_dispersion} is marked.}
    \label{fig:2017gbl_nir_spec}
\end{figure*}

The near-IR spectra obtained with GNIRS are shown in Fig.~\ref{fig:2017gbl_nir_spec}. Both the near-IR spectra and the optical spectra (discussed below) have been corrected for Milky Way reddening \citep{schlafly2011}, adopting the Cardelli extinction law \citep{cardelli1989} with $R_V$ = 3.1. The GNIRS spectrum observed at +55 days after the discovery shows strong emission lines such as Paschen and Brackett recombination lines, He\,{\sc i}, H$_2$ and [Fe\,{\sc ii}]. There are broad features visible in the Paschen and He\,{\sc i} emission lines as well as broad emission features of O\,{\sc i} at 8446\AA~ and 11287\AA. These lines, particularly the 11287\AA\ emission feature, indicate Bowen fluorescence where the O\,{\sc i} 1025\AA\ transition is pumped by Lyman~$\beta$ emission, and cascades down through these lines \citep{Bowen1947}. The spectrum obtained with the IRTF/SpeX two months later has much lower signal-to-noise. Of the broad features, only Paschen $\alpha$ is detected. The second GNIRS spectrum was obtained 448 days after the first one, +503 days after the discovery. During this period the transient decreased 2 - 2.5 mag in brightness in the near-IR. This dimming is visible in the spectra as a change in the shape of the continuum, which becomes less red as the transient has declined more in the redder bands. Additionally, the O\,{\sc i} lines visible in the first GNIRS spectrum are no longer present in the second GNIRS spectrum and the broad wings that were visible in the Paschen and helium emission lines have either disappeared or visibly decreased in strength.

\begin{figure*}
\centering
	\includegraphics[trim=0 0 0 0, clip, width=\textwidth]{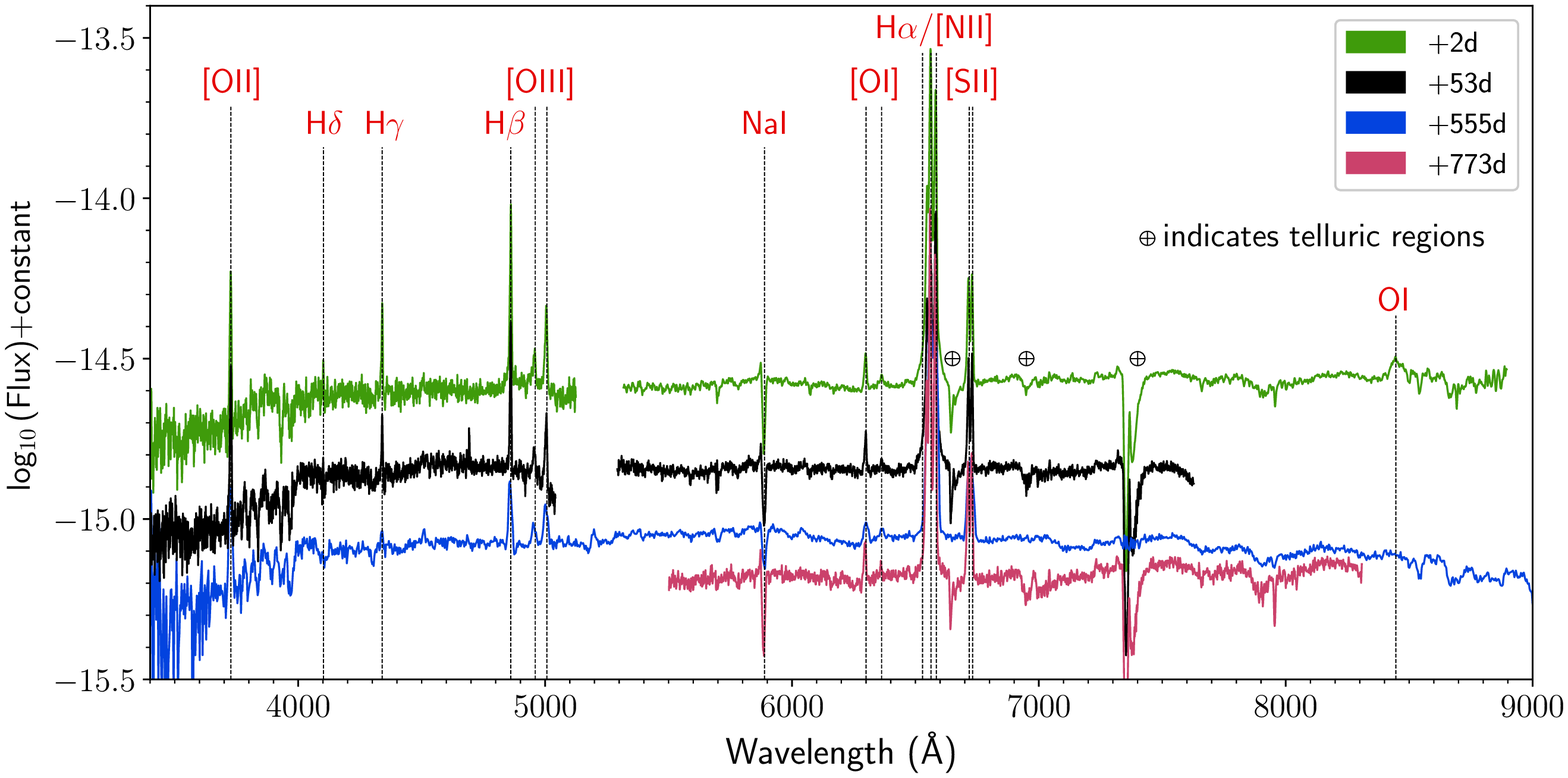}
    \caption{Optical follow-up spectra of AT~2017gbl. Spectra are corrected for Milky Way reddening. Some emission and absorption features are indicated. Note the broad O\,{\sc i} feature at 8446\AA\ that is present initially but not in late observations. Note that the broadening of the narrow lines in the spectrum taken at +555d is because the data is from a lower resolution instrument.}
    \label{fig:2017gbl_opt_spec}
\end{figure*}

The optical spectra obtained with ISIS and ALFOSC are shown in Fig.~\ref{fig:2017gbl_opt_spec}. The ISIS spectrum obtained +2 days after the discovery is dominated by narrow emission lines, such as H Balmer series, [O\,{\sc iii}], [N\,{\sc ii}] and [S\,{\sc ii}]. Additionally, the broad wings shown by the H$\alpha$/[N{\sc ii}] complex give evidence for the presence of a broad emission line corresponding to H$\alpha$. We also see the broad feature from O\,{\sc i} $\lambda$8446. The ISIS spectrum obtained at +53 days shows little evolution in the narrow features but the broad features decrease in strength. The ALFOSC spectra obtained at +555 and +773 days continue to show little change in the narrow lines along with a reduction in the strength of the broad component of H$\alpha$ and no evidence for a broad feature from O\,{\sc i} $\lambda$8446. We discuss the broad feature evolution in both optical and near-IR in Section \ref{sec:spec_analysis}.

Integral field spectroscopy of the Northern nucleus of IRAS~23436+5257 in the near-IR $K$-band is the only known pre-outburst spectrum available of the host galaxy. These data were serendipitously obtained with OSIRIS on the Keck telescope on 2016 November 18 as part of the Keck OSIRIS AO LIRGs Analysis Survey \citep{u2019}, one month after our near-IR imaging reference epoch and 1.5 months before the final pre-outburst epoch from \textit{WISE}. We simulated the slit aperture used in the GNIRS spectra to obtain a 1D-spectrum in order to compare with the post-outburst GNIRS spectra. The OSIRIS spectrum showed narrow Brackett $\delta$ and $\gamma$ in emission, with no sign of a broad component.

\subsection{Radio observations}
In the radio we observed AT~2017gbl with milliarcsecond angular resolution using the Very Long Baseline Array (VLBA) at 4.4 and 7.6 GHz (simultaneously), and with the European Very long baseline interferometry Network (EVN) at 4.9 GHz, and with the Arcminute Microkelvin Imager Large Array \citep[AMI-LA;][]{zwart2008,hickish2018} at 15.5 GHz at a typical 40\arcsec$\times$30\arcsec\ resolution. Our VLBA observations took place on 2017 August 15 \citep{2017gbl_vlba} and 2017 October 20 at epochs +38 and +99 days, with clean beam major and minor axes of (4.5$\times$1.3) milliarcsec$^2$ and (2.8$\times$0.9) milliarcsec$^2$, respectively. Our EVN observations were carried out on 2019 February 15, at epoch +587 days, and resulted in an angular resolution of (9.3$\times$4.6) milliarcsec$^2$. Three AMI epochs were obtained in the month following the discovery of AT~2017gbl, between 2017 July 12 and Aug 10 \citep{2017gbl_ami}, and three more on a longer timescale between +234 and +936 days after the discovery. We also retrieved a cutout at the position of the host from the VLA Sky Survey \citep[VLASS;][]{lacy19} at 3.0\,GHz. This image was obtained on 2019 May 2 at epoch + 663 days and has a resolution of 3\farcs1$\times$2\farcs2.

We carried out our VLBA observations recording at a bit rate of 2 Gbps using dual polarisation, and made use of the wide C-band receiver, so we simultaneously observed our target at the central frequencies of 4.4 GHz and 7.6 GHz, with a bandwidth of 128 MHz at each sub-band. We used 1 MHz width channels and an integration time of 2 sec, which resulted in negligible time- and band-width smearing. We followed standard calibration and imaging procedures within the AIPS package. We used the compact, nearby VLBA calibrator J\,2353$+$5518 as the phase-reference source. Our VLBA observations from +38 days showed one unresolved source within the 1\arcsec\ by 1\arcsec\ FOV at R.A. = $23^{\textrm{h}}46^{\textrm{m}}05.5173^{\textrm{s}}$ and Decl. = $+53^\circ 14\arcmin 01.260\arcsec$, and the astrometric uncertainty in that position is less than 1 milliarcsecond at both frequencies. Those coordinates are only 0.04\arcsec\ from the transient position in the near-IR, and are well within the near-IR astrometric uncertainties. The second VLBA epoch at +99 days showed a significant increase in flux at 4.4 GHz and 7.6 GHz, as well as a change in the spectral index. 

We scheduled our EVN observations similarly to our VLBA observations, and used seven antennas of the Western EVN array at a data rate of 2 Gbps using dual polarisation in eight sub-bands of 32 MHz each. We reduced the data following standard procedures for the EVN, and took into account ionospheric corrections for each antenna. We also performed two self-calibration steps (first in phase only and then in amplitude and phase) on the phase reference source (J\,2353$+$5518) to correct the antenna gains in the different sub-bands. This correction was especially needed for the data of a couple of antennas that had no system temperature measurements. The calibrator J\,2353$+$5518 has a compact morphology and a flux density of $0.47\pm0.04$\,Jy at 4.9\,GHz. With 64 channels per sub-band and an integration time of 2 sec, the FOV was limited by time- and bandwidth-smearing to $\sim$45\arcsec{}. A map centred on the Southern nucleus, which is only $\sim$5\arcsec away from the Northern nucleus, yields no detections. In the Northern nucleus we detected an unresolved source at a position coincident with the coordinates reported based on the VLBA observations. The EVN observation at +587 days showed that the source was still unresolved, but its 4.9 GHz flux density had decreased, compared to the early 4.4 GHz VLBA observations.

The AMI observations were taken at a central frequency of 15.5 GHz over a 5 GHz bandwidth covered by 4096 channels and measures I$+$Q polarisation. The array has baselines between 18 and 110 m leading to a characteristic resolution of between 30\arcsec\ and 50\arcsec\ depending on the number of antennas and the sky position of the target. Observations lasted between 3 and 4 hours, yielding r.m.s. values between 41 and 57 $\mu$Jy/beam. The phase calibrator J2355$+$4950 was observed interleaved with the target field for $\sim$100 s for each $\sim$10 min on source, and either 3C286 or 3C48 was used as the absolute flux calibrator. Data were calibrated and imaged in the quick look format, where data are averaged into 8 frequency channels of 0.625 GHz width at the correlator, and then flagged for radio frequency interference, and flux and phase calibrated using the custom reduction pipeline for quick look data \textsc{reduce\_dc} \citep[e.g.,][]{perrott2015,bright2018}. Data were then imported into CASA and further flagging was performed and the data were imaged using standard imaging techniques with a clean gain of 0.1 and manual masking. Fluxes were extracted using the CASA task \textsc{imfit}. The source was point like (unresolved) in all the observations and we did not fix the dimensions of the synthesised beam when fitting. The source peaked in flux density around the epoch +21 days, after which the source declined in flux up to the epoch +848 days. Between the final two epochs at +848 and +936 days no variability is observed in flux density within 1-$\sigma$. We, therefore, consider these two epochs to be transient-free, tracing the quiescent flux from the host galaxy and its Southern companion at 15.5 GHz.

\subsubsection{Host contamination correction} \label{sec:radio_host_corr}
\begin{figure}
\centering
	\includegraphics[trim=0 0 0 0, clip, width=\columnwidth]{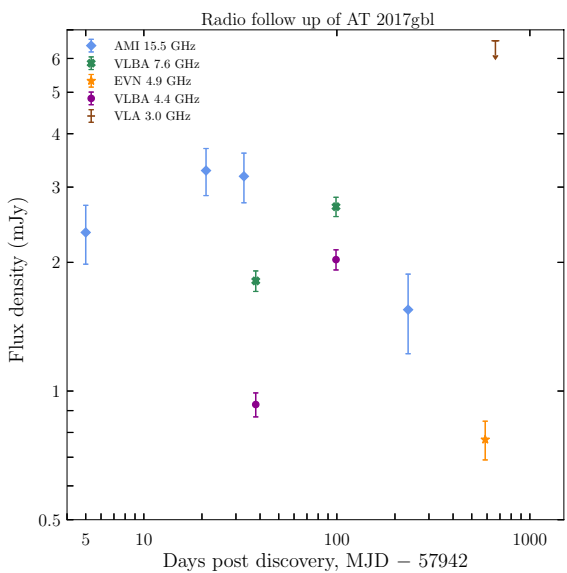}
    \caption{Radio lightcurve of AT~2017gbl. The flux densities from AMI and the VLA have been corrected for host galaxy contributions. The AMI epochs of +848 and +936 days are not shown here, as they are considered transient-free and act as the quiescent level of the host galaxy at 15.5 GHz (see text for details). The epoch at 3 GHz with the VLA has a large uncertainty, and we show the 3$\sigma$ upper limit here instead of the host-subtracted value.
    }
    \label{fig:radio_lightcurve}
\end{figure}

We list in Table \ref{tab:total_radio_flux} the properties of the pre- and post-outburst radio observations. We note that the measurements from all these observations include a contribution from the host. In the case of the AMI observations the measurements are also contaminated by emission from the Southern nucleus, which clearly emits in the radio as seen in the VLASS cutout. To alleviate this contamination, we looked for available pre-outburst radio images. The host was within the surveyed area of the NRAO VLA Sky Survey \citep[NVSS;][]{condon98} with a resolution of (45$\times$45)\,arcsec$^2$ at 1.4\,GHz. In the extracted NVSS cutout we find that IRAS\,23436$+$5257 is a compact source with a flux density of $18.70\pm1.85$\,mJy, where we have considered a 3\% uncertainty in the flux calibration \citep{condon98}, that we added in quadrature to the r.m.s. to obtain the total uncertainty in the flux density. The quiescent flux of IRAS~23436+5257 at 15.5\,GHz, taken as the mean value of the AMI epochs of +848 and +936 days, is $3.60\pm0.18$\,mJy. The angular resolution of AMI and NVSS are comparable, and therefore we can assume that they are tracing the emission of the same region. In this way, we obtained a two-point spectral index between 1.4 and 15.5 GHz of $\alpha = -0.69 \pm 0.05$ ($S_\nu \propto \nu^{\alpha}$) for IRAS\,23436$+$5257 in its quiescent state. This spectral index agrees well with the expected value for star-forming galaxies at $z<2$ \citep{delhaize17}. Therefore, most of the flux density seen by the low-resolution radio observations of AMI and the NVSS likely comes from extended, large-scale star-formation in the host. Using the spectral index between 1.4 and 15.5 GHz, we obtain a flux density of the host at 3.0\,GHz of $11.09\pm1.02$\,mJy. Having the contribution of the host to the total emission at 3.0\,GHz and 15.5\,GHz, we subtracted it from the total flux densities to obtain the radio flux densities from the AMI and VLA observations that correspond to AT~2017gbl. The VLBA and the EVN observations at milliarcsecond angular resolution trace the flux density from the innermost nuclear regions hosting AT~2017gbl. We assume that the transient dominates the compact radio emission traced by the VLBA and the EVN. The resulting transient fluxes are listed in column (7) of Table \ref{tab:total_radio_flux}, and shown in Fig.~\ref{fig:radio_lightcurve}. 

\subsection{X-ray observations}
\begin{figure}
\centering
	\includegraphics[trim=0 0 0 0, clip, width=\columnwidth]{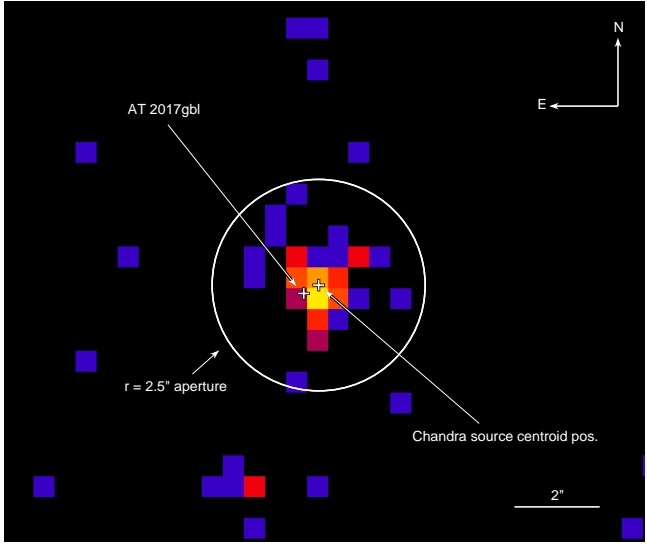}
    \caption{X-ray image of AT~2017gbl obtained with \textit{Chandra} in 0.5-7 keV band. The circle represents a 2.5\arcsec\ aperture centered on the X-ray source containing $\sim$95\% of emission from the on-axis source, which was used to extract the spectrum shown in Fig.~\ref{fig:at2017gbl_chandra_spectrum}. The position of the centroid and the position of AT~2017gbl as observed in the near-IR are also indicated.}
    \label{fig:at2017gbl_chandra}
\end{figure}

In X-rays we first observed the transient as a target-of-opportunity (ToO) on 2017 September 13 for 3 ks with the X-ray telescope on board \textit{the Neil Gehrels Swift observatory} (ObsID 00010290001). No source was found with a 3$\sigma$ luminosity upper limit of 4.7$\times10^{41}$ erg s$^{-1}$ in the 0.2-10 keV energy band, measured using an absorbed powerlaw model with \textit{N}$_{\rm H} = 3.0\times10^{21}$ cm$^{-2}$ and a photon index 0.9 (see below). A second, deeper X-ray observation with a 10 ks exposure was obtained through Director's Discretionary Time with the ACIS-S imager on \textit{Chandra X-ray Observatory}, on 2017 November 3, +118 days after the discovery. In the \textit{Chandra} observation (ObsID 20831) a point source coincident with the transient's position was detected, see Fig.~\ref{fig:at2017gbl_chandra}. The offset between the position of AT 2017gbl and the centroid position of this source is $\sim$0.4", which is less than the typical celestial location accuracy of \textit{Chandra}\footnote{For details, see \urlwofont{https://cxc.harvard.edu/proposer/POG/html/chap5.html\#tth_sEc5.4}}. 

\begin{figure}
\centering
	\includegraphics[trim=25 25 25 25, clip, width=\columnwidth]{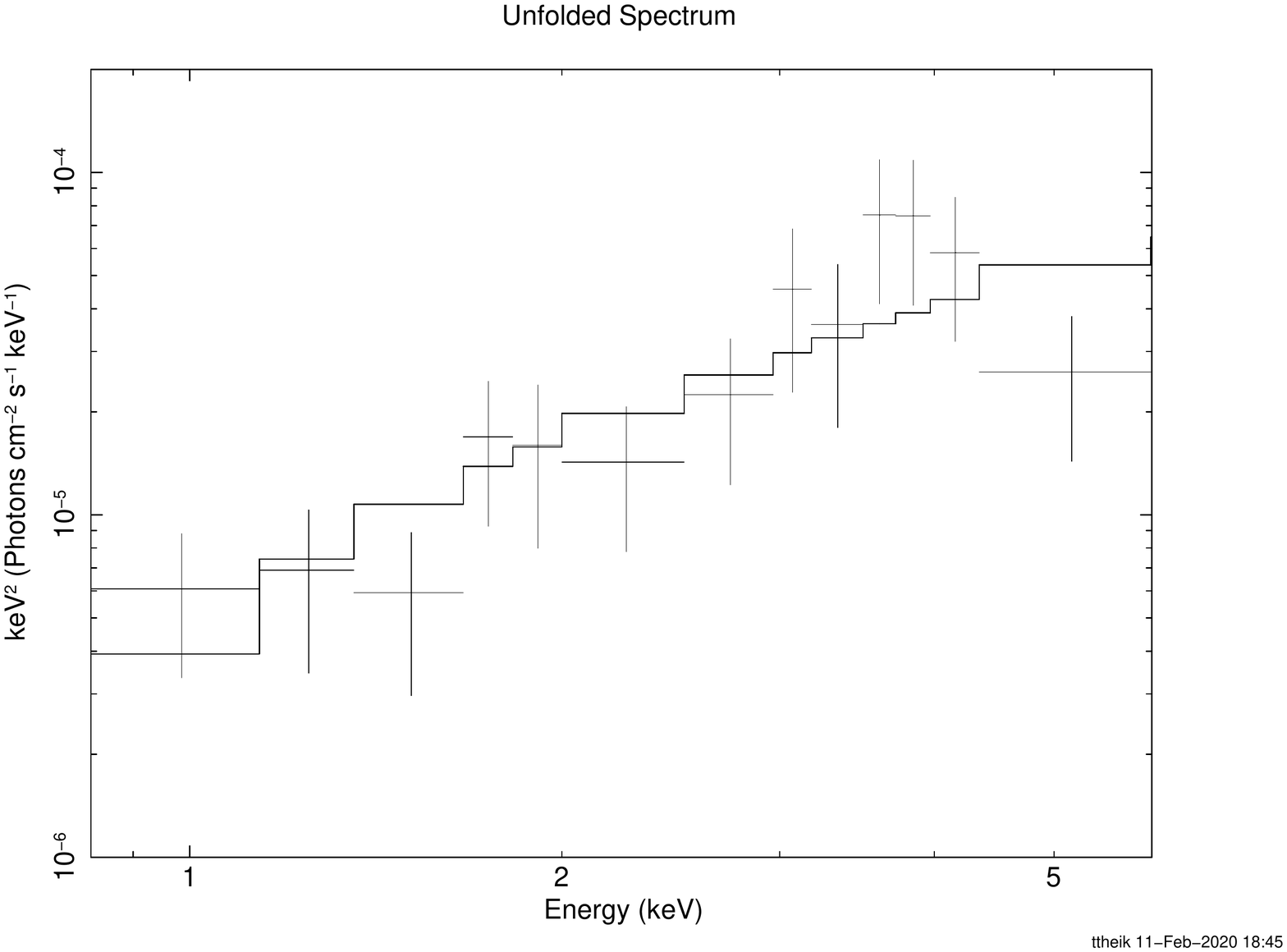}
    \caption{\textit{Chandra} X-ray spectrum of the source fitted with an absorbed power-law where \textit{N$_{\rm H}$} = 3.0$\times$10$^{21}$ cm$^{-2}$ and the fitted photon index is $\sim0.9$.}
    \label{fig:at2017gbl_chandra_spectrum}
\end{figure}

A spectrum of the source was extracted from the \textit{Chandra} observation with the {\sc specextract} tool from the CIAO 4.10 software-package, using a 2.5" aperture (containing $\sim$95\% of emission from the on-axis source) positioned on the source centroid. As any X-ray emission from a possible AGN in the nucleus would be almost entirely absorbed in the \textit{Chandra} energy band, we assume that the observed X-ray flux primarily originates from the circumnuclear population of X-ray binaries (XRBs). To test if the X-ray source can be explained by emission originating in the host galaxy of AT 2017gbl we therefore adopt a simple absorbed powerlaw model to represent the combined XRB population \citep[e.g.,][]{mitsuda1984}. To improve our fit, we also restrict the number of free parameters in our model by adopting \textit{N}$_{\rm H} = 3.0\times10^{21}$ cm$^{-2}$ from \citet{mineo2012}\footnote{We note that the Galactic \textit{N}$_{\rm H}$ in the direction of AT~2017gbl is $1.86\times10^{21}$ cm$^{-2}$ \citep{HI4PI_collab_2016}.}, to estimate the average \textit{N}$_{\rm H}$ value outside of the nucleus. In Section \ref{sec:xray_transient} we will discuss our results in the context of the empirical relationship between the star formation rate and the X-ray luminosity of a LIRG's XRB-population established in \citet{mineo2012}, and by adopting the same value for absorption, our model will better align with theirs\footnote{Minor variations in the \textit{N}$_{\rm H}$ value do not have a significant effect on our analysis.}. We therefore only fit the powerlaw photon-index, using XSPEC 12.10.0c. For the photon-index, we obtain the best fit of $0.9\pm0.3$, and for the powerlaw-component the unabsorbed model flux of $1.2 \pm 0.2\times10^{-13}$ erg cm$^{-2}$ s$^{-1}$ (0.5-8.0 keV), see Fig.~\ref{fig:at2017gbl_chandra_spectrum}. This corresponds to an unabsorbed X-ray luminosity of $L_{\textrm{X}} = 3.2_{-0.5}^{+0.6}\times10^{41}$ erg s$^{-1}$.

\section{Analysis} \label{sec:analysis}
\subsection{Host galaxy SED fitting}
\label{sec:sed_fit}
IRAS~23436+5257 is a bright LIRG with an IR luminosity of $L_{\textrm{IR}}\sim4\times10^{11}$ L$_{\odot}$ 
\citep[][adjusted to H$_0$ = 70 km s$^{-1}$ Mpc$^{-1}$]{sanders2003}. This LIRG was included in the SUNBIRD sample because the expected core-collapse SN (CCSN) rate inferred from its $L_{\textrm{IR}}$ \citep{mattila2001} is as high as $\sim$1 yr$^{-1}$. This empirical relation, however, assumes a negligible contribution to the IR luminosity of the galaxy from a potential AGN. There has not been any evidence in the literature of the presence of an AGN in IRAS~23436+5257, based on hard X-ray (14-195 keV) observations \citep{koss2013}, equivalent width of the PAH feature \citep{yamada2013}, or mid-IR $W1$ - $W2$ colours from \textit{WISE} \citep{assef2018a}.

However, the long-term pre-outburst variability in the mid-IR (see Fig.~\ref{fig:iras23436_wise_spitzer}) hints at the presence of an obscured AGN in the host galaxy of AT~2017gbl, as AGN are known to show such low-amplitude smooth variability in the mid-IR \citep{kozlowski2016}.

In order to determine the different components contributing to the total luminosity of IRAS~23436+5257, we modelled its multi-wavelength SED with a combination of libraries of starburst, AGN torus, AGN polar dust and spheroidal/cirrus component models. For this purpose we collected photometric data points available from the literature ranging from the optical to the submillimetre, from Pan-STARRS1, 2MASS, Spitzer/IRAC, IRAS \citep{sanders2003}, ISO \citep{stickel2004} and GOALS \citep{armus2009,chu2017}. In addition, we included mid-IR spectra in the range between 5 and 37 microns observed with the Spitzer IRS instrument and available through the Combined Atlas of Sources with Spitzer IRS Spectra. These observations and their reductions are described by \citet{lebouteiller2011}. The spectral resolution of the IRS data was reduced to better match the resolution of the radiative transfer models and have a wavelength grid that is separated in steps of 0.05 in the log of rest wavelength. However, in order to better constrain the AGN and starburst contributions to the SED more points were included around the 9.7 \micron\ silicate feature and the PAH features. We note that no scaling was required between the photometric points from the different sources and the mid-IR spectra which we take as an indication that the emission is dominated by one of the two galaxies.

In particular we used the library of starburst models computed with the method of \citet{efstathiou2000} as revised by \citet{efstathiou2009}, and the library of AGN torus models computed with the method of \citet{efstathiou1995}. The polar dust model was calculated in a similar way as in \citet{mattila2018}. We assume the polar dust is concentrated in discrete optically thick ($\tau_V\sim$ 100) clouds which are assumed to be spherical with no internal heating source. For each of these clouds we carry out a radiative transfer calculation to calculate their emission using the code of \citet{efstathiou1995} and assuming a normal interstellar dust mixture. However, we assume a fixed temperature of 1300 K for the dust which in this simple model is assumed to be determined by the external illumination of the clouds by the transient event. The library of spheroidal models was as described in \citet{herrero2017}. More details of the method will be given in Efstathiou et al. (2020, \textit{in prep}). The SED fits were carried out with the MCMC SED fitting code SATMC \citep{johnson2013}.

\begin{figure*}
\centering
	\begin{tabular}{cc}
        \includegraphics[angle = 0, origin = c, trim=50 130 40 270, clip, width=0.45\textwidth]{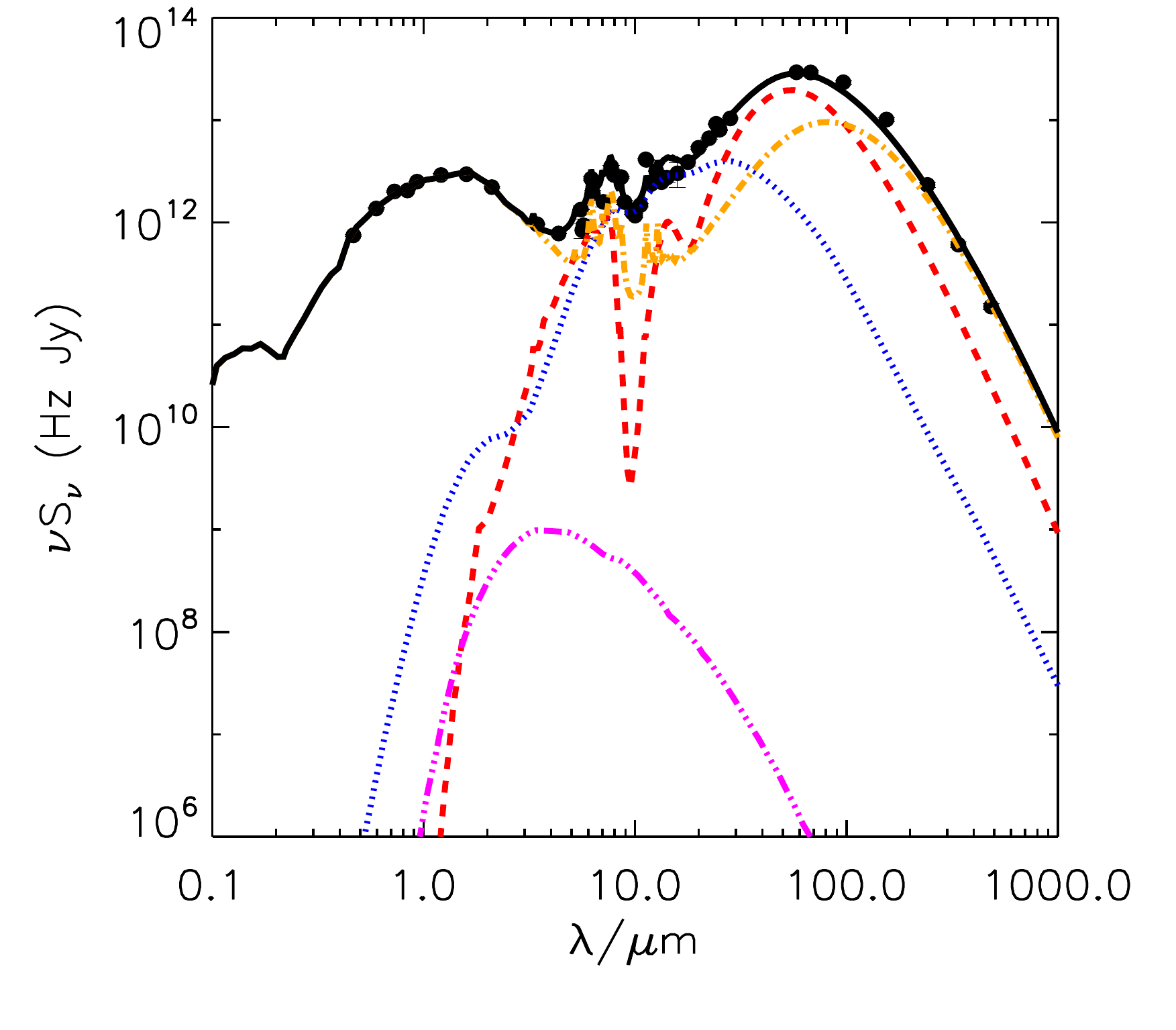} &
        \includegraphics[angle = 0, origin = c, trim=50 130 40 270, clip, width=0.45\textwidth]{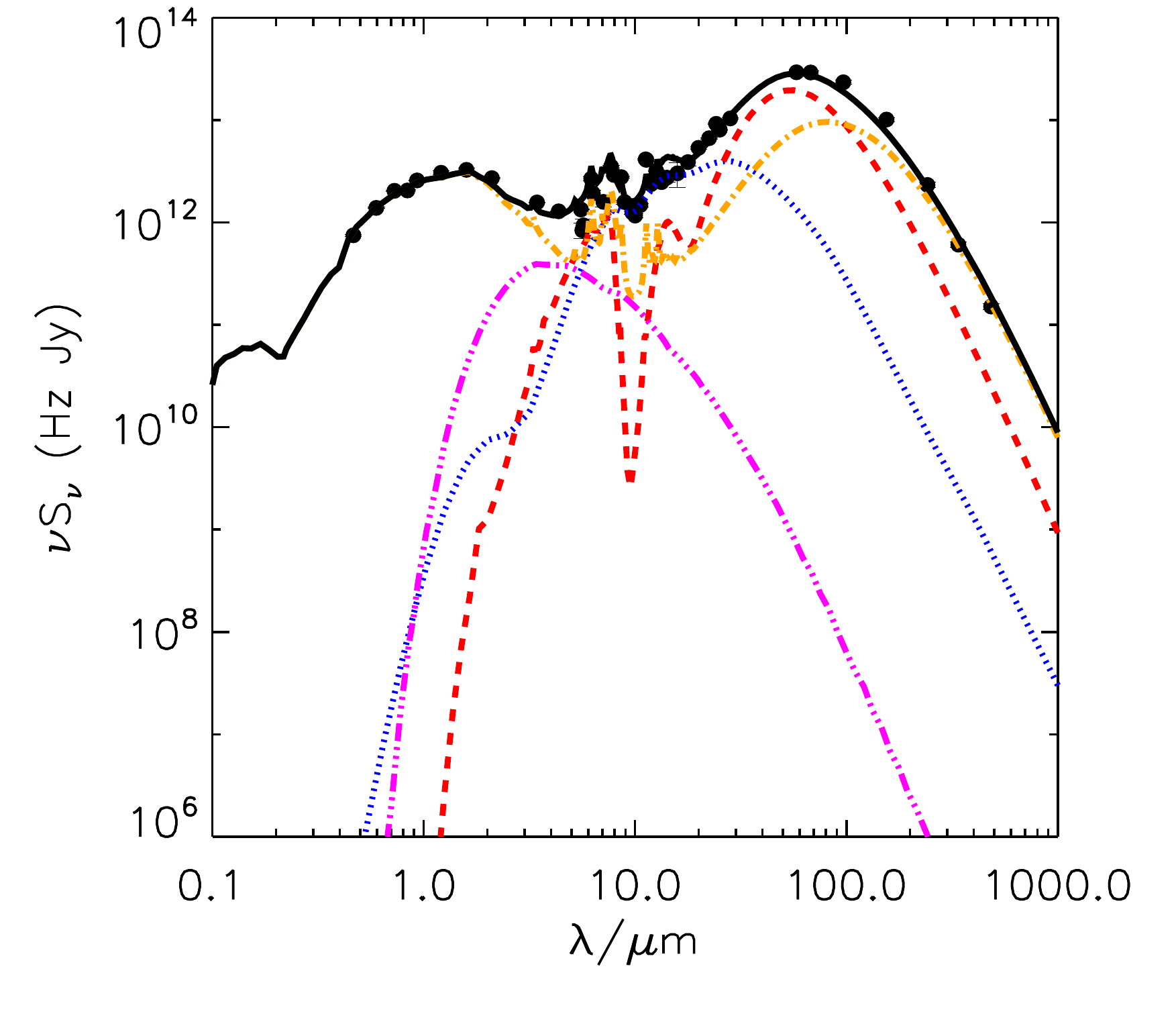} \\
	\end{tabular}
    \caption{The best fitting SED model for IRAS~23436+5257 obtained using the MCMC code SATMC \citep{johnson2013}, pre-outburst on the left and post-outburst (+10 days) on the right with the contribution from AT~2017gbl included at the wavelength range 0.5 - 4.6 \micron. The flux measurements are indicated by the black dots, and the model SED is composed of a starburst (red), an AGN torus (blue), a spheroidal host (orange) and a polar dust at 1300K (magenta) component.}
\label{fig:iras23436_sed}
\end{figure*}

Fig.~\ref{fig:iras23436_sed} shows the best fitting SED model for IRAS~23436+5257 before and 10 days after the discovery of the outburst, composed of a starburst, an AGN torus, a spheroidal host and a polar dust component at 1300 K.
The pre-outburst model is fitted to fluxes at epochs preceding AT~2017gbl. The post-outburst fit is based on the same data with the addition of the observed fluxes of AT~2017gbl in the optical, near-IR and mid-IR (0.5 - 4.6 \micron~ range). We do not expect any significant emission from the transient at longer wavelengths and assume that the archival pre-outburst flux densities describe the SED adequately at wavelengths > 5\micron, similar to the case of Arp~299-B AT1 \citep{mattila2018}.
All model parameters were fixed to values within a range of 1\% from the pre-outburst fit, with the exception of the polar dust temperature that was fixed to 1300 K and luminosity which was left as a free parameter.
The resulting model fitting parameters and the derived physical quantities are listed in Table \ref{tab:iras23436_sed}.

\begin{table}
\centering
\begin{tabular}{|l|c|c|} \hline
&Pre&Post \\ \hline
Total luminosity ($10^{11}$ L$_{\odot}$)              &4.35$_{-0.08}^{+0.09}$        &4.35$_{-0.08}^{+0.09}$   \\[5pt]
Starburst luminosity ($10^{11}$ L$_{\odot}$)          &1.61$_{-0.10}^{+0.04}$        &1.61$_{-0.10}^{+0.04}$   \\[5pt]
Spheroidal luminosity ($10^{11}$ L$_{\odot}$)         &1.34$_{-0.04}^{+0.10}$        &1.33$_{-0.04}^{+0.10}$   \\[5pt]
AGN luminosity ($10^{11}$ L$_{\odot}$)                &1.40$_{-0.09}^{+0.09}$        &1.40$_{-0.09}^{+0.09}$   \\[5pt]
Polar dust luminosity ($10^{9}$ L$_{\odot}$)          &0.01$_{-0.006}^{+0.12}$        &3.8$_{-0.2}^{+0.2}$   \\[5pt]
Polar dust covering factor (\%)                         &0.01$_{-0.01}^{+0.17}$     & -   \\[5pt] \hline
Core-collapse supernova rate ($\mathrm{SN~yr}^{-1}$)&$0.16^{+0.01}_{-0.01}$ &   \\ [5pt]
SF rate, averaged&&\\
over the past 50 Myr (M$_{\odot}~\mathrm{yr}^{-1}$) &$14.9^{+1.6}_{-1.2}$   &   \\ \hline
\end{tabular}
\caption{The model parameters obtained when fitting the SED of IRAS~23436+5257 along with the derived bolometric luminosities. In the post-outburst case all the parameters were fixed to the pre-outburst values within a range of 1\%, with the exception of the polar dust temperature and luminosity. The CCSN and SF rates are calculated based on the fitted mode parameters. The SF rate is averaged over the past 50 Myr. The SF rate averaged over the age of the starburst is higher by about a factor of 3.}
\label{tab:iras23436_sed}
\end{table}

It is noteworthy in particular that the model requires a significant AGN contribution of 32 $\pm$ 2\% to the total luminosity of the galaxy, after correcting the AGN torus luminosity for anisotropic emission. Attempts to model the data without an AGN component resulted in poorer fits (maximum log-likelihood < -3300 vs. -1166) that did not recover well the spectral range between 3-40 \micron\ around the 9.7 \micron\ silicate absorption feature. In the case of the fit including an AGN we have a total of 13 free parameters and 30 degrees of freedom whereas in the case of the fit without an AGN the number of free parameters is 8 resulting in 35 degrees of freedom. Given the small difference in the degrees of freedom the difference in log-likelihood is significant favouring the presence of an obscured AGN.

Based on this result, we inspected the \textit{Spitzer} IRS spectrum at the original resolution for the mid-IR high ionization fine structure lines that are strong in AGN, such as [Ne\textsc{v}] 14.3 \micron\ and [O\textsc{iv}] 25.9 \micron. We do not find evidence for these lines, which is in agreement with the previous results of \cite{Inami2013}, who did not detect these lines at 3$\sigma$ significance in their study which included this spectrum. While the presence of strong [Ne\textsc{v}] and [O\textsc{iv}] lines signifies an AGN, their absence does not exclude the presence of an obscured AGN significantly contributing to the IR luminosity, as shown in the sample of LIRGs optically classified as Seyfert 2 galaxies presented by \citet{alonsoherrero2012}. Due to the inclination of the torus in our model for IRAS~23436+5257, the apparent AGN luminosity is lower by a factor of $\sim$2.4, which may explain why an AGN was not detected by previous studies.

\begin{figure}
\centering
    \includegraphics[angle = 90, origin = c, trim=0 0 0 0, clip, width=\columnwidth]{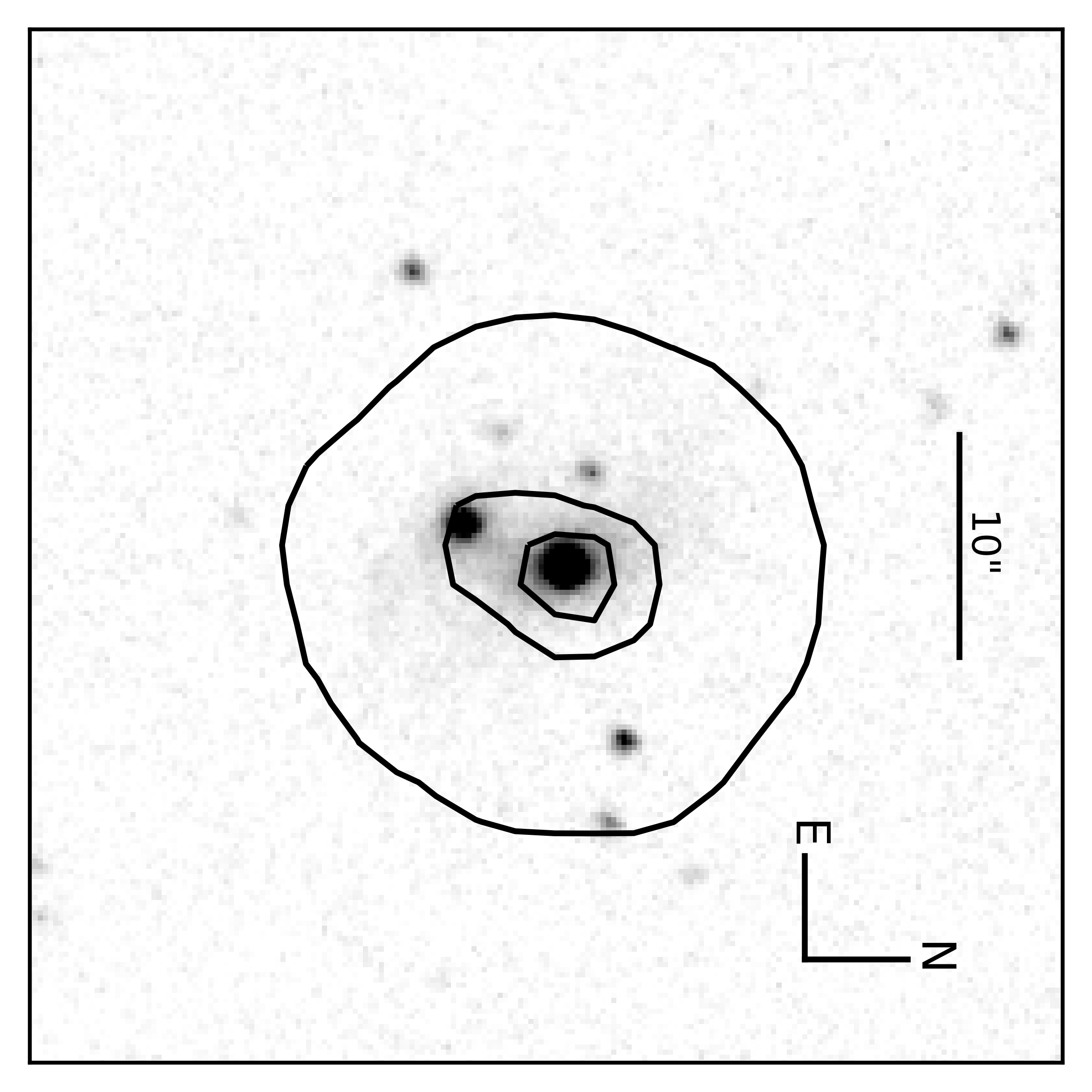}
    \caption{NOT $K_s$-band image of IRAS~23436+5257 from 2017 July 27 with 24\micron\ MIPS contours overplotted. The 24 \micron\ emission is concentrated on the Northern nucleus, supporting it as a potential AGN host.}
\label{fig:iras23436_mips}
\end{figure}

IRAS~23436+5257 consists of two nuclei, and the model fit is based on flux densities for the whole galaxy in which the two nuclei are not resolved. In order to determine the host nucleus of the possible AGN, we compare contours from a 24 \micron\ MIPS image with a near-IR $K_s$-band NOT image in Fig.~\ref{fig:iras23436_mips}. As shown, most of the 24 \micron\ emission from IRAS~23436+5257 originates in the Northern nucleus. Based on Fig.~\ref{fig:iras23436_sed}, the AGN component should dominate the emission at 24 \micron\, which would originate from a region a few hundred pc or less in size \citep[e.g.,][]{lopez2018}. Therefore we conclude that the Northern nucleus is a potential host to a dust-obscured AGN.

As a result of the contribution to the IR luminosity by the AGN, the expected CCSN rate of IRAS~23436+5257 derived from the SED fitting is much lower, at 0.16 $\pm$ 0.01 yr$^{-1}$, than expected from its IR luminosity $L_{\textrm{IR}}$. Additionally, the data are best fit by a model where the torus obscures a direct line of sight to the central AGN. The optical depth along the line of sight of the best fit model is $\tau\sim$ 90 at 1 \micron, equivalent to an extinction of A$_V$ $\sim$ 300 mag. This extinction would fully obscure the central engine at optical, near- and mid-IR wavelengths. Assuming a standard conversion to hydrogen column density from \citet{predehl1995}, this A$_V$ corresponds to $N_H \sim 5.4 \times 10^{23}$ cm$^{-2}$, which should be considered as a lower limit for the actual column density given the dust evaporation in the innermost regions close to the AGN. This corresponds to values expected from a Compton-thick AGN.

\subsubsection{Pre-outburst mid-IR variability} \label{sec:agn_variability}
The presence of a dust-obscured AGN in IRAS~23436+5257 would explain the low-amplitude mid-IR variability observed before the discovery of AT~2017gbl. As seen in Fig.~\ref{fig:iras23436_wise_spitzer}, the variability in the mid-IR shown by the Northern nucleus of IRAS~23436+5257 can be divided into two stages: a long-term decline spanning $>$10 years, followed by the steep increase and subsequent decline related to AT~2017gbl. The long-term decline as shown in the \textit{WISE} data prior to the outburst amounts to 0.20 $\pm$ 0.04 and 0.27 $\pm$ 0.05 mag at 3.4 \micron\ and 4.6 \micron, respectively, over the course of $\sim$2500 days, or 7 years. Galaxies hosting an AGN have been observed to show mid-IR variability that typically has a lower amplitude and a longer timescale than in the optical \citep[e.g.,][]{glass2004}. This difference is because the variations in the mid-IR originate in a region of a much larger extent than the optical light, and as a result any short timescale variations are smoothed out. Normal mid-IR AGN variability has been quantified by \citet{kozlowski2016} to be $<$0.3 mag over 7 years, which agrees well with the observed smooth pre-outburst mid-IR decline of the Northern nucleus of IRAS~23436+5257. The mid-IR variability connected to AT~2017gbl consists of an increase of 0.51 $\pm$ 0.04 and 0.55 $\pm$ 0.04 mag at 3.4 \micron\ and 4.6 \micron, respectively, between the last pre-outburst and the first post-outburst NEOWISE-R epochs. This increase over $\leq$198 days is not possible to reconcile with `normal' AGN variability.

\subsection{Optical and near-IR spectral analysis}
\label{sec:spec_analysis}

In our spectra of AT~2017gbl, we see narrow unresolved emission lines in the optical and near-IR associated with hydrogen, helium, [O\,{\sc iii}], [N\,{\sc ii}] and H$_2$, which are commonly found within star-forming LIRGs \citep{burston2001, valdes2005}. Given the evidence for an obscured AGN from the SED fitting of IRAS~23436+5257, we search for signs of this also in the spectra. We do not see any high ionisation coronal lines such as [Ca\,{\sc viii}] $\lambda$23218, [Si\,{\sc vi}] $\lambda$19620 or [S\,{\sc viii}] $\lambda$9915, which would indicate the presence of strong X-ray flux associated with an AGN. The near-IR galaxy spectral surveys of \citet{riffel2006,Riffle2019} show that none of the narrow lines we detect can unambiguously indicate the presence of an AGN, as they are often found in LIRGs with no evidence for an AGN. 

\begin{figure}
    \includegraphics[width=\columnwidth]{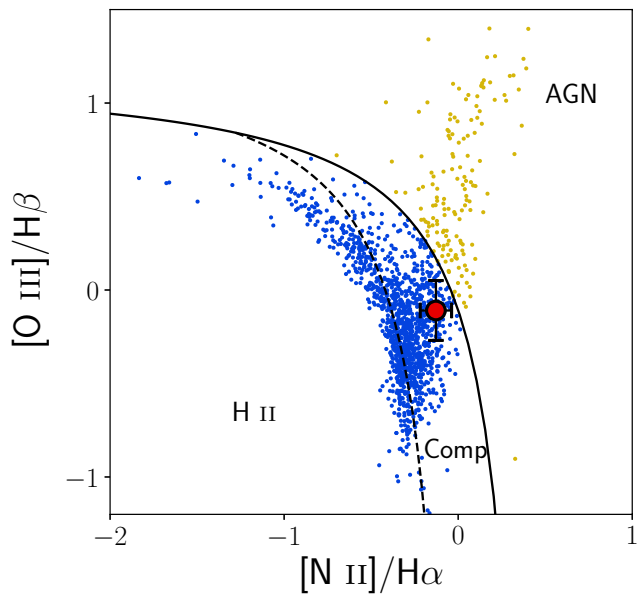}
    \caption{BPT diagram showing the position of IRAS~23436+5257 with a red point. The galaxy data shown is taken from SDSS DR7 \citep{SDSS_DR72009},  regions indicated are taken from \citet{Kewley2006}. Colouring of points indicates the region they lie within, not an independent determination of their type. Code distributed as part of \textsc{astroML} \citep{astroML} was adapted to generate this plot.}
    \label{fig:BPT}
\end{figure}

A common criterion for assessing the relative contributions of an AGN and star formation is the BPT diagram \citep{BPT1981}. For IRAS~23436+5257 we measure log$_{10}$([O\,{\sc iii}]/H$\beta$) = -0.11 $\pm$ 0.16 and log$_{10}$([N\,{\sc ii}]/H$\alpha$) = -0.128 $\pm$ 0.09 which places this galaxy in the composite AGN + SF region, as shown in Fig.~\ref{fig:BPT}. Based on the position on the BPT diagram, the presence of an AGN can not be confirmed nor ruled out. \citet{Larkin1998} suggested the line ratios [Fe\,{\sc ii}]/Pa$\beta$ and H$_{2}$/Br$\gamma$ as a diagnostic of whether galaxies are LINERs or Seyfert type AGNs using the near-IR spectral region, and this idea was further developed in \citet{Rodri2004,Rodri2005}, \citet{Riffel2013} and \citet{vaisanen2017}. \citet{Riffel2013} find 0.6 < [Fe\,{\sc ii}] / Pa$\beta$ < 2 and  0.4 < H$_{2}$ / Br$\gamma$ < 6 as determining criteria for an AGN classification based on a large sample of objects, with lower values for these ratios indicating a star-forming galaxy (SFG). In our +503d spectrum of the Northern nucleus of IRAS~23436+5257, we find [Fe\,{\sc ii}] / Pa$\beta$ = 0.47 $\pm$ 0.02 and H$_{2}$ / Br$\gamma$ = 0.80 $\pm$ 0.03, placing it within the SFG region for the former value, and AGN region for the latter. These ratios cannot provide us with a clear indication of an AGN, and again suggests that the Northern nucleus of IRAS~23436+5257 is a transitional object. \citet{Lamperti2017} find in their survey that these diagnostics are often not sufficient to diagnose AGN, as SFGs can fulfill both criteria. 

\subsubsection{Emission line fitting} 
\label{sec:line_fit}

\begin{table*}
\centering
\begin{tabular}{|l|l|l|l|l|l|} \hline
Emission line		&Epoch	    &Narrow line FWHM  &Broad line FWHM &Narrow line flux  &Broad line flux \\ \hline
					&days		    &km~s$^{-1}$               &km~s$^{-1}$   &  10$^{-15}$ $\text{erg}~\text{cm}^{-2}~\text{s}^{-1}$            &  10$^{-15}$ $\text{erg}~\text{cm}^{-2}~\text{s}^{-1}$ \\ \hline
Paschen $\alpha$    &+55   &220 $\pm$ 12          &2110 $\pm$ 120    &  23.3 $\pm$ 0.8 & 26.2 $\pm$ 2.2    \\ 
                    &+503   &200 $\pm$ 6          &2070 $\pm$ 140   &  19.2 $\pm$ 0.3   &  8.5 $\pm$ 0.9            \\ \hline
Paschen $\beta$     &+55   &189 $\pm$ 7          &1840 $\pm$ 60    &  8.7 $\pm$ 0.2  & 10.4 $\pm$ 0.6               \\ 
                    &+503   &191 $\pm$ 6          &-              &  7.6 $\pm$ 0.1   & -               \\ \hline
Paschen $\gamma$    &+55   &200 $\pm$ 25          &-              &  4.5 $\pm$ 0.3   & -             \\ 
                    &+503   &175 $\pm$ 25          &-            &  3.0 $\pm$ 0.2     &-               \\ \hline
Paschen $\delta$    &+55   &190 $\pm$ 40         &2800 $\pm$ 600  &   1.6 $\pm$ 0.2   & 1.9 $\pm$ 0.6  \\
                    &+503   &160 $\pm$ 40          &-            &  1.4$\pm$ 0.2     &-               \\ \hline
Brackett $\gamma$   &+55   &190 $\pm$ 50    &1600 $\pm$ 400  & 2.1 $\pm$ 0.3   & 1.6 $\pm$ 0.6           \\ 
                    &+503   &201 $\pm$ 27       &-        &    2.4 $\pm$ 0.2       &-               \\ \hline
He {\sc i} $\lambda$10830          &+55   &319 $\pm$ 14          &2440 $\pm$ 130  &   8.2 $\pm$ 0.3    & 13.3 $\pm$ 1.0   \\ 
                    &+503   &262 $\pm$ 14  &1910 $\pm$ 210   &  5.5 $\pm$ 0.2   & 3.8 $\pm$ 0.7    \\ \hline
O {\sc i} $\lambda$8446            &+55   &-       &1640 $\pm$ 140   &  -  & 3.4 $\pm$ 0.4             \\ 
                    &+503   &-                    &-         &         &-               \\ \hline
O {\sc i} $\lambda$11287           &+55   &-  & 1930 $\pm$ 100   &  -  &5.2 $\pm$ 0.4                   \\ 
                    &+503   &-                    &-          &        &-               \\ \hline
\hline
\end{tabular}
\caption{Emission line widths and fluxes. Where no value is given, the line was not detected. FWHM are corrected for instrumental broadening. Uncertainties listed in the table are statistical and derived from the covariance matrix of the fitting parameters. There is an additional $\sim$15\% uncertainty in the flux measurements coming from the absolute flux calibration of the spectra, that is not included in the listed uncertainties. Note that we chose not to simultaneously fit regions where multiple broad features were blended, such as Paschen $\gamma$, $\delta$ and He\,{\sc i}, with multiple gaussians and thus consider the fits in these regions to be less reliable.}
\label{tab:emission_line_fits}
\end{table*}

\begin{figure}
    \includegraphics[width=\columnwidth]{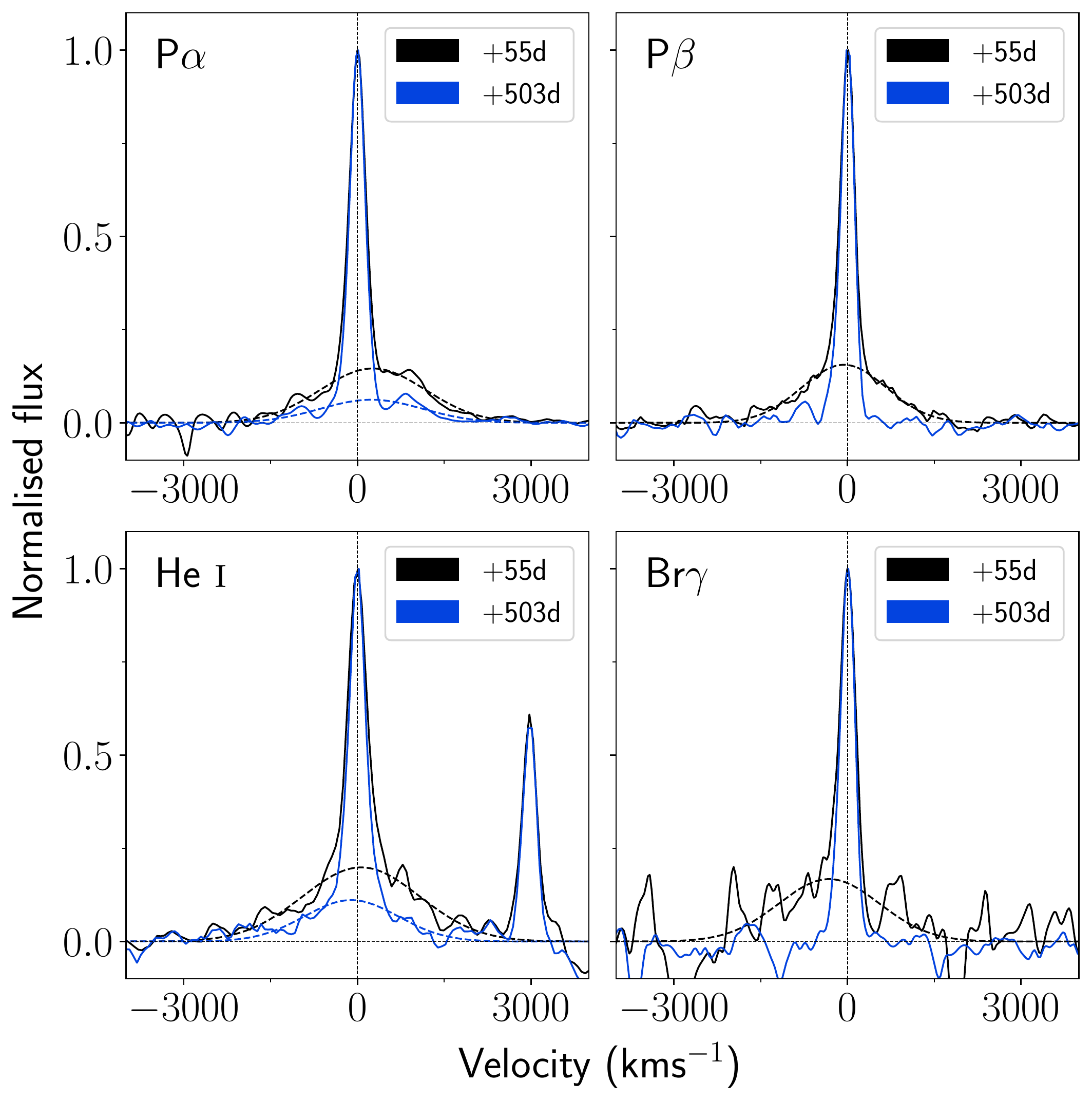}
    \caption{Fits to the emission lines that show more prominent broad features in our GNIRS spectra. The peak of the narrow feature is normalised to 1. Dashed lines show the broad gaussians that were fitted to the spectra to derive the values listed in Table \ref{tab:emission_line_fits}.}
    \label{fig:emission_line_fits}
\end{figure}

In order to quantify the widths and the apparent evolution of the broad velocity components visible in the strong emission lines such as Paschen $\alpha$ and He\,{\sc i}, we simultaneously fit the lines with a broad and a narrow Gaussian, while linearly fitting the local continuum. Measurements are corrected for instrumental broadening, the resolving powers (R) of the observations are listed in Table \ref{tab:Spectral log}. The fitting results are shown in Table \ref{tab:emission_line_fits}, with the fits to Paschen $\alpha$ and He\,{\sc i} $\lambda$10830 shown in Fig.~\ref{fig:emission_line_fits}. The line fluxes of all the broad line profile components decreased significantly between the early and late GNIRS epochs, and as such it is natural to relate this to the transient event. The detection of a broad component in both Paschen $\alpha$ and $\beta$ allows us to determine the line-of-sight extinction affecting the broad line emission, by comparing the ratio to the theoretical value assuming Case B recombination \citep{osterbrock1989,Gaskell2017}. Adopting the Cardelli extinction law \citep{cardelli1989} with $R_V$ = 3.1, this ratio implies 2.5 $\pm$ 0.6 mag of host galaxy extinction in $V$-band, where the uncertainty is derived from the co-variance matrix for the least squares fit. It is interesting to note that this value is a couple of orders of magnitude lower than estimated towards the potential AGN in Section \ref{sec:sed_fit}, implying that the broad emission lines would not originate directly from regions close to the central engine. Alternatively, the lines we observe could be scattered into our line of sight by electrons and dust in the polar regions after being produced close to the SMBH. This phenomenon has been observed in other AGN such as the well known NGC 1068 \citep{antonucci1985}.

\begin{figure}
\includegraphics[width=0.5\textwidth]{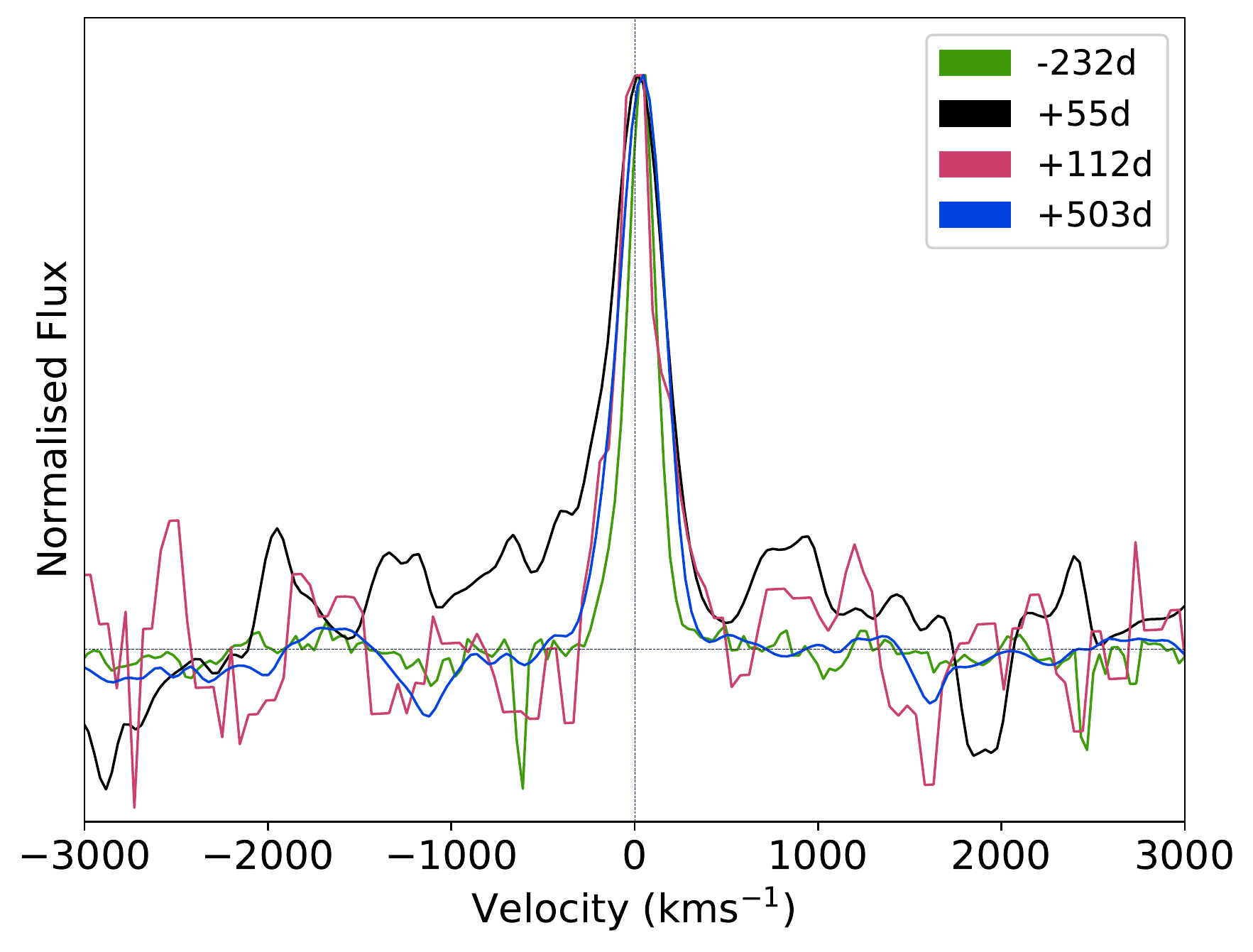}
\caption{Comparison of the Br$\gamma$ emission line evolution over time. The spectral peaks have been normalised to one after subtraction of a linear fit to the continuum. The only spectrum that shows evidence for a broad component is at +55 days (black line).}
\label{fig:Bry_comp}
\end{figure}

\begin{figure}
\includegraphics[width=0.5\textwidth]{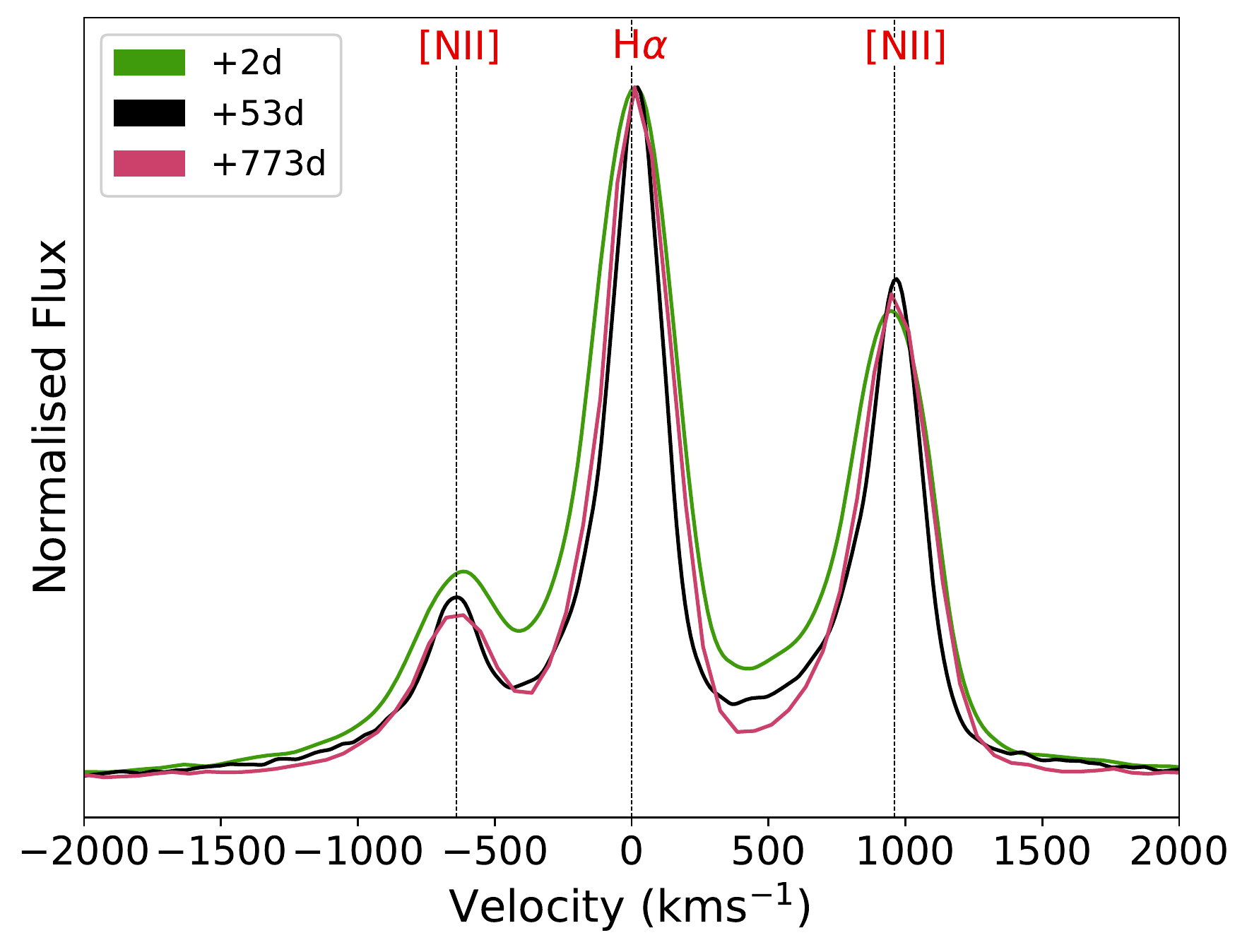} \\
\includegraphics[width=0.5\textwidth]{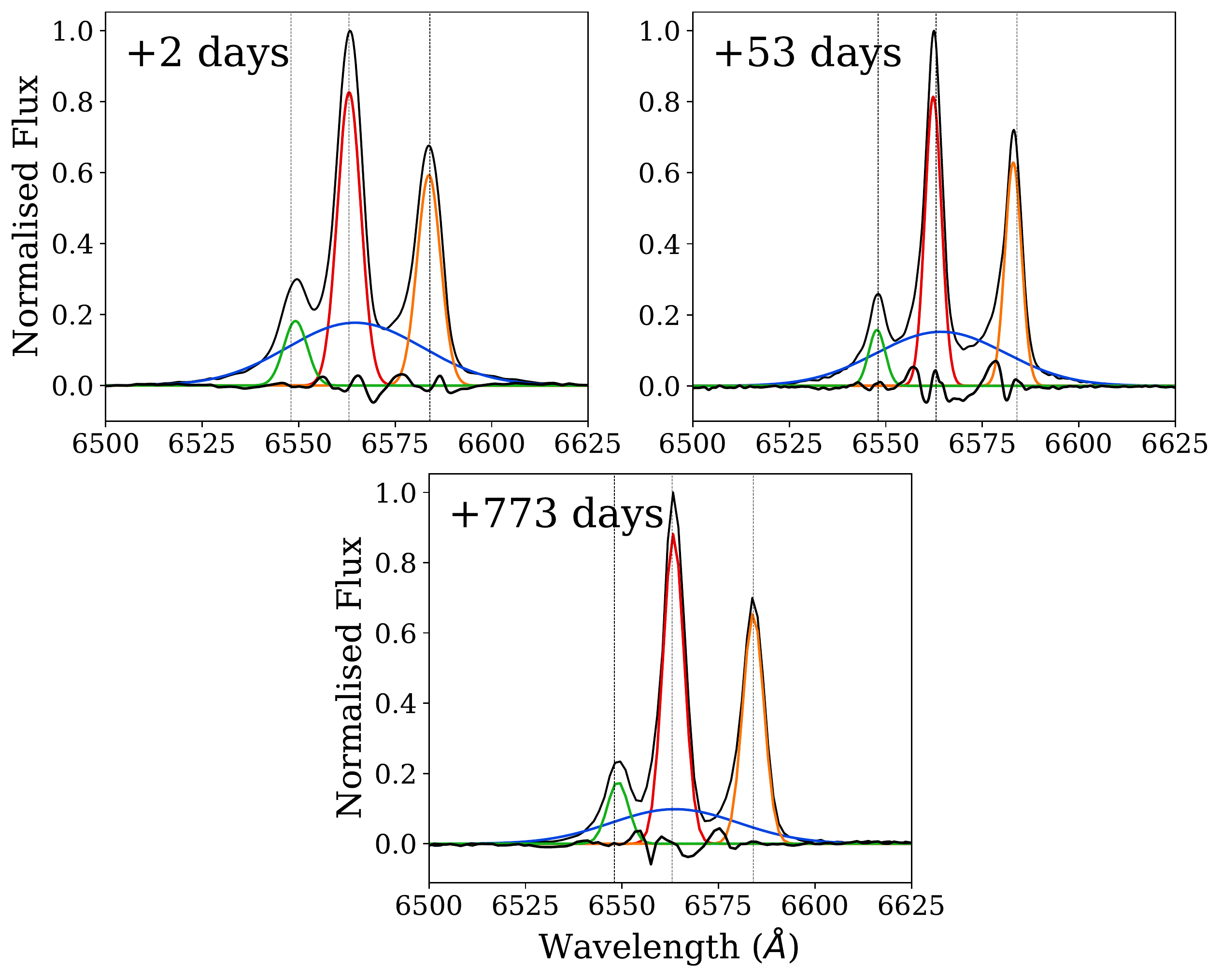}
\caption{Upper panel: Comparison of the H$\alpha$ region in our optical spectra. The peak of the H$\alpha$ line was normalised to one, after subtraction of a linear background. The earliest spectrum has a stronger contribution from the broad component than the two later observations. Lower panel: Multi-component fitting of the features in the H$\alpha$ region. Data is shown in black, and fit residuals are shown along the x axis.}
\label{fig:Halpha_comp}
\end{figure}

We inspect the evolution of the Br$\gamma$ emission line profile using the OSIRIS spectrum that was taken 232 days before AT~2017gbl was discovered. In Fig.~\ref{fig:Bry_comp} we show the evolution of this line. There is no evidence for a broad emission line component in the spectrum taken before the transient's discovery nor in the spectra taken at +112d and +503d, but the spectrum from +55d does show a broad component. This gives a pre-explosion limit of -232d for the appearance of this line, which is similar to the -260d implied by the near-IR lightcurve.

We show the region around H$\alpha$ in Fig.~\ref{fig:Halpha_comp}, with the emission line profile normalised with respect to the peak of H$\alpha$. The earliest spectrum, obtained a few days after the discovery shows evidence for an underlying broad H$\alpha$ feature. This feature appears to have very rapidly declined in flux, with the spectrum taken at +53 days already appearing similar to a spectrum taken 2 years afterwards. Also in Fig.~\ref{fig:Halpha_comp}, we show emission line fits to the spectra. Three narrow components with equal FWHM were simultaneously fitted to the data, along with a broad component. The values derived from these fits are listed in Table \ref{tab:emission_line_fits_optical}. We find that all the spectra show some evidence for a broad component, with a similar FWHM of $\sim$1800 \kms. There is no statistically significant evidence for variability in the narrow emission lines between these spectra.

\begin{table}
\centering
\begin{tabular}{|l|l|l|l|l|l|} \hline
Epoch	& Narrow line FWHM  &Broad line FWHM  \\ \hline
Days		&  km/s               &km/s         \\ \hline
+2  & 327.4 $\pm$ 1.7  &1951 $\pm$ 20  \\ \hline
+53   & 162.1 $\pm$ 2.1 &1898 $\pm$ 26  \\ \hline
+773   & 293.1$\pm$ 3.6 &1790 $\pm$ 100  \\ \hline

\end{tabular}
\caption{Emission line widths obtained from the fits to the optical spectra. The narrow line FWHM refers to the emission lines used for the H$\alpha$ and [N\,{\sc ii}] lines which were kept at the same width, and the broad line width refers to the broad H$\alpha$ emission feature. The velocity given for the narrow lines is with respect to the H$\alpha$ rest wavelength. The narrow lines are not resolved in the observations at +2 and +773 days, all other lines are corrected for instrumental broadening. Uncertainties are derived from the covariance matrix of the fitting parameters.}
\label{tab:emission_line_fits_optical}
\end{table}

\subsubsection{Velocity dispersion measurements} 
\label{sec:vel_dispersion}

In order to make a measurement of the mass of the SMBH hosted by the northern nucleus of IRAS~23436+5257, we make use of the Penalised Pixel Fitting ({\sc ppxf}) routine \citep{Cappellari2017} to measure the stellar line-of-sight velocity distribution (LOSVD) from absorption lines in our spectra. To measure the LOSVD, template stellar spectra are convolved with the corresponding LOSVD, which is paramatrised by a series of Gauss-Hermite polynomials. More details can be obtained in \citet{Cappellari2004} and \citet{Cappellari2017}.

We chose to fit the region from 23300-24750\AA~ in the observer frame which contains the CO absorption band heads (indicated in Fig.~\ref{fig:2017gbl_nir_spec}). This was performed in the GNIRS spectrum from +503 days, as the CO features have high signal to noise and are well resolved at the resolving power R$\sim$1300 of GNIRS with our instrument setup. The aperture extracted is 0.9" wide, corresponding to $\sim$640 pc at the distance of this galaxy. We made use of templates from the Gemini Near-IR Late-type stellar library \citep{Winge2009}, which contains spectra of this wavelength region for 60 stars with spectral types ranging from F7 III to M5 III obtained at spectral resolution of $\sim$3.2\AA~(FWHM). Before fitting, the template spectra are degraded to the resolution of the GNIRS spectrum. We allowed the fitting routine to include a multiplicative and an additive order one Legendre polynomial in the fit, to account for continuum emission arising from the galaxy and the contribution from the transient that is still present at this time. Otherwise, we made use of the default software parameters and in particular the bias, which controls the extent that the higher order Hermite polynomials affect the fit, was left at the default. Given the systematic sources of uncertainty we list below, this choice will not be a dominating source of uncertainty in our measurements.

The fit results in a LOSVD of $\sigma = 81\pm15$ \kms. We make use of the scaling relation given by equation 7 of \citet{Kormendy2013} to derive a BH mass of $\text{log}_{10}(\text{M}_{\text{BH}})= 6.8\pm0.4\,\text{M}_{\odot}$. This relationship has an intrinsic scatter in log(M$_{\text{BH}}$) of $0.29\pm0.03$. Note that this relation is poorly calibrated for galaxies with $\sigma$<100 \kms, as there are only a few such galaxies in the sample used to derive it. Furthermore other authors \citep[e.g.][]{Rothberg2010,Riffel2015} have noted that the value of $\sigma$ measured using the CO bandheads is systematically lower than that measured with the Ca near-IR triplet, in particular for LIRGs. \citet{Riffel2015} find a mean logarithmic offset of -$0.29\pm0.12$ for the measured $\text{M}_{\text{BH}}$ of spiral galaxies from the CO lines, compared to those measured with the Ca lines. As we can not measure the LOSVD from the Ca near-IR triplet in our spectra due to noise, we apply this correction to the BH mass. We prefer this to the relationship derived in \citet{Rothberg2010} for LIRGs, as they derived this exclusively from post merger, single nuclei LIRGs, whereas IRAS~23436+5257 is still a double nuclei system. Applying this correction, we find a BH mass of $\text{log}_{10}(\text{M}_{\text{BH}})= 7.1\pm0.4~\text{M}_{\odot}$. Using this value for the BH mass and the value for the AGN luminosity given by our SED fitting, we can derive an Eddington ratio of $\epsilon = 0.35\pm0.32$. The implications of the BH mass will be discussed in Section \ref{sec:discussion}.

\subsection{Transient SED fit}
\subsubsection{Single IR component}
\label{sec:AT2017gbl_single_comp_fit}

\begin{figure}
\centering
	\includegraphics[trim=0 0 0 0, clip, width=\columnwidth]{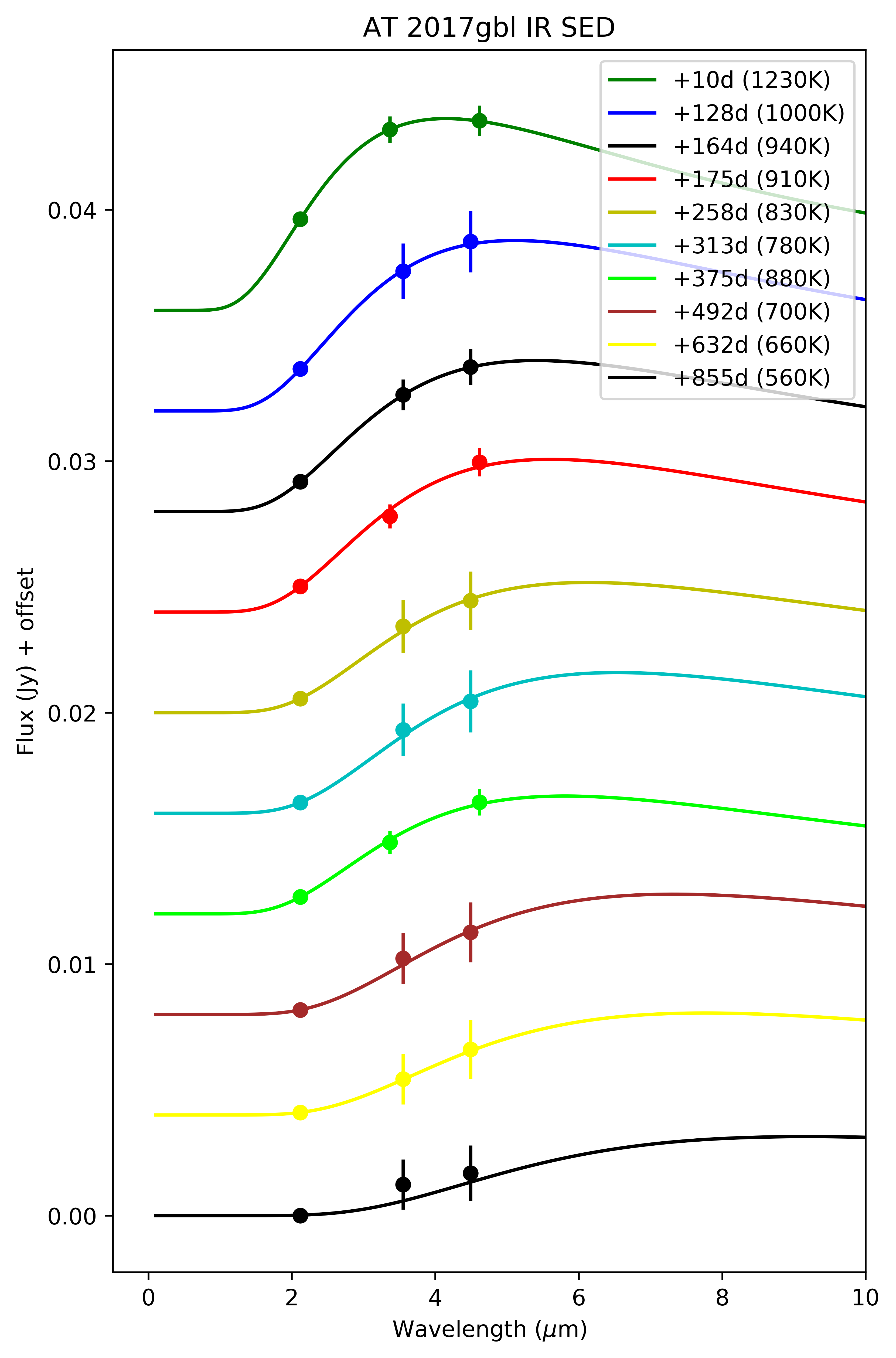}
    \caption{Blackbody fits of the mid-IR and interpolated near-IR $K_s$-band fluxes at the mid-IR epochs. Epochs are offset for clarity, and labelled by days since discovery. Fitted blackbody temperatures are given in parentheses.}
    \label{fig:at2017gbl_bb_fits}
\end{figure}

From the lightcurve of AT~2017gbl, it is clear the transient is growing redder over time, declining slowly in the mid-IR and more rapidly in the near-IR. The evolution of the IR SED is well fitted by a single warm blackbody component with a decreasing blackbody temperature and increasing radius, see Fig.~\ref{fig:at2017gbl_bb_fits}. In this figure the SEDs of the transient are shown at the epochs of the mid-IR observations, with the $K_s$ measurements from NOTCam and NIRC2 interpolated to the mid-IR epochs. Over-plotted are single blackbodies fitted to the near-IR $K_s$-band and the two mid-IR bands from either \textit{Spitzer} or \textit{WISE}. The near-IR $J$ and $H$-bands were not included in the fit, since a second optical blackbody component likely contributes significant flux at these wavelengths, see Section \ref{sec:twocompfit}. The blackbody parameters were estimated using the \textsc{emcee} python implementation of the Markov chain Monte Carlo method \citep[MCMC;][]{foreman2013}. The \textit{Spitzer} epoch of +632 days was fitted to the mid-IR fluxes and a $K_s$-band flux extrapolated from the final four $K_s$-band detections. The \textit{Spitzer} epoch of +855 days was fitted to the mid-IR fluxes and a $K_s$-band flux of zero, with an uncertainty consistent with the upper limit of $K_s$>18.0 from the +726 days near-IR epoch.

\begin{figure*}
\centering
	\begin{tabular}{cc}
        \includegraphics[trim=0 0 0 0, clip, width=0.45\textwidth]{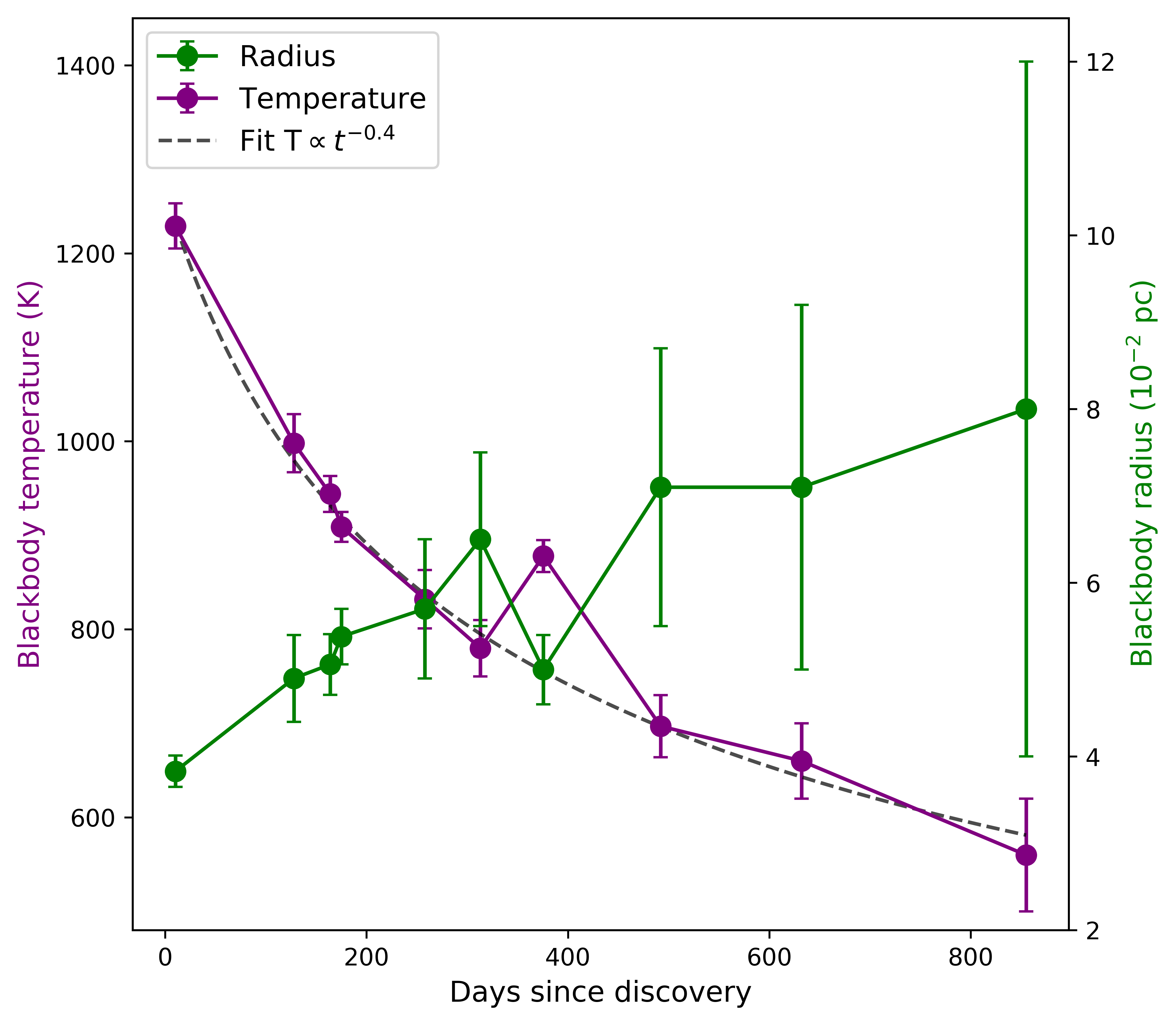} &
        \includegraphics[trim=0 0 0 0, clip, width=0.45\textwidth]{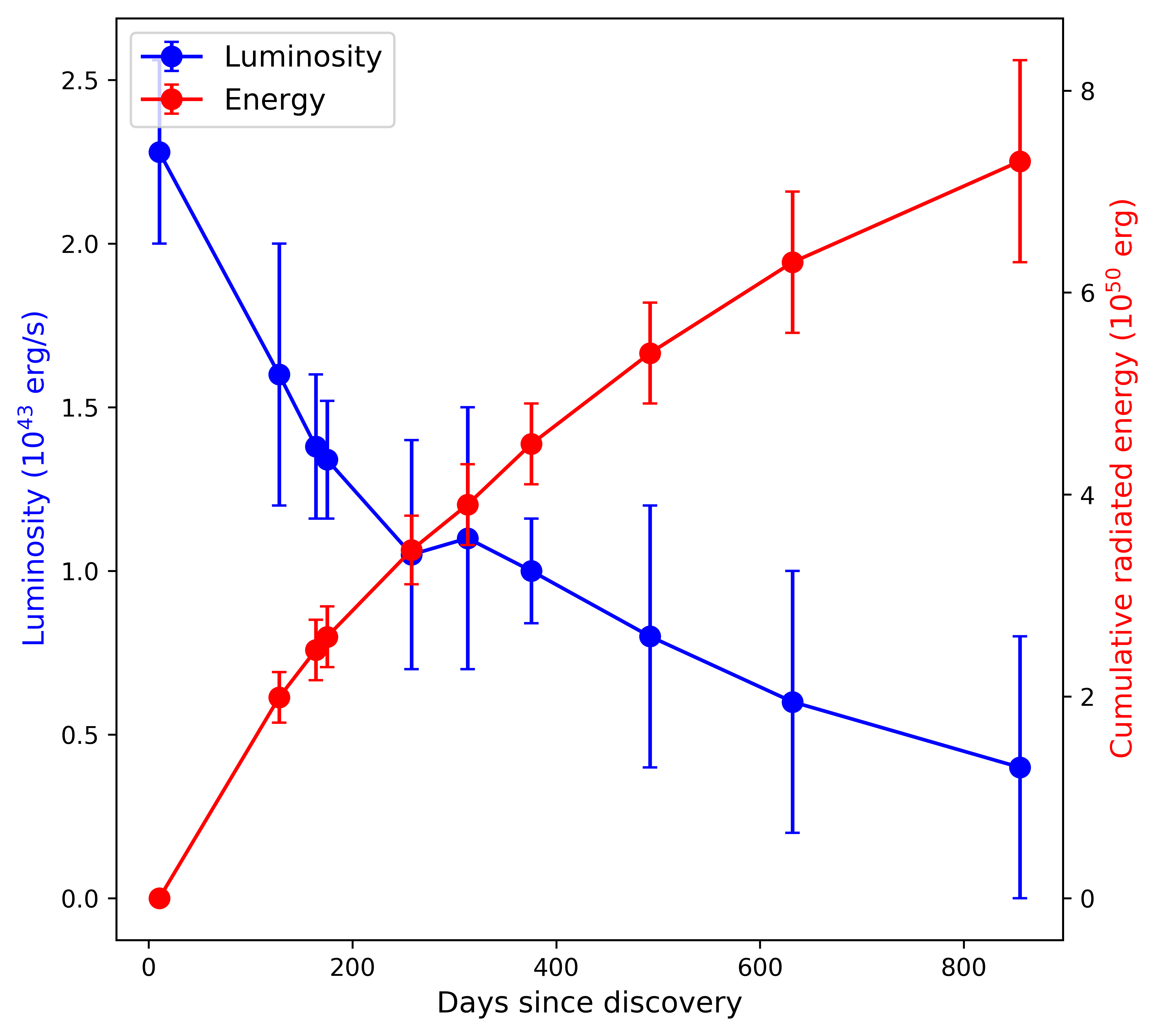} \\
	\end{tabular}
    \caption{Left panel: evolution of the IR blackbody parameters. Right panel: IR blackbody luminosity and the cumulative radiated energy.}
\label{fig:at2017gbl_bb_params}
\end{figure*}

The fitted blackbody parameters are listed in Table \ref{tab:bb_params}, given by the median value of the posterior distributions of blackbody radius and temperature realizations from the MCMC fitting, with 1$\sigma$ errors. Blackbody radius and temperature evolution is shown in the left panel of Fig.~\ref{fig:at2017gbl_bb_params}. As the host-subtracted transient source grows fainter in flux, the uncertainties in flux increase, resulting in the increasing uncertainties of the fitted blackbody parameters. We note that for the epoch of +10 days the temperature estimated from the simple blackbody fit of 1230 $\pm$ 20 K agrees within the uncertainties with the temperature estimated from the full radiative transfer model for the polar dust component at the same epoch of 1300 K. This gives confidence that the parameters obtained by fitting a simple blackbody function to the observed SED give a reasonable description of the properties of the IR emitting region.

\begin{table*}
\centering
\begin{tabular}{|c|c|r|c|c|c|c|} \hline
UT Date	        &MJD&Epoch	    &Radius			&Temperature	&Luminosity&Cumulative radiated energy\\ \hline
			    &&(days)			&($10^{-2}$ pc)	&(K)			&(10$^{43}$erg/s)&(10$^{50}$erg)  \\ \hline
2017-07-18.3    &57952.3&+10     & 3.8 $\pm$ 0.2 &1230 $\pm$ 20  &2.3 $\pm$ 0.3&- \\ 
2017-11-13.1    &58070.1&+128    & 4.9 $\pm$ 0.5 &1000 $\pm$ 30  &1.6 $\pm$ 0.4&2.0 $\pm$ 0.3 \\ 
2017-12-19.8    &58106.8&+164    & 5.1 $\pm$ 0.4 &940 $\pm$ 20   &1.4 $\pm$ 0.2&2.5 $\pm$ 0.3 \\ 
2017-12-30.0    &58117.0&+175    & 5.4 $\pm$ 0.3 &910 $\pm$ 20   &1.3 $\pm$ 0.2&2.6 $\pm$ 0.3 \\ 
2018-03-23.5    &58200.5&+258    & 5.7 $\pm$ 0.9 &830 $\pm$ 30   &1.0 $\pm$ 0.3&3.5 $\pm$ 0.3 \\ 
2018-05-17.1    &58255.1&+313    & 6.5 $\pm$ 1.0 &780 $\pm$ 30   &1.1 $\pm$ 0.4&3.9 $\pm$ 0.4 \\ 
2018-07-18.6    &58317.6&+375    & 5.0 $\pm$ 0.4 &880 $\pm$ 20   &1.0 $\pm$ 0.2&4.5 $\pm$ 0.4 \\ 
2018-11-12.6    &58434.6&+492    & 7.1 $\pm$ 1.6 &700 $\pm$ 30   &0.8 $\pm$ 0.4&5.4 $\pm$ 0.5 \\ 
2019-04-01.5    &58574.5&+635    & 7.1 $\pm$ 2.1 &660 $\pm$ 40   &0.6 $\pm$ 0.4&6.3 $\pm$ 0.7 \\
2019-11-10.3    &58797.3&+855    & 8.1 $\pm$ 3.9 &560 $\pm$ 60   &0.4 $\pm$ 0.4&7.3 $\pm$ 1.0 \\ 
\hline
\end{tabular}
\caption{Blackbody parameters obtained for the 2.2 - 4.5 \micron\ SEDs of AT~2017gbl.}
\label{tab:bb_params}
\end{table*}

Table \ref{tab:bb_params} also lists the luminosity in erg s$^{-1}$ associated with the blackbody using the Stefan-Boltzmann law: 
\begin{equation}
L = 4 \pi R^2 \sigma T^4,
\end{equation}
\noindent where $R$ is the radius of the blackbody in cm, $T$ its temperature in Kelvin and $\sigma$ the Stefan-Boltzmann constant ($5.670\times10^{-5}$ erg cm$^{-2}$ s$^{-1}$ K$^{-4}$). The total energy of AT~2017gbl radiated in the IR can be estimated by integrating the luminosity over the time between the epochs. As shown in in Table \ref{tab:bb_params} and the right panel of Fig.~\ref{fig:at2017gbl_bb_params}, the radiation emitted as a blackbody between +10 and +855 days after discovery totals 7.3 $\pm$ $1.0 \times10^{50}$ erg. This value is a lower limit for the total radiated energy, since the transient was discovered after (or at) the peak, and the transient was still faint but ongoing at +855 days, so the rise to the peak and the late time evolution are not included, and neither is the contribution of a potential cooler component at longer IR wavelengths. The implications of the energy budget of AT~2017gbl on its nature are discussed in Section \ref{sec:discussion}.

\subsubsection{IR echo}
There has been considerable research in recent years on the variability of extra-galactic sources in the mid-IR, made possible to a large extent by the observing strategy and sampling of the AllWISE and NEOWISE-R surveys, as exemplified in Fig.~\ref{fig:iras23436_wise_spitzer}. These studies have resulted in the discovery of multiple mid-IR outbursts with some similarities to AT~2017gbl \citep{vanvelzen2016a,jiang2016,dou2016,dou2017,jiang2017,mattila2018,jiang2018,assef2018b,wang2018,yan2019,yang2019,gromadzki2019,sun2020}, which have been interpreted as thermal emission of dust following absorption of a fraction of the UV/optical light from an energetic transient event. Such `IR echoes' have often been shown to be well described by a single component blackbody with temperatures ranging from $\sim$500 K to $\sim$1500~K, with the upper limit being dictated by the temperature at which dust sublimates \citep[e.g.,][]{draine1984}. A sufficiently energetic optical outburst will sublimate the dust out to a radius which depends on the peak luminosity of the outburst \citep{koshida2014}. As a consequence, the IR emission will only start rising after a time delay, and the IR peak is delayed with respect to the optical peak. The luminosity and timescale of an IR echo depends on the luminosity of the event, the dust geometry and on the dust covering factor, i.e., the fraction of UV-optical radiation from the event intercepted by dust in the surrounding few parsecs \citep{lu2016}.

As the IR fluxes of AT~2017gbl are well fitted with a single component blackbody throughout its evolution, with the temperature decreasing from 1230 K down to 560 K, these IR fluxes are consistent with being dominated by an IR echo following the event's UV/optical display. We note that at the epoch of +375 days the blackbody parameters in Fig. \ref{fig:at2017gbl_bb_params} show a break in their trend. This is related to the bump in the near-IR lightcurve observed at +370 days, see Fig. \ref{fig:at2017gbl_lc}. This could be the result of an influx of UV/optical continuum photons from a secondary outburst following the main outburst. We further discuss this bump in the context of the nature of AT~2017gbl in Section \ref{sec:discussion}.

\subsubsection{Implications of IR echo properties} \label{sec:outburst_epoch}
The IR echo is a response to the underlying transient event, and as such its properties can be linked to that of the transient by making some basic assumptions. We assume a spherically symmetric system where a short pulse of bolometric luminosity $L_{\rm bol}$ is generated from the outburst at $t=0$. As a result of intensive heating, dust particles will sublimate within certain radius $R_{\rm sub}$ where the dust is heated to the sublimation temperature $T_{\rm sub}\simeq 1600\rm\,K$ \citep[the exact value depends mildly on the grain composition and size, e.g.,][]{lu2016}. Beyond the sublimation radius, the temperature $T$ of a grain of radius $a$ at radius $R$ from the source is determined by the equilibrium between heating and radiative cooling:

\begin{equation} \label{eq:equilibirum}
e^{-\tau} \frac{L_{\textrm{bol}}}{4\pi R^2} \pi a^2 Q_{\textrm{abs}} = \lara{Q_{\textrm{abs}}}_{\rm P}\ 4\pi a^2 \sigma T^4.
\end{equation}

\noindent Here $Q_{\textrm{abs}}\simeq1$ is the absorption efficiency factor of of the incident radiation \citep{draine2011} and $\lara{Q_{\textrm{abs}}}_{\rm P}$ is the Planck-averaged absorption efficiency factor \citep{draine1984}. Although dust grains are efficient absorbers of UV/optical photons, the absorption (and hence emission) efficiency drops as the photon wavelength becomes comparable or smaller than the grain size. For the range of temperatures $500\lesssim T\lesssim 1500\,$K and dust radii $a\lesssim 1\rm \,\mu m$, we take the approximation $\lara{Q_{\textrm{abs}}}_{\rm P}\simeq a_\mu (T/1000\rm\, K)$ (where $a_\mu = a/(\rm \mu m)$) appropriate for graphite grains\footnote{Optical properties of interstellar dust grain models have been calculated by \citet{draine1984}, and the tabulated data can be found at \href{https://www.astro.princeton.edu/~draine/dust/dust.html}{https://www.astro.princeton.edu/\~{}draine/dust/dust.html}}. For the astrophysical silicate model, the approximation is $\lara{Q_{\textrm{abs}}}_{\rm P}\simeq 0.3 a_\mu$ without the temperature dependence.

The factor $\mathrm{e}^{-\tau}$ accounts for the flux attenuation by the surviving dust below radius $R$. If the system is optically thin for UV/optical photons, then this factor can be ignored. In the other extreme limit where the reprocessing system is highly optically thick, then all the UV/optical photons from the central explosion are absorbed in a thin layer just beyond the sublimation radius. The IR echo of AT~2017gbl with temperature $T\in(500, 1200)$K comes from radii that are a factor of $(T_{\rm sub}/T)^{2.5}\gtrsim 2$ larger than the sublimation radius. Dust particles at these radii are heated by the re-emitted IR photons from the UV-absorbing layers. We infer from the H recombination lines that the reprocessing system is optically thin to IR photons, so the $\mathrm{e}^{-\tau}$ can also be ignored. The only difference from the UV optically thin case is that the bolometric luminosity $L_{\rm bol}$ should be understood as the IR emission near the sublimation radius, which may be a factor of a few less than the original UV/optical luminosity due to smearing on a timescale of $R_{\rm sub}/c\lesssim 100$ days (as we show below).

In the following, we adopt $R = c t$ and ignore the $\mathrm{e}^{-\tau}$ factor based on the above argument. Taking graphite grain as our fiducial model, Equation \ref{eq:equilibirum} can be rewritten as:

\begin{equation}
T^5 = \frac{L_{\textrm{bol}}}{a_{\mu}} \frac{1000K}{16\pi\sigma c^2} \frac{1}{t^2}.
\end{equation}
This relation means that $T\propto A t^{-0.4}$ where $A$ is a constant:

\begin{equation}
A = \left(\frac{L_{\textrm{bol}}}{a_{\mu}} \frac{1000K}{16\pi\sigma c^2}\right)^{1/5}.
\end{equation}

\noindent As the UV/optical radiation propagates outwards with time, the temperature of the dust re-emitted IR emission drops with time as a power-law. Since $t = t_{\textrm{epoch}} - t_0$ is the time since the outburst, this allows us to determine the outburst epoch $t_0$. In Fig. \ref{fig:at2017gbl_bb_params} is shown the result from fitting $T = A\ t^{-0.4}$ to the observed blackbody temperature evolution using \textsc{emcee}, with the normalization factor $A$ and $t_0$ as free parameters. We do not include the epoch at +375 days in the fit, since this bump is likely due to extra emission of hotter dust on the opposite side from the observer with longer light-path than $ct$ (basically breakdown of our spherical assumption) or a secondary outburst from the continuum source. We note that fitting the temperature evolution only before the bump gives the same result as the complete fit shown here. 

The resulting best estimate of $t_0$ is 142 $\pm$ 11 days. Furthermore, the fitted value of $A$ allows us to estimate $L_{\textrm{bol}}/a_{\mu}$, resulting in $L_{\textrm{bol}}/a_{\mu} = 1.2\pm0.1\times10^{45}$ erg s$^{-1}$. The range for the grain radius $a$ is typically 0.1 - 1 $\mu$m, so $L_{\textrm{bol}}\sim 0.1 - 1 \times10^{45}$ erg s$^{-1}$. The determination of $t_0$ and $L_{\textrm{bol}}$ assumed the outburst to be well described as a short pulse of constant luminosity, while in reality the outburst will have a rising and fading phase. However, the blackbody temperature evolution is well described by the model, so we assume that on the timescales considered here (100s of days for the IR echo), the outburst can be approximated by a short-lived pulse, where the inferred $t_0$ is the peak epoch of the pulse at luminosity $L_{\textrm{bol}}$.

In Section \ref{sec:AT2017gbl_single_comp_fit} we established a peak luminosity of the IR echo of $2.3\times10^{43}$ erg s$^{-1}$, and an energy budget of $\gtrsim7.3\times10^{50}$ erg. A comparison with the peak outburst luminosity of $L_{\textrm{bol}}\sim 0.1 - 1 \times10^{45}$ erg s$^{-1}$ implies a small dust covering factor ($L_{\textrm{IR}}$ / $L_{\textrm{bol}}$) of $\sim1-10$\%, which is consistent with the pre-outburst SED fit of the host, where the contribution to the IR emission by polar dust was small, see Fig. \ref{fig:iras23436_sed}. Furthermore, the energy budget inferred from the IR echo constrains the grain radius $a$ to $\gtrsim0.1\mu$m, as otherwise the luminosity would not be sufficient to fulfill the energy budget within the short timescale ($<$100 days) of the pulse.

Finally, the inferred transient luminosity can be inserted back into Equation \ref{eq:equilibirum} to solve for the sublimation radius where $T(R_{\rm sub}) = T_{\rm sub}$, and we obtain:
\begin{equation}
    R_{\rm sub} = (1.8\times10^{17}\mathrm{\, cm})\, \left(L_{\rm bol}/a_\mu \over 10^{45}\mathrm{erg\, s^{-1}}\right)^{1/2} \left(T_{\rm sub}\over 1600\mathrm{\, K}\right)^{5/2},
\end{equation}
which corresponds to a light crossing time of $R_{\rm sub}/c\simeq 70$ days.

\subsubsection{Two component fit of optical/IR epoch}
\label{sec:twocompfit}

\begin{figure}
\centering
        \includegraphics[trim=0 0 0 0, clip, width=0.45\textwidth]{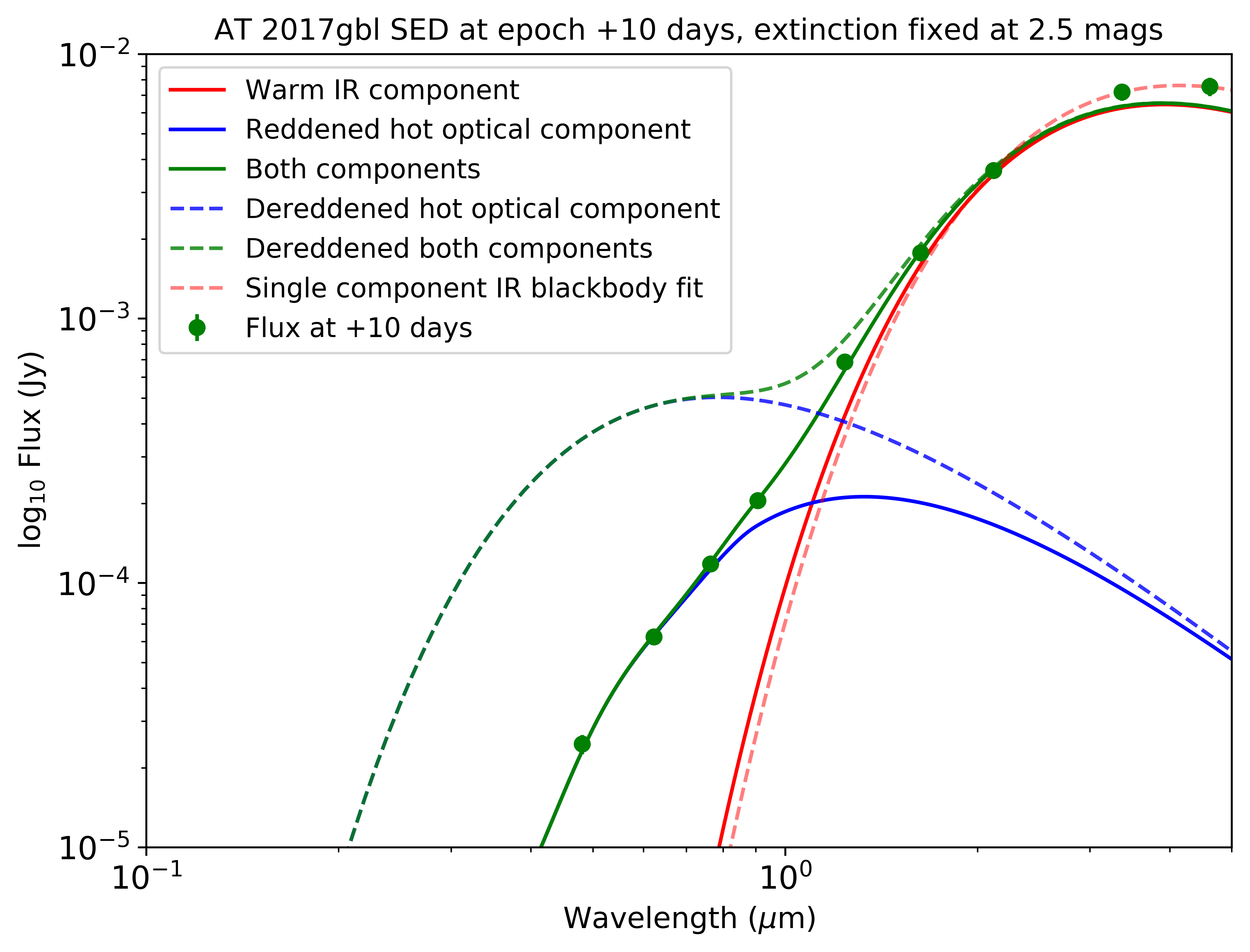}
    \caption{MCMC two components fit to all optical, near-IR and mid-IR fluxes of the epoch +10 days after discovery, with a fixed extinction at A$_V$ = 2.5 mag affecting the optical blackbody. In red is shown the warm IR component, in blue the dust-extincted optical component and in green the sum of the blackbodies and the observed fluxes. The dashed lines indicate the de-reddened intrinsic optical blackbody in blue, and in red the single component fit of the IR blackbody.}
\label{fig:twocomp_plot_20170718}
\end{figure}

In addition to the bright IR detection of AT~2017gbl, the transient has also been detected at optical wavelengths. Fig.~\ref{fig:twocomp_plot_20170718} shows the optical $g$, $r$, $i$ and $z$-band fluxes, as well as the IR fluxes, at the time of the first mid-IR epoch of 2017 July 18, +10 days after discovery, after correcting for Milky Way extinction. In $i$- and $z$-band the fluxes have been derived by interpolating the first two detections. With only a single $g$ and $r$ detection available, it was assumed that in the 8 days between the optical detection and the first mid-IR epoch, the transient had declined in flux in $g$- and $r$-band by the same fraction as in $i$-band. The figure also shows with the dashed red line the earlier derived blackbody fitted to the near-IR $K_s$-band and mid-IR fluxes, as shown in Fig.~\ref{fig:at2017gbl_bb_fits}. It is clear that at shorter wavelengths there is flux in excess to the fitted warm blackbody.

At the longer wavelengths not only the first mid-IR epoch is well fitted with a warm blackbody, but so are all the subsequent epochs. Therefore, we assume that the optical excess at early times has a different origin, and explore whether the excess may be explained by a second blackbody peaking at shorter wavelengths, attenuated by some level of dust extinction. From the ratio of the broad emission lines, which we have connected to the transient event, we have inferred a line of sight extinction to the source of the broad line emission of $\sim$2.5 magnitudes in $V$ (see Section. \ref{sec:line_fit}). Therefore we fit the optical and IR fluxes of the first epoch simultaneously with two blackbody components, one warm component and one hot component, the latter of which we assume is dust-extincted by A$_V$ = 2.5 mag, where we adopt the Cardelli extinction law \citep{cardelli1989} with $R_V$ = 3.1. This assumption leaves four free parameters: the IR blackbody radius and temperature and the optical blackbody radius and temperature. We draw the IR priors from a Gaussian distribution centered on the earlier derived best estimates, with limits of $500 - 2500$~K in temperature and 10$^{-4}$ to 10$^{-1}$ pc in radius. The optical blackbody radius and temperature are drawn from a flat prior ranging between 10$^{-6}$ and 10$^{-2}$ pc, and 3000 and 50000 K, respectively.

Fig.~\ref{fig:twocomp_plot_20170718} shows the resulting blackbody fit in log-log space. In red the warm IR component is shown, in blue the dust-extincted optical component and in green the sum of the blackbodies and the observed fluxes as green points. Also plotted is the de-reddened intrinsic optical blackbody. Fig.~\ref{fig:twocomp_plot_20170718_corner} shows the corresponding corner plot of the two-component fit, with well behaved posterior distributions of all free parameters. The warm IR blackbody parameters from the two-component fit are very similar to the single component fitting results, with a radius of 3.24 $\pm$ 0.1 $\times\ 10^{-2}$ pc and a temperature of 1300 $\pm$ 12 K. The best estimates of the optical blackbody parameters are 0.081 $\pm$ 0.011 $\times\ 10^{-2}$ pc for the radius and 6500$^{+1700}_{-1100}$ K for the temperature, with the uncertainties estimated by varying the line-of-sight extinction between +/- 1$\sigma$. The resulting optical blackbody luminosity is 0.8$^{+0.7}_{-0.4}$ $\times\ 10^{43}$ erg s$^{-1}$.

\begin{figure}
\centering
	\includegraphics[trim=0 0 0 0, clip, width=\columnwidth]{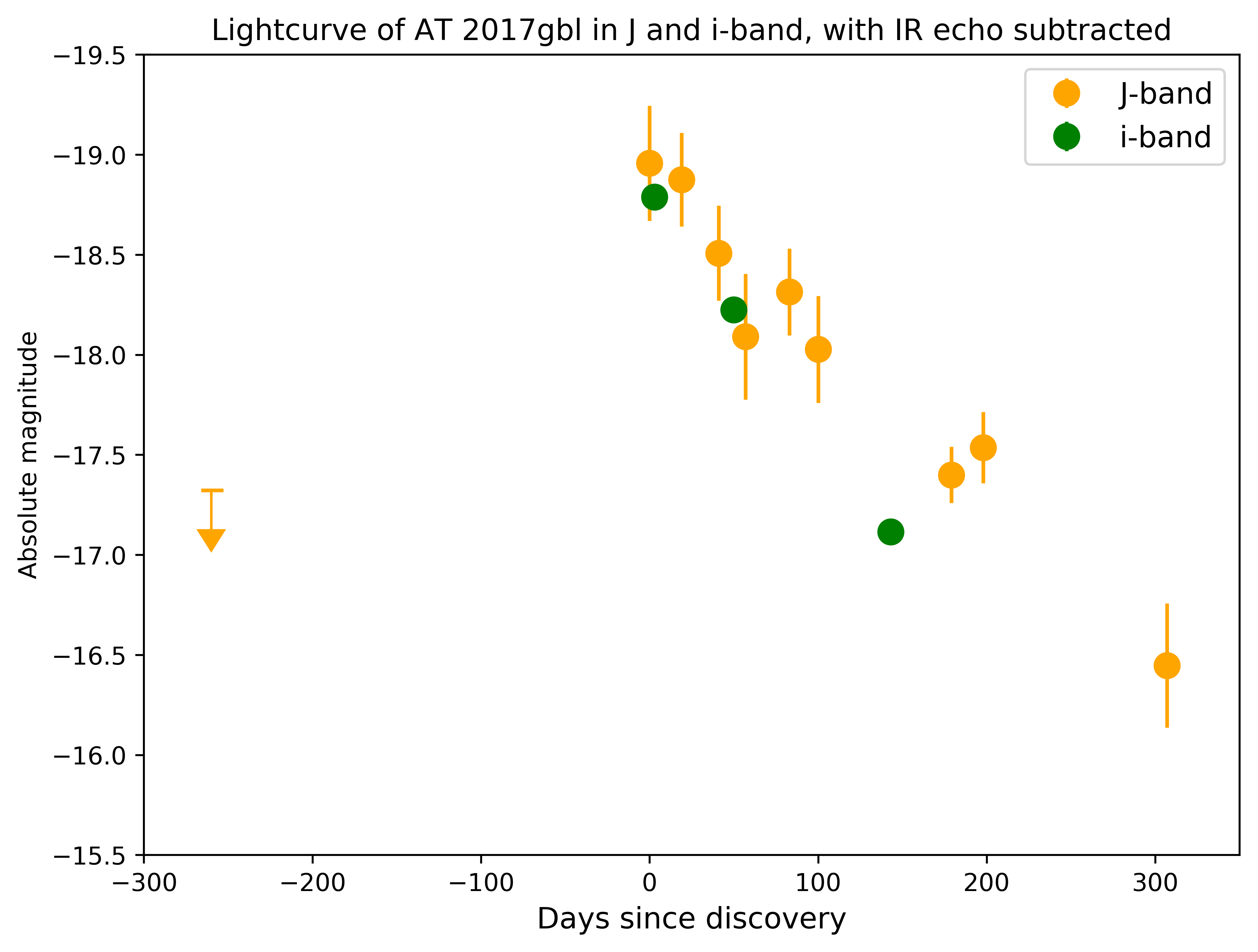}
    \caption{Evolution of AT~2017gbl in $i$ and $J$ after subtraction of the IR component, in absolute magnitudes. The magnitudes are dereddened for a line of sight extinction of A$_V$ = 2.5 mag as determined through Paschen line ratios in Section \ref{sec:line_fit}, and corrected for the Milky Way extinction. Both filters show a similar continuous decline, from which we conclude the IR echo subtracted $J$-band flux traces the evolution timescale of the optical blackbody component well. The $J$-band upper limit at epoch -260 days from 2016 October 21 is the latest pre-outburst epoch available to constrain the evolution of the optical component.}
    \label{fig:at2017gbl_excess}
\end{figure}

The remainder of the mid-IR epochs lack the optical photometry to accurately constrain a double component fit with four free parameters, as the transient declined rapidly in the optical. Only in the near-IR $J$-band is there a significant excess in flux over the warm IR blackbody, with the transient remaining visible for 307 days. In order to trace the evolution of the optical blackbody component, in Fig.~\ref{fig:at2017gbl_excess} we plot the $i$ and $J$-band absolute magnitudes after subtracting the IR blackbody flux inferred from the single component fits, and dereddening for a line of sight extinction of A$_V$ = 2.5 mag and correcting for the Milky Way extinction. Here we assume the single component fits to the $K_s$ and mid-IR fluxes describe the IR blackbody well, as was shown to be the case in the first epoch. As can be seen, both $i$ and $J$-band show a similar continuous decline. As a result we conclude the $J$-band magnitude, with the IR echo contribution subtracted, traces the evolution timescale of the optical blackbody well. This means the optical blackbody has faded below the detection limit at $<$370 days after the discovery. The epoch from -260 days before discovery constrains the rise of the optical blackbody, within the upper limit, while the peak epoch was estimated to be at -142 $\pm$ 11 days in Section \ref{sec:outburst_epoch}.

\subsection{Radio properties of AT~2017gbl}\label{sec:radio_properties}

We detected radio emission from AT~2017gbl at both milli-arcsec (with the EVN and VLBA arrays) and arcsec scales (with VLA and AMI). Using the values from Table \ref{tab:total_radio_flux} we can obtain the spectral behaviour of AT\,2017gbl at different epochs. At +38 days past discovery, we find a 4.4 to 7.6\,GHz spectral index of $1.22\pm0.16$, which becomes less inverted by +99 days, with a value of $0.52\pm0.14$. Additionally, by +33 days the 15.5\,GHz emission had started to decline and was already optically thin, while at 4.4 and 7.6\,GHz it was still optically thick. This is shown by the less inverted spectral index in the same epoch between 4.4 and 15.5\, GHz of $0.98\pm0.12$ and between 7.6 and 15.5\,GHz of $0.79\pm0.20$. We estimate by eye the peak of the radio lightcurve between 13-15 days at 15.5 GHz, and between 150-200 days at 4.4/4.9 GHz and 7.6 GHz.
As seen in Figure \ref{fig:radio_lightcurve}, the radio emission is transparent first at high frequencies (15.5\,GHz) and later at lower frequencies. The detection of the mJy source at milli-arcsec scales indicates a high brightness temperature. These characteristics correspond to a non-thermal, synchrotron origin of the radio emission.

The radio lightcurve of AT\,2017gbl consists of only 9 data points. We therefore tried to fit the radio lightcurve first considering only synchrotron self-absorption \citep[SSA, see][]{weiler02}, and then considering pure free-free absorption (FFA). None of the fits reproduced the observed flux densities.
The dust-obscured AGN potentially present in IRAS~23436+5257 could contribute to the compact emission observed with VLBI, which we have assumed is most likely dominated by AT~2017gbl in Table~\ref{tab:total_radio_flux} and Fig.~\ref{fig:radio_lightcurve}, see Section \ref{sec:radio_host_corr}. The observed luminosity at 4.4/4.9 GHz is minimal at the EVN epoch of +587 days, which puts an upper limit on the quiescent AGN luminosity at 4.4/4.9 GHz of <1.96 $\pm$ 0.20 $\times\ 10^{28}$ erg s$^{-1}$ Hz$^{-1}$, or <40\% of the observed peak luminosity at 4.4/4.9 GHz. Given the timescales of our observations, it is likely that AT~2017gbl is still contributing at the lower frequencies and the EVN epoch does not represent the quiescence level of the host. Additional VLBI observations over the next years will help establish if the compact radio source disappears completely and, if not, what the level of the quiescent AGN contribution is.

\subsubsection{Implications of radio lightcurve}
The radio emission comes from synchrotron emission by electrons accelerated by the shock driven by the outflow into the surrounding medium. The shock radius typically expands as a power-law function with time. For instance, in the Sedov-Taylor regime (appropriate after the shock has decelerated to non-relativistic speeds), the shock radius increases as $R\propto t^{2/(5-k)}$ and the shock speed decreases as $v\propto t^{-(3-k)/(5-k)}$ for a density profile $n(R)\propto R^{-k}$.

The radio spectrum evolution requires significant low-frequency absorption at early time due to either FFA or SSA. In either the FFA or SSA case, the characteristic absorption frequency at which the optical depth equals to unity evolves as a decreasing power-law function of time $\nu_{\rm a}\propto t^{-p}$. Since the flux at a given frequency $\nu$ reaches the maximum when $\nu_{\rm a}(t) \simeq \nu$, we can infer the peak times (since outburst) at two different frequencies as follows:
\begin{equation}\label{eq:power-law}
    {\nu_1\over \nu_2} \sim \left(t_{\rm peak, \nu_1}\over t_{\rm peak,\nu_2}\right)^{-p}.
\end{equation}
For $\nu_1 = 15.5\rm\, GHz$ and $\nu_2=7.6\rm\, GHz$, we have $t_{\rm peak,\nu_1} \simeq t_0 + 14\,$d and $t_{\rm peak,\nu_2} \simeq t_0 + 175\,$d (where $t_0$ is the pre-discovery explosion time), and hence $t_0 \simeq 150/(2^{1/p}-1)\,$ days. This is in rough agreement with that inferred from the dust temperature evolution in Section \ref{sec:outburst_epoch}, provided that $p\sim 1$ (for $p\in (0.5, 1.5)$, we have $t_0\simeq 50$ to 250 d). In the following, we discuss the two absorption scenarios for which we calculate the power-law index $p$.

In the FFA case, the optical depth is given by $\tau_{\rm ff}(\nu) \simeq 3.4\times10^{-28} T_4^{-1.3} \nu_{15.5\rm\,GHz}^{-2.1}\ \mathrm{EM}$, where $\mathrm{EM} = \int n^2 \mathrm{d}\ell\sim n^2 R$ is the emission measure \citep{draine2011} in units of $\rm cm^{-5}$. For electron temperature $T\gtrsim 10^4\rm\, K$ and shock radius $R\lesssim200$ light days, we obtain a lower limit on the density of the surrounding medium $n\gtrsim 10^5\,\mathrm{cm^{-3}}$. It is possible to estimate EM by the luminosity of Paschen $\alpha$ recombination line, by assuming a uniform H\textsc{ii} cloud whose size is given by the linewidth according to Keplerian motion. The resulting EM is indeed much greater than $10^{28}\rm\, cm^{-5}$, meaning that the low-frequency radio emission is likely free-free absorbed if the shock is buried below the HII cloud. However, the estimate is subjected to uncertainties of clumpiness and electron temperature. More generally, for a power-law density profile $n(R)\propto R^{-k}$, we have $\mathrm{EM}\propto R^{1-2k}\propto t^{2(1-2k)/(5-k)}$, so we obtain $\nu_{\rm a}\propto t^{(1-2k)/(5-k)}$. Thus, we have $p\in (0.5, 1.5)$ for density profile $k\in (1.4, 2.4)$.

In the SSA case \citep{chevalier1998}, the specific luminosity near the self-absorption frequency $\nu_{\rm a}$ is given by $L_{\nu_{\rm a}} \simeq 4\pi^2 R^2 (2\nu_{\rm a}^2/c^2) \gamma_{\rm a}m_{\rm e} c^2$, where $\gamma_{\rm a}$ is the Lorentz factor corresponding to a characteristic synchrotron frequency of $\nu_{\rm a}$. For shock speed $v = \beta c$ and pre-shock medium density $n$, the magnetic field strength in the shocked region is given by $B \simeq \sqrt{16\pi \epsilon_{\rm B}n m_{\rm p} v^2} = (0.87 \mathrm{\,G})\, \epsilon_{\rm B,-1}^{1/2} n_4^{1/2} \beta_{-1}$, where $\epsilon_{\rm B}$ is the fraction of energy in B-fields. Then, the Lorentz factor is given by $\gamma_{\rm a} \simeq 80\, \nu_{\rm a,15.5\,GHz}^{1/2} \epsilon_{\rm B,-1}^{-1/4} n_4^{-1/4} \beta_{-1}^{-1/2}$. Thus, the shock radius is well constrained by the observed $L_{\nu_{\rm a}}$ and $\nu_{\rm a}$,
\begin{equation}\label{eq:SSA-radius}
    R \simeq (8.7\times10^{15}\mathrm{\,cm})\, L_{\nu_{\rm a},29}^{1/2} \nu_{\rm
    a, 15.5}^{-5/4} \epsilon_{\rm B,-1}^{1/8} (n_4\beta_{-1}^2)^{1/8}.
\end{equation}
Ignoring the weak dependencies on $\epsilon_{\rm B}$, density $n$ and shock speed $\beta$, we obtain $R\propto L_{\nu_{\rm a}}^{1/2} \nu_{\rm a}^{-5/4} \propto t^{2/(5-k)}$. We plug in $\nu_1 = 15.5\rm\, GHz$, $L_{\rm peak, \nu_1}=10^{29}\rm\, erg\,s^{-1}\, Hz^{-1}$, $\nu_2=7.6\rm\, GHz$, and $L_{\rm peak, \nu_2}=8\times 10^{28}\rm\, erg\,s^{-1}\, Hz^{-1}$, and then obtain $(t_{\rm peak, \nu_1}/t_{\rm peak, \nu_2})^{2/(5-k)} \simeq 0.5\simeq \nu_2/\nu_1$, and hence $p\simeq 2/(5-k)$. Thus, the power-law index in Equation (\ref{eq:power-law}) is $p\in (0.5, 1.5)$ for density profile $k\in (1, 3.7)$.

Another potentially useful quantity that can be derived from the radio data is the total energy of the outflow. Regardless of the absorption scenario, the number of electrons radiating at frequency $\nu = 15.5\rm\, GHz$ near the flux peak (when the system is marginally optically thin) can be estimated by $N_{\rm e} = L_{\nu}/P_\nu$, where $P_\nu \simeq e^3B/m_{\rm e} c^2$ is the synchrotron specific power. One way of estimating the minimum energy of the outflow is $E_{\rm min}^{(1)}\sim \epsilon_{\rm e}^{-1} N_{\rm e} \gamma m_{\rm e} c^2$, where $\epsilon_{\rm e}\sim 0.1$ is the fraction of energy in electrons. Thus, we obtain
\begin{equation}
    E_{\rm min}^{(1)} \sim (5\times 10^{47}\mathrm{\,erg})\, \epsilon_{\rm B,-1}^{-3/4} (n_4\beta_{-1}^2)^{-3/4}.
\end{equation}
Our fiducial value for $\epsilon_{\rm B}$ is conservative, because it is usually inferred to be much less than 0.1 (in that case, the total energy is much larger). Another way of estimating the outflow energy is to add up the total energy of protons behind the shock $E_{\rm min}^{(2)} \sim (4\pi/3)R^3 n \beta^2 m_{\rm p}c^2/2 \simeq (3\times10^{47}\mathrm{\,erg})\, R_{16}^3 n_4\beta_{-1}^2$. From Equation \ref{eq:SSA-radius}, we know that $R\gtrsim 10^{16}\rm\, cm$, because otherwise the emission at 15.5 GHz will be self-absorbed. Increasing the product $n\beta^2$ will decrease $E_{\rm min}^{(1)}$ but increase $E_{\rm min}^{(2)}$. Overall, we constrain the total energy of the outflow to be more than $3\times10^{47}\rm\, erg$.

Additionally, we provide a lower limit for the total number of electrons in the shocked region\footnote{The number of electrons with Lorentz factor order unity is at least a factor of $\gamma$ larger than that of the emitting electrons $N_{\rm e}$ (near Lorentz factor $\gamma$).} $N_{\rm tot}\gtrsim N_{\rm e}\gamma$. And then, in the SSA scenario, the density of the surrounding medium is constrained to be $n\sim N_{\rm tot}/R^3\gtrsim 10^{4}\mathrm{\,cm^{-3}} \epsilon_{\rm B, -1}^{-0.5} \beta_{-1}^{-1}$. Since we also obtain $n\gtrsim 10^{5}\rm\, cm^{-3}$ in the FFA scenario, we conclude that the circumnuclear medium of AT~2017gbl is unusually dense, which may be the physical reason why the source is radio bright.

\subsection{X-ray properties of AT 2017gbl} \label{sec:xray_transient}
The luminosity of the X-ray source observed at the position of AT~2017gbl at +118 days was $L_{\text{0.5-8 keV}} = 3.2 \times10^{41}$ erg s$^{-1}$. In addition to a potential contribution by the transient, several emission components of the host galaxy (X-ray binaries, hot gas and AGN) can also contribute to the observed luminosity. Lacking a quiescent X-ray epoch, we estimate the host galaxy contribution using an empirical relationship between the X-ray luminosity of a LIRG and its star formation rate: $L_{\textrm{0.5-8 keV}}$  $\approx$ 3.7$\times10^{39\pm0.4}$ $\frac{SFR}{({\rm M}_{\odot} {\rm yr}^{-1})}$ erg s$^{-1}$ \citep[e.g.,][using the scaling factor for unresolved galaxies]{mineo2012}. Here the X-ray emission originates primarily from the galaxy's population of X-ray binaries and is expected to remain constant on the timescales we consider.
Following \citet{mineo2012} we estimated the SFR following their Equations 9 and 10, assuming a negligible fraction of the IR luminosity being due to the old stellar population and a negligible amount of SFR observable in the near-UV. In this way we obtain a SFR of 69 $M_{\odot}$ yr$^{-1}$ using exactly the same approach as \citet{mineo2012} in their analysis. This corresponds to $L_{\text{X}} \sim 3 \times10^{41}$ erg s$^{-1}$. Therefore, the observed X-ray emission may be entirely explained by the host galaxy.

\section{Discussion} \label{sec:discussion}
In this section we discuss the nature of AT~2017gbl implied by various key observations. We consider the scenarios of a SN, a CLAGN, and a TDE. We aim to explain the following:

\begin{itemize}
\setlength\itemsep{1em}
	\item The IR SED of AT~2017gbl is consistent with a single component blackbody throughout its evolution, indicating dust emitting at a narrow range of temperatures and a negligible amount of foreground extinction to the IR emitting region. Its integrated radiation emitted from +10 to +855 days after discovery is $7.3\times10^{50}$ erg. This value is a lower limit to the total radiated energy of the underlying transient, as AT~2017gbl was discovered at, or post, the IR peak brightness, and at mid-IR wavelengths has not fully faded yet.
	
	\item The GNIRS spectrum obtained at +55 days past discovery shows prominent, broad Paschen, Brackett, He\,{\sc i} and O\,{\sc i} emission features with FWHM $\sim$2000~km s$^{-1}$, which have diminished or disappeared in the GNIRS spectrum at +503 days. The pre-outburst near-IR $K$-band OSIRIS spectrum does not show signs of broad emission features in the Brackett lines. Similarly, the optical ISIS spectrum obtained at +2 days past discovery shows a prominent broad H$\alpha$ feature, which diminished over time.
	
	\item The discovery epoch shows flux at optical wavelengths that is not explained by the IR echo. Assuming the optical emission is affected by a dust extinction of A$_V$ = 2.5 mag, inferred from the line ratio of the broad Paschen emission lines, it is well fitted with a blackbody of $\sim$6500 K.
	
	\item The radio emission is of synchrotron origin, peaking around 13-15 days past discovery at 15.5 GHz with observed host-subtracted peak luminosity of $8.4\times10^{28}$ erg s$^{-1}$ Hz$^{-1}$, and 150-200 days past discovery at 4.4/4.9 GHz with an observed peak luminosity of 5.2 $\times10^{28}$ erg s$^{-1}$ Hz$^{-1}$.
	
	\item The rise time to the peak IR luminosity is constrained to $\lesssim$140 days, as inferred from the IR blackbody temperature evolution. The subsequent decline of the IR echo is traced by the mid-IR observations, where the transient has faded close to the detection limit at +855 days. The evolution of the optical component is loosely constrained by the IR echo subtracted $J$-band lightcurve with a rise time of <260 days and decline to the detection limit within <370 days.
\end{itemize}

\begin{figure*}
\centering
	\includegraphics[angle = 0, origin = c, trim=0 0 0 0, clip, width=0.8\textwidth]{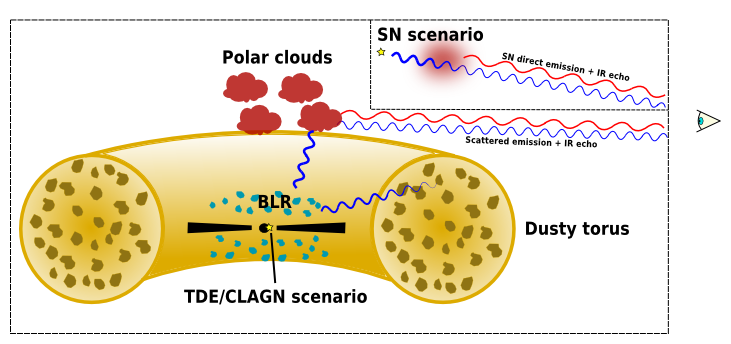}
    \caption{A schematic diagram showing the sources of IR and optical emission (and absorption) in the scenarios for AT~2017gbl either as a SN, or a TDE/CLAGN. Note the components are not to scale. In all scenarios the IR emission is interpreted as an IR echo, where UV/optical emission from the transient is absorbed and re-radiated at IR wavelengths. The optical emission in the case of a SN is considered direct dust-attenuated emission, whereas in the case of a TDE/CLAGN it is considered scattered light originating in the BLR. Direct emission from a TDE/CLAGN is not visible due to the presence of the dusty torus along the line of sight. It must be noted that the potential SN location is not constrained, however, there is 2.5 mag of extinction along the line of sight.}
    \label{fig:at2017gbl_schematic}
\end{figure*}

To aid the discussion, Fig.~\ref{fig:at2017gbl_schematic} shows a schematic of the geometry and origin of the different emission and absorption components in the three different scenarios discussed here.

\subsection{Supernova}\label{sec:sn_discussion}
SN types other than Type II are ruled out, as the IR spectra show prominent broad hydrogen lines, such as Paschen $\alpha$ and $\beta$, which we have associated with the transient through its evolution. Based on the energy budget of AT~2017gbl inferred from its integrated IR luminosity, normal Type II SNe are also excluded as a viable scenario, as they typically have a total bolometric radiated energy output of less then 10$^{49}$ erg \citep{utrobin2007}.

Notable exceptions are the most luminous and long-lasting Type IIn SNe such as SN 2010jl \citep{fox2013, andrews2011, fransson2014}, SN 2015da \citep{tartaglia2019} and SN 1988Z \citep{aretxaga1999}, which have been observed to emit $>10^{50}$ erg and display prominent hydrogen emission lines from the interaction of the SN ejecta with H-rich circumstellar gas. Furthermore, many Type IIn SNe have been observed to show prominent long-lasting mid-IR emission, most likely due to re-radiation by pre-existing dust \citep{fox2011}.

\begin{figure}
\centering
	\includegraphics[trim=0 0 0 0, clip, width=\columnwidth]{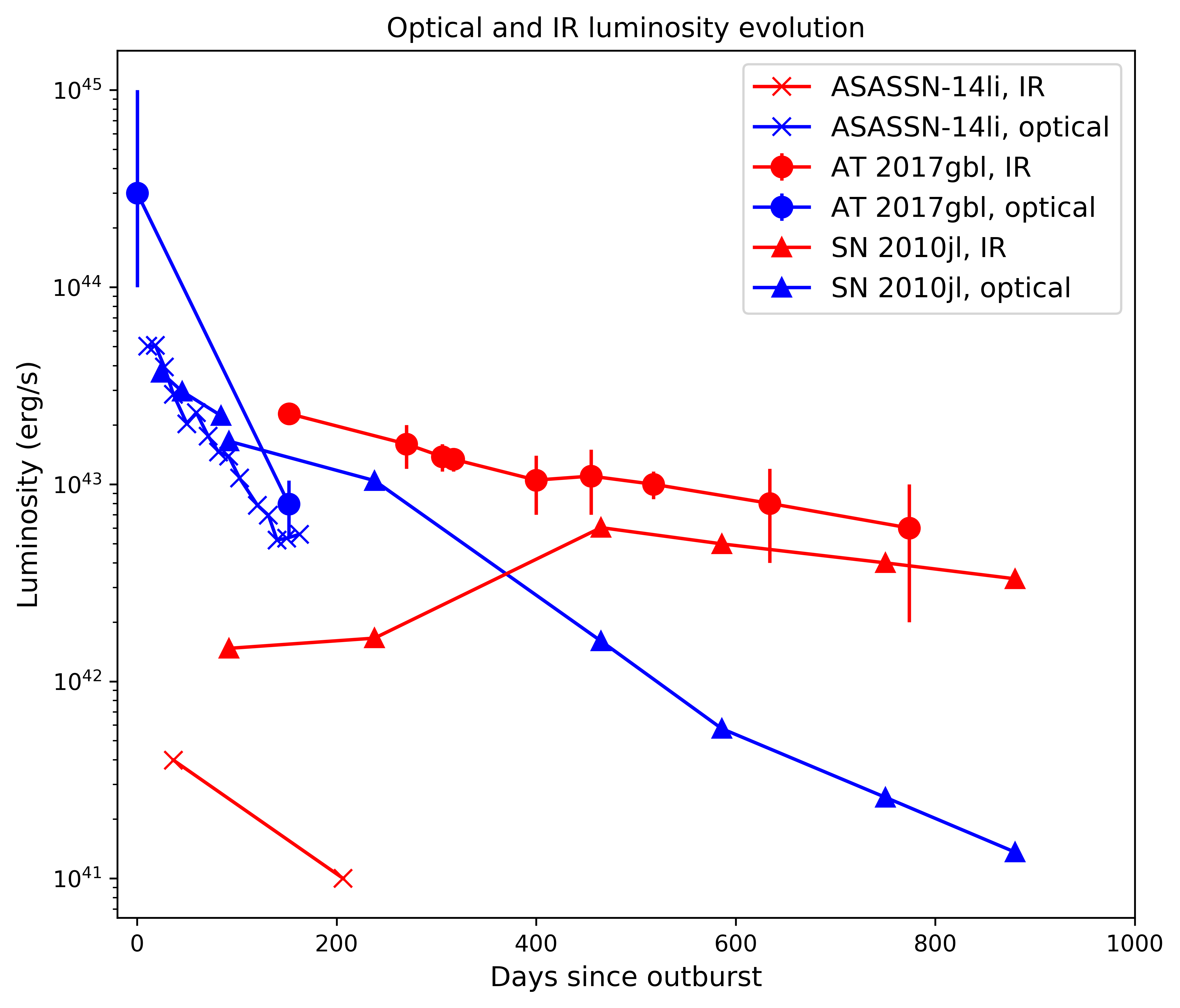}
    \caption{Evolution of the blackbody luminosities of the optical and IR components of AT~2017gbl, SN 2010jl \citep{zhang2012,fransson2014}, and ASASSN-14li \citep{holoien2016,jiang2016}. The first three optical epochs of SN 2010jl are based on bolometric luminosities from \citet{zhang2012}, which should be close to its optical blackbody luminosity because the IR component was negligible at early times. The x-axis start at the outburst epoch (-142 days) of AT~2017gbl inferred in Section \ref{sec:outburst_epoch}, the explosion epoch (MJD 55478) of SN~2010jl \citep{fransson2014}, and the discovery epoch (MJD 56983) of ASASSN-14li \citep{holoien2016}. The optical luminosity of AT~2017gbl at day 0 is the range inferred from fitting the blackbody temperature evolution in Section \ref{sec:outburst_epoch}.}
    \label{fig:bb_lum_comp}
\end{figure}

The coverage of SN 2010jl and SN 2015da spanned several years, and extended from the optical to the mid-IR, which allows for a detailed comparison with AT~2017gbl. The SEDs of SN 2010jl \citep{fransson2014} and SN 2015da \citep{tartaglia2019} were shown to be well represented by a combination of a hot blackbody peaking in the optical and a warm blackbody peaking in the IR, with the IR components lagging the optical components in luminosity by 400-600 days, see Fig.~\ref{fig:bb_lum_comp} for the IR and optical blackbody luminosity lightcurves of SN 2010jl. For both SN 2010jl and SN 2015da, the optical component was considered direct photospheric emission from the SN and emission originating in the interaction between the ejecta and the circumstellar gas, and the IR component interpreted as an IR echo. We assume a similar scenario for AT~2017gbl, see Fig.~\ref{fig:at2017gbl_schematic}, where the optical emission from AT~2017gbl is direct SN emission, affected by a line-of-sight extinction of A$_V$ = 2.5 mag inferred from the Paschen line ratio. 

\subsubsection{Energetics}
As shown in Fig.~\ref{fig:bb_lum_comp}, the optical blackbody luminosity of SN 2010jl peaked around discovery at $\sim4 \times 10^{43}$ erg s$^{-1}$ (similar to SN 2015da), and in the first 100 days declined by $\sim$50\%. The total radiated energy of SN 2010jl, excluding the IR echo, was estimated to be $6.5 \times 10^{50}$ erg, very similar to SN 2015da at $6.3 \times 10^{50}$ erg. This already makes SNe 2010jl and 2015da among the most energetic Type IIn SNe observed to date. In both cases the very large energetics were explained by the efficient conversion of the kinetic energy into radiation requiring extremely large mass-loss rates of $\sim$0.1 and 0.6~M$_{\odot}$~yr$^{-1}$, respectively \citep{fransson2014,tartaglia2019}. In comparison, the optical luminosity of AT~2017gbl at the peak, inferred from the IR blackbody temperature evolution in Section \ref{sec:outburst_epoch}, was $10^{44}$ - $10^{45}$ erg s$^{-1}$. According to our observations it declined by at least an order of magnitude down to 0.8$^{+0.7}_{-0.4}$ $\times\ 10^{43}$ erg s$^{-1}$ at +10 days after the discovery (or +152 days after the peak epoch inferred in Section \ref{sec:outburst_epoch}). This results in a total radiated energy for AT~2017gbl, excluding the IR echo, of $7 \times 10^{50}$ to $7 \times 10^{51}$ erg over the first 150 days since the peak, which corresponds to $\sim$1 to 10 times the energy budget of SN 2010jl and SN 2015da in the optical.

The energetics of the IR echo of AT~2017gbl are better constrained, and clearly exceed those of SN 2010jl and SN 2015da. The IR components of SN 2010jl and SN 2015da peaked at luminosities well below that observed for AT~2017gbl, by factors of at least $\sim$4 and 15, respectively, see Fig.~\ref{fig:bb_lum_comp}. As a consequence, the total energy emitted by their IR echoes was significantly less than observed in the case of AT~2017gbl, at $2.7 \times 10^{50}$ erg for SN 2010jl and $8.5 \times 10^{49}$ erg for SN 2015da (up to day 1233). The energy radiated by the IR echo of AT~2017gbl is $7.3 \times 10^{50}$ erg, which as discussed in Section \ref{sec:AT2017gbl_single_comp_fit} is considered a lower limit.

The total radiated energies of the IR echoes of SNe 2010jl and 2015da amount to $\sim$40\% and $\sim$10\%, respectively, of their optical components. Here it is noteworthy that SN 2010jl is one of the most luminous Type IIn SNe observed to date, and brighter at 4.5 \micron\ than the brightest mid-IR detections of the \citet{fox2011} sample of Type IIn SNe. Assuming that most of the energy of AT~2017gbl in the optical was radiated in the first 150 days given its rapid J-band decline, the energy radiated by its IR echo lies between 10 and 100\% of the energy radiated in the optical. If this fraction was similar to that of SN 2010jl and SN 2015da, the total energy associated with AT~2017gbl would be substantially greater than the total energies of SNe~2010jl and 2015da.

\subsubsection{IR emission lines}
The emission lines that we can unambiguously associate with AT~2017gbl based on their evolution are Paschen $\alpha$ - $\beta$, He\,{\sc i} $\lambda$10830, O\,{\sc i} $\lambda\lambda$8445 and 11290, all with a width of $\sim$ 2000 km s$^{-1}$, see Table \ref{tab:emission_line_fits}. Both SN 2010jl and SN 2015da show the same emission lines, but they are narrower. The Paschen lines of SN 2015da evolve from $\sim$ 600 km s$^{-1}$ at epoch +26 days to 1100 - 1300 km s$^{-1}$ at epoch +607 days. Intermediate width lines of a few 1000 km~s$^{-1}$ are seen in 1988Z-like SNe such as 2005ip and 2006jd \citep{Stritzinger2012}, but they are accompanied by much stronger H emission lines with FWHM of $\sim$10$^{4}$~km~s$^{-1}$~for the first few hundred days, which we would likely see in our spectra.

\subsubsection{Radio luminosity and timescale}
The radio peak luminosity and the time to peak of AT~2017gbl agree well with those observed for the radio emitting SNe of Type IIn and, marginally also, for SNe of Type Ibc, whose time to peak is usually well below 100 days \citep{pereztorres15}.
However, in the radio, SN 2010jl was fainter than other Type IIn SNe and was only detected with the VLA after $\sim$500 days from the explosion. The 5 GHz emission was estimated to have peaked around 900 days from the explosion, implying that FFA was likely the dominant process resulting in the slow rise to the maximum \citep{chandra2015}. Similarly, in the case of the extremely luminous and long-lasting Type IIn SN 1988Z that was observed to radiate 2 $\times 10^{51}$ erg \citep{aretxaga1999} the radio emission was observed to peak at 5 GHz at an epoch of $\sim$900 days as a result of FFA arising in the dense circumstellar medium (CSM) \citep{williams2002}. 

It is clear that the radio evolution of AT~2017gbl differs significantly from that of SNe 2010jl and 1988Z. First, the lightcurve was not fitted well by SSA or FFA models, see Section \ref{sec:radio_properties}. A possible explanation is that the transient is not well explained by spherically expanding material, as expected in the case of, e.g., a SN. Second, assuming an outburst epoch of -142 days before discovery, at 15.5 GHz the radio lightcurve was observed to peak very early, around $\sim$150 days, with a peak luminosity of $\sim1\times10^{29}$\,erg\,s$^{-1}$\,Hz$^{-1}$. At 4.4/4.9 GHz our VLBI observations indicate the transient peaked and became optically thin some time around $\sim$300-350 days. Following \citet{PT2009b} we find that if AT\,2017gbl was a Type II SN, then it would have a mass-loss rate of the order of $5\times10^{-5}$ M$_{\odot}$\,yr$^{-1}$, assuming a wind velocity of 10 km\,s$^{-1}$. Such a mass-loss rate is more typical of less massive progenitors that, on the other hand, should not have yielded such an energetic event as observed in the IR. Therefore, the radio properties of AT~2017gbl are not consistent with those observed for the most luminous and slowly declining Type IIn SNe that are powered by interaction with a massive and dense CSM resulting from extremely high mass loss rates.

We note that also Type II superluminous SNe (SLSN) have been observed to radiate up to $5\times10^{51}$ erg of energy \citep{nicholl2020}, and display prominent hydrogen emission in their spectra \citep[e.g.,][]{inserra2018}. The Type II class is dominated by so-called Type IIn SLSNe \citep{galyam2019}, which technically includes SN 2010jl. Other types of Type II SLSNe are rare, and as such we deem our analysis of Type IIn SNe as a scenario for AT~2017gbl to be also applicable for Type II SLSNe.

In summary, based on the combination of both the energetics inferred from our IR observations, as well as the luminosity and evolution timescale of the radio counterpart, we do not deem a SN scenario a plausible explanation for AT~2017gbl.

\subsection{Changing look AGN}
Since AT~2017gbl is coincident at the AO resolution with the Northern nucleus of IRAS~23436+5257, which potentially hosts a dust-obscured AGN, it is natural to consider their relation. As argued in Section \ref{sec:agn_variability}, the observed variability in the IR is difficult to reconcile with what would be considered as normal AGN variability. However, mid-IR variability with similar amplitude and luminosities as AT~2017gbl have been observed in CLAGN \citep{sheng2017,sheng2019,stern2018}, where it was interpreted as hot-dust emission echoing drastic drops or rises in accretion rate of the central AGN. If AT~2017gbl is related to a dramatic change in the accretion rate by the central SMBH, no direct line of sight is available to the transient event at optical and IR wavelengths due to the obscuration of $\sim$300 mag in A$_V$ through the dusty torus, see Section \ref{sec:sed_fit}.

\subsubsection{Geometry of the system}
In the CLAGN scenario we interpret the IR echo to have originated in dust clouds along the polar axis to which we have a direct line of sight, a geometry which has been observed in dust-obscured AGN \citep{asmus2019}. The broad spectral lines associated with AT~2017gbl in the near-IR (Paschen $\alpha$ - $\delta$, He\,{\sc i} $\lambda$10830, and O\,{\sc i} $\lambda\lambda$8445 and 11290) are common to the BLR of AGN \citep[e.g.,][]{landt2008}. This leads us to conclude the broad spectral features, as well as the optical emission, originate in the BLR, the flux of which responds to an increase from continuum photons from the AGN, lagged by some period due to light travel time delays \citep{peterson2004}. However, the BLR is fully obscured by the dusty torus, so the observed BLR emission would have to have been scattered into our line of sight, after being initially emitted in the polar direction, unobscured by the dusty torus. Such a geometry has also been observed in, e.g., NGC 1068, a Seyfert 2 galaxy in which polarization spectra of the nucleus exhibited broad Balmer lines indicative of a Type 1 AGN hidden behind the torus \citep{antonucci1985,marin2018}. The origin of the scattered light in this case would be clouds along the polar axis, where we also expect the IR echo to have originated from. The dust extinction of A$_V$ = 2.5 mag inferred from the Paschen line ratios is then interpreted as the combined extinction along the line of sight from the AGN BLR via the scattering regions to the observer. This geometry is visualized in Fig. \ref{fig:at2017gbl_schematic}.

\subsubsection{Timescales in the optical and IR}
CLAGN are typically identified by comparing archival spectra from years ago with recent data \citep[e.g.,][]{macleod2016}. Only in recent years, due to widefield synoptic surveys, has imaging data become available to construct (sparsely sampled) lightcurves of the outbursts related to CLAGN. The observed optical timescales involved are typically of order of 1000s of days \citep{yang2018,macleod2019,graham2020}. Based on the IR echo subtracted $J$-band lightcurve of AT~2017gbl (see Fig.~\ref{fig:at2017gbl_excess}), which acts as a tracer of its optical blackbody, the timescale (incl. rise and decline) of the optical variability of AT~2017gbl is limited to <630 days. This timescale is very short for a CLAGN, although there have been recent discoveries of unusual fast evolving CLAGN \citep{trakhtenbrot2019,frederick2019}.

Similarly, the IR echoes associated with CLAGN have typical timescales that are much slower than observed for AT~2017gbl. The timescale of mid-IR variability observed by \citet{sheng2017} in a sample of 10 known CLAGN ranged between 3.5 to 4.5 years, both in fading and brightening. Similar mid-IR fading timescales of 4-6 years were observed by \citet{sheng2019} for a sample of 6 quasars, spectroscopically classified as ``turn-off'' CLAGN. The fastest evolving IR echo associated with a CLAGN is a bright mid-IR flare in SDSS1115+0544 discovered by \citet{yan2019}, which they interpreted to result from a ``turn-on'' AGN rather than a TDE or Type IIn SN, based on the lightcurve evolution and a UV detection at late times. Its \textit{WISE} mid-IR lightcurve peaked $\sim$1 and $\sim$1.5 years after the last pre-outburst epoch in $W1$ and $W2$, respectively, at luminosities similar to AT~2017gbl. Inspection of the NEOWISE-R archive shows the transient in SDSS1115+0544 is still steadily declining in the mid-IR, four years after the peak.

These timescales are in contrast with the rise time of the IR echo of AT~2017gbl, which is limited to $\lesssim$140 days. The \textit{Spitzer} mid-IR lightcurve of the host nucleus of AT~2017gbl has declined close to the pre-outburst levels at +855 days (2.3 years) after the discovery. This timescale would suggest AT~2017gbl also evolved faster in its mid-IR decline compared to the sample of CLAGN, although the contrast is less distinct than for the rise. 

Recently, \citet{cannizzaro2020} laid out a detailed analysis of the timescales involved with AGN accretion rate changes, in order to explain the timescales involved with nuclear transient Gaia16aax, which rose to peak in the optical in $\sim$200 days. They concluded that most of the proposed accretion disk variability mechanisms are on timescales longer than observed for Gaia16aax, and/or not resulting in a sufficiently large amplitude. As with SDSS1115+0544, the observed rise to the peak with \textit{WISE} in the mid-IR of Gaia16aax at $\sim$1 year was significantly slower than the observed optical rise to the peak. The lightcurve of an IR echo lags the optical lightcurve due to light travel time effects, and will thus show slower evolution, which depends on the dust geometry \citep[e.g.,][]{lu2016}. The IR echo of AT~2017gbl rose to the peak in $\lesssim$140 days, which implies an even faster evolution in the optical.

The only CLAGN for which radio variability is reported in the literature is Mrk 590 \citep{koay2016}, a Seyfert galaxy which has been declining in accretion rate over the past decade \citep{denney2014}. The CLAGN was observed to increase in flux at 1.4 GHz by 28\% over 12 years, followed by a decrease in flux of 48\% over the following 20 years. These amplitudes and timescales are again in stark contrast with AT~2017gbl, which at 4.4 GHz doubled in flux in 63 days, followed by a drop of >60\% in the following 488 days.

In summary, if AT~2017gbl is a result of a dramatic change in accretion by the dust-obscured AGN, the timescale of these changes is limited by the observed timescales of the mid-IR, optical and radio lightcurves. The pre-outburst $J$-band limit and subsequent decline of the optical emission constrains the period of optical variability to <630 days, including the pre-outburst limit of -260 days. The observed rise to the peak of the IR echo in $\lesssim$140 days implies the optical rise was much faster, further constraining the period of intensified accretion which ionized the BLR. Similarly, the radio evolution of AT~2017gbl showed variability of amplitudes and timescales which have not been observed for CLAGN. In conclusion, if AT~2017gbl is due to a CLAGN, it would be one of the fastest evolving CLAGN, with a unique radio lightcurve.

\subsection{Tidal disruption event}
TDEs have been observed to radiate up to $10^{51} - 10^{52}$ erg \citep[e.g.,][]{holoien2016,mattila2018} with peak luminosities up to $\sim10^{45}$ erg s$^{-1}$ \citep[e.g.,][]{vanvelzen2020}, which satisfies the energy budget and the inferred peak luminosity of AT~2017gbl. Furthermore, TDEs evolve in the optical over timescales of 200-250 days \citep[e.g.,][]{vanvelzen2020}, within the limits imposed by the timescale of the optical emission of AT~2017gbl. Assuming the presence of an AGN the interpretation of the IR emission as an IR echo and the optical component of AT~2017gbl being dominated by the BLR emission scattered into our line of sight by polar clouds is the same as in the CLAGN scenario (see Fig. \ref{fig:at2017gbl_schematic}).

The optical blackbody temperature of 6500 K derived in Section \ref{sec:twocompfit} from the early optical emission of AT~2017gbl is not consistent with direct emission of optical/UV selected TDE candidates, since their early-time emission is characterised by blackbody temperatures of around 2 $\times$ 10$^4$ K \citep[e.g.,][]{hung2017}. However, recent simulations of TDEs occurring in AGN have shown that the spectrum of the resulting transient may not look like a standard TDE and it is even uncertain in which energies the radiation emerges from the event \citep{chan2019}. Furthermore, substantial reprocessing of the TDE emission may have occurred within the dense BLR clouds. In Fig. \ref{fig:bb_lum_comp} we show the optical luminosity evolution of the prototypical TDE ASASSN-14li \citep{holoien2016}. Within uncertainties, the optical peak luminosity and decline rate of AT~2017gbl is similar to that of ASASSN-14li. Notably, ASASSN-14li occurred in a galaxy where the presence of an AGN is suspected based on radio observations \citep{vanvelzen2016b,bright2018}.

\subsubsection{IR echoes}
The presence of IR echoes in TDEs has been established only during the past few years. \citet{lu2016} showed through model calculations that depending on the total radiated energy in the UV-optical and the sky covering factor of the dust clouds the dust emission following a TDE peaks at 3-10 \micron\ with typical luminosity between $10^{42}$ and $10^{43}$ erg s$^{-1}$. By now such IR echoes have been observed for several TDEs. For example, \citet{jiang2016} discovered a significant mid-IR increase in the \textit{WISE} data of ASASSN-14li 36 days after the discovery. By fitting the quasi-simultaneous broadband SED with a two-component blackbody, they determined an integrated IR luminosity of 2.5 $\times\ 10^{41}$ erg s$^{-1}$, see Fig. \ref{fig:bb_lum_comp}. Similarly, \citet{dou2016} estimated mid-IR luminosities in the range of 0.4 - 2.0 $\times\ 10^{43}$ erg s$^{-1}$ for four TDE candidates, and \citet{vanvelzen2016a} reported on 3.4 \micron\ emission of $\sim 10^{42}$ erg s$^{-1}$ for three previously known TDE candidates.

The observed IR echoes arising from the TDE candidates span a wide range both in terms of IR luminosity, from $10^{41}$ erg s$^{-1}$ for ASASSN-14li to $10^{44}$ erg s$^{-1}$ in the case of the flare in IRAS F01004-2237 \citep{dou2017}; as well as in duration, from $\sim$500 days for TDEs discussed in \citet{vanvelzen2016a} to $>$ 4000 days for Arp~299-B AT1 \citep{mattila2018}. In comparison, the IR echo of AT~2017gbl was relatively bright with a peak IR luminosity of $\sim2.3\times10^{43}$ erg s$^{-1}$. In terms of timescale, the evolution of the IR echo of AT~2017gbl was comparable to those TDEs with fast-paced IR echoes, such as PTF-09ge and PTF-09axc \citep{vanvelzen2016a} and ASASSN-14li \citep{jiang2016}. The IR echo of AT~2017gbl rose to peak in $\lesssim$140 days, and was only marginally detected in the mid-IR in the last epoch of +855 days, having already faded in the near-IR after 590 days. We note that the timescales of IR emission likely reflect the geometry of the dust responsible for the re-radiation of the UV/optical emission (through light travel time effects) rather than the actual timescales of that emission \citep[e.g.,][]{lu2016}. PTF-09ge and PTF-09axc were detected at or after peak in the mid-IR some 200 days after the optical peak of the TDE, and the IR echo of ASASSN-14li peaked only 36 days after the optical TDE discovery. All three faded in the mid-IR in <500 days.

Based on our IR observations, the total radiated energy of AT~2017gbl was at least $7.3\times10^{50}$ erg. Following \citet{metzger2016} and assuming 10\% as the accreted fraction and 10\% as the radiative efficiency we find that a star with a mass above 0.1 M$_{\odot}$ can explain the energetics of AT~2017gbl. Stars with such a mass can be disrupted by a non-spinning BH at the lower end of the mass range of $\text{log}_{10}(\text{M}_{\text{BH}})= 7.1\pm0.4~\text{M}_{\odot}$ estimated for the SMBH at the centre of IRAS 23436+5257 \citep{lawsmith2017}. Outflows, which are responsible for the radio emission, can be generated by the self-crossing of the fallback stream \citep{lu2020} or the accretion disk wind \citep{strubbe09}.

\subsubsection{TDEs in LIRGs}
Although TDEs are still a relatively rare phenomenon with observed rates of $\sim10^{-8}$ yr$^{-1}$ Mpc$^{-3}$ \citep[e.g.,][]{vanvelzen2018}, there have now also been two TDE (candidates) discovered in (U)LIRGs, both showing long-lasting IR emission. \citet{tadhunter2017} reported the discovery of a bright nuclear transient in the ULIRG IRAS F01004-2237. They interpret the transient as a TDE based on strong and variable broad He\,{\sc i} $\lambda$5876 and He\,{\sc ii} $\lambda$4686 lines in their spectrum. More recently, however, \citet{trakhtenbrot2019} favoured an origin of intensified accretion into the SMBH. The optical transient in IRAS F01004-2237 was followed by an IR echo spanning several years with a peak mid-IR luminosity of 2-3 $\times 10^{44}$ erg s$^{-1}$ \citep{dou2017}.

\citet{mattila2018} reported the discovery and follow-up of an extremely energetic transient Arp~299-B AT1 in the LIRG Arp 299, with the IR luminosity peaking at $\sim$ $6 \times 10^{43}$ erg s$^{-1}$. Over a decade of VLBI observations a resolved and expanding off-axis radio jet was detected coincident with the position of Arp~299-B AT1, showing that the transient had arisen from a TDE. The IR emission was interpreted to originate from dust, in the polar regions of the AGN torus viewed almost edge-on, responsible for absorbing and re-radiating much of UV-optical emission of the transient.

Similar to AT~2017gbl, Arp~299-B AT1 was observed in the near- and mid-IR and its IR SED was well described by an expanding and cooling blackbody. In both cases the IR SED was well described by a single component blackbody indicating a narrow range of temperatures and only a small amount (in the IR) of extinction in the foreground to the IR emitting region. In the first $\sim$ 560 days its IR blackbody parameters evolved from $(4.23 \pm 0.06) \times\ 10^{-2}$~pc to $(8.83 \pm 0.09) \times\ 10^{-2}$~pc, and from $1045 \pm 7$ to $902 \pm 5$~K, in radius and temperature respectively. These values are remarkably similar to those of AT~2017gbl, see Table \ref{tab:bb_params}. However, in the case of AT~2017gbl the blackbody temperature declined much more rapidly resulting in a decreasing luminosity, while Arp~299-B AT1 increased in IR luminosity for over 2000 days. Consequently, the energy budget of Arp~299-B AT1 inferred from its IR emission is much larger at $>$ $1.5\times10^{52}$ erg. Initially, no optical counterpart was observed for Arp~299-B AT1 and only deep archival \textit{HST} observations allowed a possible optical counterpart to be identified through image subtraction. The optical depth through the dusty torus towards the central engine in Arp 299-B1 is extremely large, corresponding to an optical extinction of $\sim10^3$ mag. Therefore, the optical emission (if related to Arp~299-B AT1) can only be explained as scattered light from the gas and dust in the polar regions. The differences in the IR and optical properties between Arp~299-B AT1 and AT~2017gbl could arise from differences in the geometry and extent of the dusty regions surrounding the UV-optical transient, as well as the properties of the SMBH and the disrupted star.

\subsubsection{Radio}
The peak radio luminosity of AT~2017gbl is three to four times higher than the peak luminosities of the TDEs ASASSN-14li and Arp~299-B AT1. However, the radio evolution of AT~2017gbl appears to be more similar to that of ASASSN-14li, whose 5.0 GHz emission peaked at about 150 days, and at $\sim$50 days at 15.7 GHz \citep{bright2018}. On the contrary, the radio emission from Arp~299-B AT1 peaked much later,  close to $\sim$1000 days after the event. The nature of the radio emission in ASASSN-14li is still under debate, and is unclear whether it was powered by a jet or by an expanding, non-relativistic outflow \citep[e.g.,][]{vanvelzen2016b,alexander2016, decolle19}. The nature of the radio emission in Arp~299-B AT1 is clear: a relativistic jet that decelerated due to its huge surrounding density \citep{mattila2018}. The bright radio emission in these events is likely due to high densities in the surrounding medium. Our VLBI observations could not resolve the radio morphology of AT 2017gbl, and therefore we cannot at this stage confirm, or rule out, whether the radio emission of AT 2017gbl is powered by a TDE jet, or a non-relativistic outflow.

Finally, TDEs have displayed bumps or plateaus in their lightcurves at UV wavelengths, with a timescale of 100-200 days due to a relativistic orbital pericenter of a TDE with a massive $10^7$ - $10^8$ M$_{\odot}$ SMBH \citep{leloudas2016,wevers2019}, and at optical wavelengths over timescales of 300-600 days due to potential multiple interactions between a dense debris stream and the accretion disk \citep{cannizzaro2020}. Such a temporary influx of UV/optical photons would be echoed by reprocessed emission in the IR, which might explain the plateau/bump feature in the IR lightcurve of AT~2017gbl and the resulting increase in temperature of the blackbody fitted to the IR data at +375 days past discovery shown in Fig.~\ref{fig:at2017gbl_bb_fits}.

\subsubsection{Rate estimate of AT~2017gbl-like events in LIRGs}
In conclusion, based on the constraints imposed by the energy budget, peak outburst luminosity, and the observed timescales of the radio lightcurve and IR echo, we favor a TDE scenario for AT~2017gbl as opposed to a CLAGN. The discovery of AT~2017gbl resulted from the SUNBIRD programme \citep{kool2018}, where we monitored a sample of $\sim$40 nearby LIRGs in the near-IR for SNe. If we assume all AT~2017gbl-like TDEs would have been detected in the near-IR for $\sim$1 year, the total control time (e.g., two epochs six months apart for a given LIRG would result in 1.5 years control time) covered as part of SUNBIRD is 72.7 years. Assuming a Poisson process and adopting the confidence limits of \citet{gehrels1986}, we obtain an estimate for the rate of AT~2017gbl-like TDEs in LIRGs of $10^{n}$ LIRG$^{-1}$ year$^{-1}$ with $n = -1.9^{+0.5}_{-0.8}$, with $1\sigma$ errors. While this estimate has significant uncertainties and further work is required to constrain it, it agrees with the TDE rate estimate of $10^{-2}$ LIRG$^{-1}$ year$^{-1}$ obtained by \citet{tadhunter2017} based on their detection of one TDE candidate as a result of two epochs of spectroscopic observations of a sample of 15 LIRGs. We note that our rough estimate should also be considered as a lower limit for the actual TDE rate in LIRGs assuming that not all the TDEs show such luminous and long-lasting IR emission. This rate estimate supports the notion that LIRGs exhibit an elevated TDE rate compared to field galaxies.

\subsection{Alternative scenarios}
We consider also a scenario where the source of AT~2017gbl is a sufficiently energetic SN in the BLR, as was suggested for a nuclear IR-luminous transient in W0948+0318 \citep{assef2018b}. This would explain the observed spectral features of AT~2017gbl, since they would originate from the ionized BLR in the same way as in the TDE/CLAGN scenario, but this scenario is constrained by the radio properties in the same way as the SN scenario discussed in Section \ref{sec:sn_discussion}. Luminous nuclear transients with strong IR excess have been discovered in Seyfert galaxies. \citet{kankare2017} offered two alternative interpretations for such an event, PS1-10adi: either an extremely energetic SN or a tidal disruption of a star, powered by shock interaction between expanding material and large quantities of surrounding dense matter. Additionally, \citet{moriya2017} proposed a scenario where superluminous transients in AGN, such as the TDE candidate PS16dtm (sharing similarities with PS1-10adi), are powered by interaction between accretion-disk winds and clouds in the BLR, leading to luminosities of $\sim$10$^{44}$ erg s$^{-1}$ on timescales of $\sim$100 days. However, with a direct line of sight to the transient event fully obscured, it is difficult to ascertain the viability of these less common scenarios.


\section{Summary and Conclusions}  \label{sec:conclusion}
In July 2017 we discovered that the Northern nucleus of IRAS~23436+5257 had significantly brightened in near-IR Keck images compared to the previous epoch from October 2016. This transient event AT~2017gbl was also detected in the optical, albeit much fainter, and steadily declined in brightness in the near-IR and optical during the first $\sim$300 days. Additionally, AT~2017gbl was detected in the mid-IR, both by comparing follow-up \textit{Spitzer} imaging with archival epochs, as well as through inspection of archival AllWISE and NEOWISE-R data from the \textit{WISE} telescope. The mid-IR lightcurves not only revealed AT~2017gbl, but also a small but significant decline in flux of the nucleus prior to AT~2017gbl. We showed that the host-subtracted IR SED of AT~2017gbl is well represented by a single expanding and cooling blackbody component. From this IR echo, a lower limit to the energy budget could be inferred of $7.3\times10^{50}$ erg. Fitting the smooth evolution of the IR blackbody temperature also allowed us to constrain the peak outburst epoch of the underlying transient to $\sim$140 days before discovery, with a peak luminosity of 0.1 - 1$\times 10^{45}$ erg s$^{-1}$. Inspection of optical and near-IR spectroscopy obtained at early epochs showed a $\sim$2000 km s$^{-1}$ broad component to the strong hydrogen, He\,{\sc i} and O\,{\sc i} emission lines. This broad component disappeared or had visibly decreased at later epochs, so we attribute this component to the transient event. In addition to the IR echo, it is clear that at early times there is optical emission in addition to the IR blackbody. We show that after correcting for a V-band extinction of 2.5 mag, which is inferred from the line ratio of the broad (transient) components of strong Paschen lines, this is well fitted by a blackbody with a temperature of 6500 K peaking at optical wavelengths. At radio wavelengths, AT~2017gbl was still rising at the discovery epoch, reaching the peak at 15.5 GHz $\sim$13-15 days past the discovery, while at 4.4/4.9 GHz it took 150-200 days. Our VLBI imaging showed an evolving compact source coincident with the host galaxy nucleus (to within 0.04\arcsec) as observed in the near-IR with AO.

We compared the observed properties of AT~2017gbl with three different scenarios that have sufficiently large energy budgets: an energetic SN, a CLAGN due to a dramatic accretion rate change of the AGN, and a TDE. In all scenarios the IR blackbody is well explained by an IR echo, where dust absorbs UV/optical light from the transient and reradiates in the IR. In the following, we summarise the main observational properties of AT~2017gbl that lead us to conclude a TDE as the most likely scenario for this transient:

\begin{itemize}
\setlength\itemsep{1em}
    \item Based on radiative transfer modelling of the host galaxy IR SED, we expect the Northern nucleus of IRAS~23436+5257 to host an obscured AGN, where the line of sight towards the central engine is fully obscured at optical, near- and mid-IR wavelengths. The amplitude and timescale of the pre-outburst mid-IR variability of the Northern nucleus can be explained by the presence of an AGN, but the outburst related to AT~2017gbl does not fit with `normal' AGN variability.

    \item The energy budget, the luminosity of the IR echo, and the spectral features of AT~2017gbl can be explained by a (super)luminous Type IIn SN such as SN 2010jl and SN 2015da. These supernovae also show prominent IR echoes, but require extreme mass-loss rates of order $\sim0.1$ M$_{\odot}$~yr$^{-1}$ to explain their properties. However, based on the the evolution and luminosity of the radio counterpart of AT~2017gbl, such a high mass-loss rate is ruled out. As such, we conclude that AT~2017gbl is unlikely to be a SN. 

    \item If AT~2017gbl is related to the central SMBH, either as a TDE of a star or a dramatic change in the accretion rate of an AGN (as seen in CLAGN), we should not be able to observe the event directly in the optical, near- and mid-IR due to the strong obscuration by the dusty torus. In this case the observed optical emission is interpreted as scattered light from polar clouds, having originated from the transient itself and in the BLR. This is supported by the fitted optical blackbody temperature of 6500 K and the $\sim$2000 km s$^{-1}$ broad emission lines, which are not compatible with those of optically discovered TDE candidates, but are consistent with being dominated by emission originated in the BLR. Similarly, the spectral signatures (Paschen $\alpha$/$\beta$/$\delta$, He\,{\sc i}, O\,{\sc i}) we have attributed to the transient event are observed in AGN BLR.

    \item As TDEs and CLAGN are governed by similar physics, it is challenging to discern between the two scenarios. In the case of AT~2017gbl, the main ways to differentiate are the evolution timescales in the optical, mid-IR and radio. The rise to peak of the IR echo of AT~2017gbl is constrained to $\lesssim$140 days, and the following fast optical decline ($\sim$1 mag in 100 days) is traced by near-IR $J$-band, after subtraction of the IR echo contribution, to <370 days. In comparison with known CLAGN, these timescales are extremely fast, while they match well with the optical timescales of TDEs. In particular ASASSN-14li evolved in the optical and IR over timescales similar to AT~2017gbl, while also showing a strikingly similar evolution and luminosity at radio wavelengths. As such we find a TDE as the most plausible scenario to explain the observed properties of AT~2017gbl.
\end{itemize}

AT~2017gbl is the third TDE candidate to be discovered in a LIRG, after Arp~299-B AT1 in Arp 299 and the TDE candidate in IRAS F01004-2237. Like the TDE candidate discovered by \citet{tadhunter2017} in IRAS F01004-2237, the discovery of AT~2017gbl is the product of a monitoring programme of only a modest sample of LIRGs. Finding a TDE in such a sample supports the suggestion that the rate of TDEs in LIRGs is orders of magnitudes larger than that in field galaxies. Furthermore, the discovery of TDEs in actively starforming galaxies such as LIRGs is in contrast with the apparent preference of optical TDE discoveries for post-starburst E+A galaxies \citep{arcavi2014,french2016}. It is therefore possible that the current optical TDE sample is affected by a strong observational bias against detecting dust-obscured transients in galactic nuclei. The high fraction of E+A galaxies that have undergone a recent galaxy merger has been offered as an explanation for their TDE overabundance \citep{french2016}. Interestingly, LIRGs are often interacting systems \citep{vaisanen2012} and the morphology of IRAS~23436+5257 hosting two nuclei suggests a recent or an ongoing interaction and major merger between two galaxies. Furthermore, the LIRG host of Arp~299-B AT1 is also undergoing a major merger. This is noteworthy as the fraction of galaxies undergoing mergers below a redshift z$<$1 is very low \citep{conselice2009}. This would suggest that galaxy mergers may be the leading cause of elevated TDE rates both in E+A galaxies as well as in LIRGs. Both systematic monitoring of galaxy mergers and LIRGs and detailed galaxy merger simulations are therefore required to constrain their TDE rate to test this scenario.

\section*{Acknowledgements}
We thank the anonymous referee for insightful comments and suggestions. We would like to thank Petri V\"ais\"anen and Seppo Laine for useful discussions on LIRG properties and \textit{WISE} data analysis, respectively. We thank David Williams for scheduling the final AMI-LA epoch and Chi-Ho Chan for comments on the submitted version of the manuscript.
ECK acknowledges support from the G.R.E.A.T. research environment, financial support from the visitor and mobility program of the Finnish Centre for Astronomy with ESO (FINCA), funded by the Academy of Finland grant nr 306531, and support from The Wenner-Gren Foundations.
TMR acknowledges the financial support of the Jenny and Antti Wihuri foundation and the Vilho, Yrj{\"o} and Kalle V{\"a}is{\"a}l{\"a} Foundation of the Finnish academy of Science and Letters.
MPT acknowledges financial support from the State Agency for Research of the Spanish MCIU through the ``Center of Excellence Severo Ochoa'' award to the Instituto de Astrof\'{\i}sica de Andaluc\'{\i}a (SEV-2017-0709) and through grant PGC2018-098915-B-C21 (MCI/AEI/FEDER).
CRC acknowledges support by the Chinese Academy of Sciences (CAS), through grant CAS16013 of the CAS South America Center for Astronomy (CASSACA) and Programa de Astronom\'{\i}a CONICYT, Chile.
PGJ and GC acknowledge support from European Research Council Consolidator Grant 647208.
MG is supported by the Polish NCN MAESTRO grant 2014/14/A/ST9/00121.
GEA is the recipient of an Australian Research Council Discovery Early Career Researcher Award (project DE180100346) and acknowledges partial support through the Australian Research Council's \textit{Discovery Projects} funding scheme (project DP200102471).
ST acknowledges financial support from the Russian Foundation for Basic Research project 17-52-80139 BRICS-a. 
RMcD is the recipient of an Australian Research Council Future Fellowship Award (project number FT150100333).
WL is supported by the David and Ellen Lee Fellowship at Caltech.

NUTS2 is funded in part by the Instrument Center for Danish Astronomy.

This work is based in part on observations made with the Nordic Optical Telescope, operated by the Nordic Optical Telescope Scientific Association at the Observatorio del Roque de los Muchachos, La Palma, Spain, of the Instituto de Astrofisica de Canarias.

Some of the data (PI: S. Ryder; program IDs Z229N2L, Z271N2L) presented herein were obtained at the W. M. Keck Observatory, which is operated as a scientific partnership among the California Institute of Technology, the University of California and the National Aeronautics and Space Administration. The Observatory was made possible by the generous financial support of the W. M. Keck Foundation. The authors wish to recognize and acknowledge the very significant cultural role and reverence that the summit of Maunakea has always had within the indigenous Hawaiian community.  We are most fortunate to have the opportunity to conduct observations from this mountain.

We thank the Mullard Radio Astronomy Observatory for carrying out the AMI-LA observations.

This work is based in part on observations (PIs: E. Kool, K. Maeda; program IDs GN-2017B-DD-2, GN-2018B-FT-109) obtained at the Gemini Observatory, which is operated by the Association of Universities for Research in Astronomy, Inc., under a cooperative agreement with the NSF on behalf of the Gemini partnership: the National Science Foundation (United States), National Research Council (Canada), CONICYT (Chile), Ministerio de Ciencia, Tecnolog\'{i}a e Innovaci\'{o}n Productiva (Argentina), Minist\'{e}rio da Ci\^{e}ncia, Tecnologia e Inova\c{c}\~{a}o (Brazil), and Korea Astronomy and Space Science Institute (Republic of Korea).

The scientific results reported in this article are based in part on observations (PI: T. Heikkil\"a; program ID 18208589) made by the \textit{Chandra X-ray Observatory}, and this research has made use of the CIAO software package provided by the Chandra X-ray Center (CXC). This article includes results based on observations made by the \textit{Neil Gehrels Swift observatory}. The authors thank \textit{Chandra} and \textit{Swift} teams for the approval and rapid scheduling of our observations.

This work is based in part on observations made with the William Herschel Telescope. The WHT is operated on the island of La Palma by the Isaac Newton Group of Telescopes in the Spanish Observatorio del Roque de los Muchachos of the Instituto de Astrof\'isica de Canarias. The ISIS and ACAM data were obtained as part of (17A)N4/N6, (18A)N4 and (18B)N5.

This work is based in part on observations (PI: S. Mattila; program IDs 13226, 14054) made with the Spitzer Space Telescope, which is operated by the Jet Propulsion Laboratory, California Institute of Technology under a contract with NASA. This publication makes use of data products from the Wide-field Infrared Survey Explorer, which is a joint project of the University of California, Los Angeles, and the Jet Propulsion Laboratory/California Institute of Technology, funded by the National Aeronautics and Space Administration. This publication also makes use of data products from NEOWISE, which is a project of the Jet Propulsion Laboratory/California Institute of Technology, funded by the Planetary Science Division of the National Aeronautics and Space Administration.

This work is based in part on observations from   programs RP028B and BP225 (PI: M. P\'erez-Torres) obtained with the European VLBI Network (EVN) and the Very Long Baseline Array (VLBA), respectively. The European VLBI Network is a joint facility of independent European, African, Asian, and North American radio astronomy institutes. The National Radio Astronomy Observatory is a facility of the National Science Foundation operated under cooperative agreement by Associated Universities, Inc. 

This publication makes use of data products from the Two Micron All Sky Survey, which is a joint project of the University of Massachusetts and the Infrared Processing and Analysis Center/California Institute of Technology, funded by the National Aeronautics and Space Administration and the National Science Foundation.

Funding for the SDSS and SDSS-II has been provided by the Alfred P. Sloan Foundation, the Participating Institutions, the National Science Foundation, the U.S. Department of Energy, the National Aeronautics and Space Administration, the Japanese Monbukagakusho, the Max Planck Society, and the Higher Education Funding Council for England. The SDSS Web Site is \urlwofont{http://www.sdss.org/}.

The SDSS is managed by the Astrophysical Research Consortium for the Participating Institutions. The Participating Institutions are the American Museum of Natural History, Astrophysical Institute Potsdam, University of Basel, University of Cambridge, Case Western Reserve University, University of Chicago, Drexel University, Fermilab, the Institute for Advanced Study, the Japan Participation Group, Johns Hopkins University, the Joint Institute for Nuclear Astrophysics, the Kavli Institute for Particle Astrophysics and Cosmology, the Korean Scientist Group, the Chinese Academy of Sciences (LAMOST), Los Alamos National Laboratory, the Max-Planck-Institute for Astronomy (MPIA), the Max-Planck-Institute for Astrophysics (MPA), New Mexico State University, Ohio State University, University of Pittsburgh, University of Portsmouth, Princeton University, the United States Naval Observatory, and the University of Washington.

\section*{Data availability}
The photometric data underlying this article are available in the article. The spectroscopic data will be made available on WISeREP\footnote{\urlwofont{https://wiserep.weizmann.ac.il/}} under the designation AT 2017gbl.



\bibliographystyle{mnras}
\bibliography{17gbl_paper} 



\appendix

\section{Photometric and spectroscopic observations}
\begin{table*}
\centering
\begin{tabular}{|c|c|r|c|c|c|c|} \hline
UT Date&MJD&Epoch&Telescope/Instrument&\multicolumn{3}{|c|}{Magnitude} \\ \hline
&&(days)&&$J$&$H$&$K_s$\\ \hline
2016-10-21.4    &57682.4&-260    &Keck/NIRC2&	template					&	template				 &	template\\
2017-07-08.5    &57942.5&discovery    &Keck/NIRC2	&16.03	$\pm$ 0.03	&14.45	$\pm$ 0.04	&13.22	$\pm$ 0.07\\
2017-07-27.2    &57961.2&+19	&NOT/NOTCam	&16.18	$\pm$ 0.04	&14.61	$\pm$ 0.03	&13.27	$\pm$ 0.04\\
2017-08-19.2    &57983.2&+41	&NOT/NOTCam	&16.50	$\pm$ 0.03	&14.89	$\pm$ 0.03	&13.48	$\pm$ 0.04\\
2017-09-03.9    &57999.9&+57	&NOT/NOTCam	&16.81	$\pm$ 0.05	&14.97	$\pm$ 0.03	&13.58	$\pm$ 0.04\\
2017-09-29.9    &58025.9&+83	&NOT/NOTCam	&16.90	$\pm$ 0.04	&15.18	$\pm$ 0.05	&13.68	$\pm$ 0.04\\
2017-10-17.1    &58042.1&+100	&NOT/NOTCam	&17.19	$\pm$ 0.06	&15.35	$\pm$ 0.04	&13.87	$\pm$ 0.04\\
2017-12-05.2    &58092.2&+150    &Keck/NIRC2 &                   &                   &14.26	$\pm$ 0.04\\
2018-01-03.9    &58121.9&+179	&NOT/NOTCam	&18.14	$\pm$ 0.09	&16.39	$\pm$ 0.08	&14.68	$\pm$ 0.05\\
2018-01-22.8    &58140.8&+198	&NOT/NOTCam	&18.08	$\pm$ 0.14	&16.64	$\pm$ 0.07	&14.84	$\pm$ 0.05\\
2018-05-12.2    &58249.2&+307	&NOT/NOTCam	&19.23	$\pm$ 0.26	&17.13	$\pm$ 0.12	&15.61	$\pm$ 0.09\\
2018-07-14.1    &58312.1&+370	&NOT/NOTCam	&-                  &17.10	$\pm$ 0.07	&15.02	$\pm$ 0.03\\
2018-09-04.0    &58364.0&+422	&NOT/NOTCam	&-                  &17.52	$\pm$ 0.10	&15.59	$\pm$ 0.05\\
2018-09-24.1    &58384.1&+442	&NOT/NOTCam	&-                  &17.98	$\pm$ 0.09	&15.80	$\pm$ 0.04\\
2018-10-16.1    &58406.1&+464	&NOT/NOTCam	&-                  &18.32	$\pm$ 0.17	&16.12	$\pm$ 0.06\\
2018-11-03.9    &58425.9&+483	&NOT/NOTCam	&-                  &18.59	$\pm$ 0.26	&16.50	$\pm$ 0.09\\
2018-11-24.0    &58445.0&+503	&NOT/NOTCam	&-                  &>19.2              &16.48	$\pm$ 0.08\\
2018-12-10.9    &58462.9&+520	&NOT/NOTCam	&-                  &-                  &16.43	$\pm$ 0.05\\
2018-12-30.9    &58482.9&+540	&NOT/NOTCam	&-                  &-                  &16.57	$\pm$ 0.09\\
2019-01-10.9    &58493.9&+551	&NOT/NOTCam	&-                  &-                  &16.61	$\pm$ 0.09\\
2019-02-18.9    &58532.9&+590	&NOT/NOTCam	&-                  &-                  &16.85	$\pm$ 0.11\\
2019-07-04.2	&58668.2&+726  &NOT/NOTCam	&                   &                   &>18.0\\
2019-07-22.9	&58686.9&+744   &NOT/NOTCam	&template           &template           &\\
2019-09-14.1	&58740.1&+798    &NOT/NOTCam	&                   &                   &template\\
\hline
\end{tabular}
\caption{AT~2017gbl near-IR photometry in apparent magnitudes in the Vega system. NIRC2 and NOTCam template epochs are indicated. A hyphen indicates the transient was observed, but not detected. An empty entry indicates the transient was not observed in the filter for that epoch.}
\label{tab:nir_mags}
\end{table*}

\begin{table*}
\centering
\begin{tabular}{|c|c|r|c|c|c|c|c|} \hline
UT Date&MJD&Epoch&Telescope/Instrument&\multicolumn{4}{|c|}{Magnitude} \\ \hline
&&(days)&&$g$&$r$&$i$&$z$  \\ \hline
2017-07-11.1    &57945.1&+3   &WHT/ACAM   &21.22 $\pm$ 0.09&19.93 $\pm$ 0.02   &19.08 $\pm$ 0.02   &	18.36 $\pm$ 0.03\\
2017-08-27.9    &57992.9&+50   &WHT/ACAM   &               &    		        &19.66 $\pm$ 0.05   &	18.96 $\pm$ 0.05\\
2017-11-28.9    &58085.9&+143 &NOT/ALFOSC &               &                    &20.79 $\pm$ 0.04   &	20.45 $\pm$ 0.03\\
2018-07-09.2    &58308.2&+366   &WHT/ACAM   &               &-    		        &-                  &-\\
2018-09-08.1    &58369.1&+427   &NOT/ALFOSC &               &                    &template           &template\\
2019-01-29.9    &58512.9&+570   &WHT/ACAM   &template       &template            &template           &template\\
\hline
\end{tabular}
\caption{AT~2017gbl optical photometry in apparent magnitudes in the AB system. ACAM and ALFOSC template epochs are indicated. A hyphen indicates the transient was observed, but not detected. An empty entry indicates the transient was not observed in the filter for that epoch.}
\label{tab:opt_mags}
\end{table*}

\begin{table*}
\centering
\begin{tabular}{|c|c|r|c|c|c|} \hline
UT Date&MJD&Epoch&Telescope&\multicolumn{2}{|c|}{Magnitude} \\ \hline
&&(days)&&3.6 \micron&4.5 \micron\\ \hline
2004-12-15.0    &53354.0&-4589	&\textit{Spitzer}	&11.14	$\pm$ 0.05	&10.68	$\pm$ 0.05\\
2011-01-26.6    &55587.6&-2355	&\textit{Spitzer}	&11.24	$\pm$ 0.05	&10.84	$\pm$ 0.06\\
2017-11-13.1    &58070.1&+128	&\textit{Spitzer}	&10.75	$\pm$ 0.05	&10.19	$\pm$ 0.05\\
2017-12-19.8    &58106.8&+164	&\textit{Spitzer}	&10.82	$\pm$ 0.05	&10.26	$\pm$ 0.05\\
2018-03-23.5    &58200.5&+258	&\textit{Spitzer}	&10.92	$\pm$ 0.05	&10.36	$\pm$ 0.05\\
2018-05-17.1    &58255.1&+313	&\textit{Spitzer}	&10.93	$\pm$ 0.05	&10.36	$\pm$ 0.06\\
2018-11-12.6    &58434.6&+492	&\textit{Spitzer}	&11.03	$\pm$ 0.05	&10.46	$\pm$ 0.06\\
2019-04-01.5    &58574.5&+632	&\textit{Spitzer}	&11.11	$\pm$ 0.05	&10.52	$\pm$ 0.06\\
2019-11-10.3    &58797.3&+855	&\textit{Spitzer}	&11.13	$\pm$ 0.05	&10.61	$\pm$ 0.05\\
\hline
\end{tabular}
\caption{Mid-IR photometry of Northern nucleus of IRAS~23436+5257, observed with \textit{Spitzer}.}
\label{tab:iras_23436_spitzer_mags}
\end{table*}

\begin{table*}
\centering
\begin{tabular}{|c|c|r|c|c|c|} \hline
UT Date&MJD&Epoch&Mission&\multicolumn{2}{|c|}{Magnitude} \\ \hline
&&(days)&&3.4 \micron&4.6 \micron\\ \hline
2010-01-12.8    &55208.8&-2733	&AllWISE	&10.87	$\pm$ 0.03	&10.20	$\pm$ 0.03\\
2010-07-18.4    &55395.4&-2547	&AllWISE	&10.89	$\pm$ 0.03	&10.25	$\pm$ 0.03\\
2011-01-10.6    &55571.6&-2371	&AllWISE	&10.94	$\pm$ 0.03	&10.32	$\pm$ 0.03\\
2014-01-14.4    &56671.4&-1271	&NEOWISE-R	&10.96	$\pm$ 0.03	&10.35	$\pm$ 0.03\\
2014-07-22.0    &56860.0&-1083	&NEOWISE-R	&10.98	$\pm$ 0.03	&10.40	$\pm$ 0.03\\
2015-01-12.8    &57034.8&-908	&NEOWISE-R	&11.00	$\pm$ 0.03	&10.42	$\pm$ 0.03\\
2015-07-19.2    &57222.2&-720	&NEOWISE-R	&11.00	$\pm$ 0.03	&10.44	$\pm$ 0.03\\
2016-01-07.9    &57394.9&-548	&NEOWISE-R	&11.04	$\pm$ 0.03	&10.43	$\pm$ 0.03\\
2016-07-17.4    &57586.4&-356	&NEOWISE-R	&11.04	$\pm$ 0.03	&10.43	$\pm$ 0.03\\
2017-01-01.5    &57754.5&-188	&NEOWISE-R	&11.07	$\pm$ 0.03	&10.47	$\pm$ 0.03\\
2017-07-18.3    &57952.3&+10	&NEOWISE-R	&10.56	$\pm$ 0.03	&9.92	$\pm$ 0.03\\
2017-12-30.0    &58117.0&+175	&NEOWISE-R	&10.76	$\pm$ 0.03	&10.01	$\pm$ 0.03\\
2018-07-18.6    &58317.6&+375	&NEOWISE-R	&10.83	$\pm$ 0.03	&10.11	$\pm$ 0.03\\
\hline
\end{tabular}
\caption{Mid-IR photometry of IRAS~23436+5257, observed with \textit{WISE}.}
\label{tab:iras_23436_wise_mags}
\end{table*}

\begin{table*}
\centering
\begin{tabular}{|c|c|r|c|c|c|} \hline
UT Date&MJD&Epoch&Telescope&\multicolumn{2}{|c|}{Magnitude} \\ \hline
&&(days)&&3.6 \micron&4.5 \micron\\ \hline
2017-11-13.1    &58070.1&+128	&\textit{Spitzer}	&11.8	$\pm$ 0.13	&11.11	$\pm$ 0.12\\
2017-12-19.8    &58106.8&+164	&\textit{Spitzer}	&12.0	$\pm$ 0.15	&11.28	$\pm$ 0.13\\
2018-03-23.5    &58200.5&+258	&\textit{Spitzer}	&12.32	$\pm$ 0.18	&11.56	$\pm$ 0.15\\
2018-05-17.1    &58255.1&+313	&\textit{Spitzer}	&12.36	$\pm$ 0.19	&11.56	$\pm$ 0.18\\
2018-11-12.6    &58434.6&+492	&\textit{Spitzer}	&12.79	$\pm$ 0.25	&11.9	$\pm$ 0.23\\
2019-04-01.5    &58574.5&+632	&\textit{Spitzer}	&13.28	$\pm$ 0.37	&12.14	$\pm$ 0.27\\
2019-11-10.3    &58797.3&+855	&\textit{Spitzer}	&13.43	$\pm$ 0.42	&12.61	$\pm$ 0.32\\ \hline
&&&&3.4 \micron&4.6 \micron\\ \hline
2017-07-18.3    &57952.3&+10	& \textit{WISE}	&11.64	$\pm$ 0.07	&10.94	$\pm$ 0.08\\
2017-12-30.0    &58117.0&+175	& \textit{WISE}	&12.33	$\pm$ 0.12	&11.2	$\pm$ 0.09\\
2018-07-18.6    &58317.6&+375	& \textit{WISE}	&12.65	$\pm$ 0.15	&11.51	$\pm$ 0.11\\
\hline
\end{tabular}
\caption{AT~2017gbl host-subtracted mid-IR photometry.}
\label{tab:mir_mags}
\end{table*}

\begin{table*}
\centering
\begin{tabular}{|c|c|r|c|c|c|c|c|c|} \hline
 UT Date & MJD &Epoch&Telescope & Instrument &  Range & Resolving power\\ \hline
  &  &(days)& &  &($\mu$m)\\ \hline
 2016$-$11$-$18 &57710&-232  &Keck & OSIRIS & 1.96$-$2.30   & \\ 
 2017$-$07$-$10 &57944&+2  &WHT & ISIS& 0.35$-$0.80   & 900\\ 
 2017$-$08$-$30 &57995&+53  &WHT & ISIS& 0.35$-$0.80   & 1800\\ 
 2017$-$09$-$01 &57997&+55  &Gemini N & GNIRS   & 0.85$-$2.50  & 1300-1400\\      
 2017$-$10$-$28 &58054&+112  &IRTF & SpeX  & 0.8-2.40   & \\ 
 2018$-$11$-$23 &58445&+503  &Gemini N & GNIRS   & 0.85$-$2.50   & 1300-1400\\ 
 2019$-$01$-$14 &58497&+555  &NOT & ALFOSC   & 0.32$-$0.96   & 360\\ 
 2019$-$08$-$20 &58715&+773  &NOT & ALFOSC   & 0.57$-$0.86   & 1000\\ 
 \hline

\end{tabular}
\caption{Log of spectroscopic observations of AT~2017gbl or its host IRAS~23436+5257. Resolving powers were taken from the instrument descriptions available online, and were compared against skylines in our data to check for consistency.}
\label{tab:Spectral log}
\end{table*}

\begin{table*}
\centering
\begin{tabular}{|c|c|r|c|r|c|c|c} \hline
UT Date         & MJD&Epoch&Array & Frequency & Total flux density& AT~2017gbl Flux density & AT~2017gbl Luminosity \\ 
&&(days)&&(GHz)&(mJy)&(mJy)&($\times 10^{28}$\,erg\,s$^{-1}$\,Hz$^{-1}$)\\ 
(1)&(2)&(3)&(4)&(5)&(6)&(7)&(8)\\ \hline
1995-03-12  &49788&-8154  & VLA   &  1.4        & $18.70\pm1.85$          &-&- \\ 
2017-07-13  &57947&+5  & AMI   & 15.5        & $5.95\pm0.30$           & $2.35\pm0.37$    &   $5.99\pm0.94$\\
2017-07-29  &57963&+21  & AMI   & 15.5        & $6.88\pm0.35$           & $3.28\pm0.41$    &   $8.36\pm1.05$\\
2017-08-10  &57975&+33  & AMI   & 15.5        & $6.78\pm0.34$           & $3.18\pm0.42$    &   $8.11\pm1.08$\\
2017-08-15  &57980&+38  & VLBA  & 4.4         & $0.93\pm0.06$           & $0.93\pm0.06$    &   $2.37\pm0.15$\\
2017-08-15  &57980&+38  & VLBA  & 7.6         & $1.81\pm0.10$           & $1.81\pm0.10$    &   $4.62\pm0.26$\\
2017-10-15  &58041&+99  & VLBA  & 4.4         & $2.03\pm0.11$           & $2.03\pm0.11$    &   $5.18\pm0.28$\\
2017-10-15  &58041&+99  & VLBA  & 7.6         & $2.70\pm0.14$           & $2.70\pm0.14$    &   $6.89\pm0.36$\\
2018-02-27  &58176&+234 & AMI  & 15.5        & $5.15\pm0.26$           & $1.55\pm0.33$    &   $3.95\pm0.83$\\
2019-02-15  &58529&+587  & EVN  & 4.9         & $0.77\pm0.08$           & $0.77\pm0.08$    &   $1.96\pm0.20$\\
2019-05-02  &58605&+663  & VLA   &  3.0        & $11.63\pm1.95$          & $<6.60$    &   $<16.83$       \\ 
2019-11-03  &58790&+848  & AMI  & 15.5         & $3.56\pm0.18$           &-&-\\
2020-01-30  &58878&+936  & AMI  & 15.5         & $3.64\pm0.19$           &-&-\\

\hline
\end{tabular}
\caption{Properties of the radio observations toward IRAS\,23436$+$5257. Column (1-3) list the observation date in UTC and MJD formats, and the days after discovery, (4) the radio array, (5) the central frequency, (6) the measured total flux density, (7) the flux density of AT~2017gbl after correcting the AMI and the VLA observations for the host contribution (see text), and (8) the corresponding luminosity.}
\label{tab:total_radio_flux}
\end{table*}

\begin{figure}
\centering
        \includegraphics[trim=0 0 0 0, clip, width=0.45\textwidth]{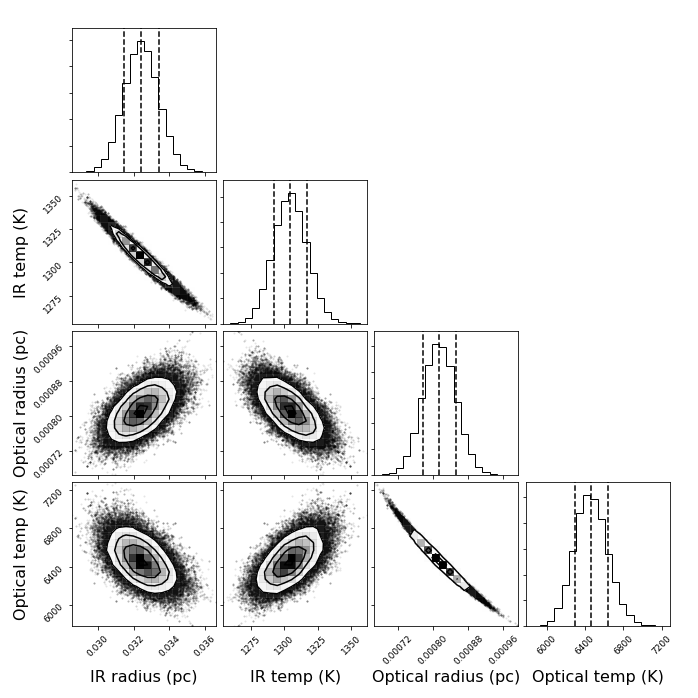} \\
    \caption{Corner plot of of the posterior distributions of the MCMC two components fit to all optical, near-IR and mid-IR fluxes of the epoch at +10 days, see Fig \ref{fig:twocomp_plot_20170718}. The fitted values are indicated by the center lines, with the 1$\sigma$ confidence interval indicated on either side.}
\label{fig:twocomp_plot_20170718_corner}
\end{figure}

\onecolumn
\noindent $^{9}$International Centre for Radio Astronomy Research, Curtin University, GPO Box U1987, Perth, WA 6845, Australia\\
$^{10}$Finnish Centre for Astronomy with ESO (FINCA), University of Turku, Vesilinnantie 5, FI-20014 University of Turku, Finland\\
$^{11}$Aalto University Mets{\"a}hovi Radio Observatory, Mets{\"a}hovintie 114, FI-02540 Kylm{\"a}l{\"a}, Finland\\
$^{12}$Department of Physics, University of Oxford, OX1 3RH, Oxford, United Kingdom\\
$^{13}$Department of Astrophysics/IMAPP, Radboud University, P.O. Box 9010, 6500 GL Nijmegen, the Netherlands\\
$^{14}$SRON, Netherlands Institute for Space Research, Sorbonnelaan 2, 3584 CA Utrecht, The Netherlands\\
$^{15}$School of Physics, O'Brien Centre for Science North, University College Dublin, Belfield, Dublin 4, Ireland\\
$^{16}$Astronomical Observatory, University of Warsaw, Al. Ujaz-dowskie 4, 00-478 Warszawa, Poland\\
$^{17}$Department of Astronomy, Kyoto University, Kitashirakawa-Oiwake-cho, Sakyo-ku, Kyoto 606-8502, Japan\\
$^{18}$Ritter Astrophysical Research Center, University of Toledo, Toledo, OH 43606, USA\\
$^{19}$ARC Centre of Excellence for All Sky Astrophysics in 3 Dimensions (ASTRO 3D)\\
$^{20}$INAF Osservatorio Astronomico di Padova, Vicolo dell'Osservatorio 5, 35122 Padova, Italy\\
$^{21}$Millennium Institute of Astrophysics (MAS), Nuncio Monse\~nor Sotero Sanz 100, Providencia, Santiago, Chile\\
$^{22}$Departamento de Ciencias Fisicas, Universidad Andres Bello, Fernandez Concha 700, Las Condes, Santiago, Chile\\
$^{23}$Department of Physics, Florida State University, 77 Chieftan Way, Tallahassee, FL 32306, USA\\
$^{24}$Space Research Institute of the Russian Academy of Sciences, Profsoyuznaya Str. 84/32, Moscow 117997, Russia\\
$^{25}$Department of Physics and Astronomy, 4129 Frederick Reines Hall, University of California, Irvine, CA 92697, USA\\
$^{26}$Institute of Astronomy, University of Cambridge, Madingley Road, Cambridge CB3 0HA, UK

\bsp	
\label{lastpage}
\end{document}